\documentclass{emulateapj}
\usepackage{natbib}
\newcommand{\tx}[1]{\textrm{#1}}

\newcommand{\kms}{km~$\tx{s}^{-1}$}

\slugcomment{ApJ, in prep.}
\shorttitle{Cluster SN rate at $z\sim0.1$}
\shortauthors{Sand et al.}

\begin{document}
 \title{The Multi-Epoch Nearby  Cluster Survey: type Ia supernova rate measurement in $z\sim0.1$ clusters and the late-time delay time distribution}

\author{David J. Sand,$\!$\altaffilmark{1,2,3}  Melissa L. Graham,$\!$\altaffilmark{2,3} Chris Bildfell, $\!$\altaffilmark{4} Dennis Zaritsky, $\!$\altaffilmark{5} Chris Pritchet, $\!$\altaffilmark{4} Henk Hoekstra, $\!$\altaffilmark{6} Dennis W. Just, $\!$\altaffilmark{5} St\'{e}phane Herbert-Fort, $\!$\altaffilmark{5} Suresh Sivanandam, $\!$\altaffilmark{7,8} Ryan Foley, $\!$\altaffilmark{9,10} Andisheh Mahdavi $\!$\altaffilmark{11}}  \email{dsand@lcogt.net}

\begin{abstract}
We describe the Multi-Epoch Nearby Cluster Survey (MENeaCS), designed to measure the cluster Type Ia supernova (SN Ia) rate in a sample of 57 X-ray selected galaxy clusters, with redshifts of $0.05 < z < 0.15$.  Utilizing our real time analysis pipeline, we spectroscopically confirmed twenty-three cluster SN Ia, four of which were intracluster events.  Using our deep CFHT/Megacam imaging, we measured total stellar luminosities in each of our galaxy clusters, and we performed detailed supernova detection efficiency simulations.  Bringing these ingredients together, we measure an overall cluster SN Ia rate within $R_{200}$ (1 Mpc) of $0.042^{+0.012}_{-0.010}$$^{+0.010}_{-0.008}$ SNuM ($0.049^{+0.016}_{-0.014}$$^{+0.005}_{-0.004}$ SNuM) and a SN Ia rate within red sequence galaxies of $0.041^{+0.015}_{-0.015}$$^{+0.005}_{-0.010}$ SNuM ($0.041^{+0.019}_{-0.015}$$^{+0.005}_{-0.004}$ SNuM).  The red sequence SN Ia rate is consistent with published rates in early type/elliptical galaxies in the `field'.  Using our red sequence SN Ia rate, and other cluster SNe measurements in early type galaxies up to $z\sim1$, we derive the late time ($>2$ Gyr) delay time distribution (DTD) of SN Ia assuming a cluster early type galaxy star formation epoch of $z_f=3$.  Assuming a power law form for the DTD, $\Psi(t)\propto t^s$, we find $s=-1.62\pm0.54$.  This result is consistent with predictions for the double degenerate SN Ia progenitor scenario ($s\sim-1$), and is also in line with recent calculations for the double detonation explosion mechanism ($s\sim-2$).  The most recent calculations of the single degenerate scenario delay time distribution predicts an order of magnitude drop off in SN Ia rate $\sim$6-7 Gyr after stellar formation, and the observed cluster rates cannot rule this out.

\end{abstract}
\keywords{later}

\altaffiltext{1}{Harvard Center for Astrophysics and Las Cumbres Observatory Global Telescope Network Fellow}
\altaffiltext{2}{Las Cumbres Observatory Global Telescope Network, 6740
Cortona Drive, Suite 102, Santa Barbara, CA 93117, USA}
\altaffiltext{3}{Department of Physics, Broida Hall, University of
California, Santa Barbara, CA 93106, USA}
\altaffiltext{4}{Department of Physics and Astronomy, University of
Victoria, PO Box 3055, STN CSC, Victoria BC V8W 3P6, Canada}
\altaffiltext{5}{Steward Observatory, University of Arizona, Tucson AZ 85721}
\altaffiltext{6}{Leiden Observatory, Leiden University, Niels Bohrweg 2, NL-2333 CA Leiden,
The Netherlands}
\altaffiltext{7}{Dunlap Institute for Astronomy and Astrophysics, 50 St. George St., Toronto, ON Canada M5S 3H4}
\altaffiltext{8}{Dunlap Fellow}
\altaffiltext{9}{Harvard-Smithsonian Center for Astrophysics, 60 Garden Street, Cambridge MA 02138}
\altaffiltext{10}{Clay Fellow}
\altaffiltext{11}{Department of Physics and Astronomy, San Francisco State University, San Francisco CA 94132}
\section{Introduction}

Type Ia supernovae (SNe Ia) play a central role in astrophysics because of their use as standard candles for cosmography \citep[e.g.][]{Riess98,Perlmutter99}, as well as being a prime source for the chemical enrichment of the Universe.  While it is widely accepted that SNe Ia are the thermonuclear explosions of carbon oxygen white dwarfs near the Chandrasekhar mass \citep[e.g.][]{Hillebrandt00}, it is still unclear by what mechanism the white dwarf accretes the necessary mass, or if there are multiple pathways to a SN Ia explosion \citep[for a recent review, see][]{Maozreview}.  This knowledge about the SN~Ia progenitor, and by extension the types of stellar populations that house them, will aid next generation cosmological measurements since any systematic change in SN~Ia properties as a function of stellar population age, mass or star formation history will be reflected in the changing galaxy population surveyed at higher redshift.  An understanding of the progenitor(s) of SN~Ia will similarly inform the location and epoch of iron enrichment.

There are currently three main progenitor/explosion scenarios being actively studied.  The first is the single degenerate (SD) scenario, where a single white dwarf accretes material from a companion, overfilling its Roche lobe until it reaches the  Chandrasekhar mass \citep{Whelan73,Nomoto82}.  The double degenerate (DD) scenario postulates the merger of two carbon oxygen white dwarfs, with a total mass which exceeds the Chandrasekhar limit, through gravitational inspiral and subsequent carbon ignition \citep{Iben84,Webbink84}.    Finally there is the double detonation scenario, a specific variant of the SD scenario, in which an outer He-burning shell, being fed from a companion, undergoes several flashes \citep[e.g.][]{Bildsten07}, ultimately leading to a final event which ignites the underlying carbon oxygen white dwarf before the Chandrasekhar mass is reached \citep[e.g][]{Woosley94} -- a sub-Chandrasekhar mass event.  

One route to distinguishing between the various SN Ia progenitor scenarios is via the delay time distribution (DTD).  The DTD is the expected SN Ia rate as a function of time following a burst of star formation, and each potential progenitor will have a different DTD.  For instance, in the SD scenario, the DTD is driven by the main-sequence lifetime of the secondary star, while in the DD scenario the delay time is associated with the time necessary for gravitational wave emission to cause the orbit to decay \citep{Greggio05}.

One approach to recovering the DTD has been through comparing the SN Ia rate in stellar populations of different characteristic ages.  By teasing out the relative SN rates in galaxies ranging from actively star forming to long dead, several pioneering studies found evidence for both `prompt' SNe Ia which explode $\lesssim$500 Myr after star formation, and `delayed' SNe Ia which explode $\sim 1-10$ Gyr after explosion \citep{Scannapieco05,Mannucci05,Mannucci06,Sullivan06}.  This initially suggested that there may be two separate populations of SNe Ia (the so-called `A+B' model), but most theoretical DTD functions predict SN~Ia occurring over a broad range of ages in any case without invoking multiple populations \citep{Greggio08}.  Much recent observational  work has gone into deriving the SN Ia DTD through a variety of methods \citep[e.g.][]{Totani08,Brandt10,Horiuchi10,MaozBad,Maoz11,Graur11}. Theoretically, both analytic \citep[e.g.][]{Greggio05} and  binary population synthesis model results \citep[e.g.][]{Yungelson00,Mennekens10,Ruiter09,Ruiter11} have been used to predict the functional form of the DTD for different progenitor models.

A relatively clean way of probing the DTD, at least for stellar populations $\gtrsim 1$ Gyr old, is by measuring the cluster SN Ia rate from low to high ($z\sim1.5$) redshifts.  Cluster red sequence (or, more loosely, early type) galaxies have colors and spectra which are consistent with having formed at high redshift ($z\sim3$) in a single burst of star formation, followed by passive evolution \citep[e.g.][]{Stanford98,Eisenhardt08}.  Given this, the observed SN Ia rate in cluster red sequence galaxies as a function of time after this initial burst of star formation is a direct measure of the shape of the DTD, without any need to model the star formation history of a complex galaxy population.  

The cluster SN Ia rate has been measured from $0 \lesssim z \lesssim1.5$, although the rate uncertainties continue to be dominated by small number statistics.  At $z\lesssim0.2$, several groups measure consistent cluster rates, but have noticed the SN rate among cluster ellipticals is a factor of $\sim$2-3 higher than in the field at the $\sim1-2\sigma$ level \citep{Sharon07,Mannucci08,Dilday10}.  If true, this rate enhancement may be due to a sprinkling of recent star formation in cluster ellipticals, since the SN~Ia rate increases dramatically in the presence of star formation \citep[e.g.][]{Sullivan06}.  Cluster rate measurements at moderate to high redshift are sparser, and are based on $\lesssim10$ SN~Ia, not always with spectroscopic confirmation.  These pioneering measurements include Hubble Space Telescope (HST) reimaging of clusters at $0.5 < z < 0.9$ by \citet{Galyam02,Sharon10}, and results taken from a subset of the Supernova Legacy Survey at $z\sim0.5$ \citep{Graham08}.  The HST Cluster Supernova Survey has calculated the SN Ia rate in clusters at $0.9 < z < 1.45$ \citep{Barbary10}.  Putting all of the above results together lead to constraints on the late time DTD in SN~Ia, under the assumption that the bulk of cluster stars formed at high redshift and have evolved passively since \citep{Barbary10,Maoz10clus}.  Note that these earlier works used SN Ia rates over the entire cluster galaxy population in their DTD measurements, rather than focusing in on the red sequence/early type population.

In our first paper on the Multi-Epoch Nearby Cluster Survey (MENeaCS) we described our intracluster SN Ia, and used their relative numbers to constrain the intracluster stellar mass fraction in our cluster sample \citep{Sand11}.
In this paper, we describe a new measurement of the cluster SN Ia rate based on 23 SNe Ia found in 57 galaxy clusters at $0.05 < z < 0.15$ during the course of our $\sim$2 year survey.  From these measurements and others in the literature we constrain the late-time ($>2$ Gyr) SN~Ia DTD, specifically utilizing those measurements of the SN Ia rate in red sequence or early type galaxies.  

A plan of the paper follows.  In \S~\ref{sec:techdesign} we discuss the MENeaCS survey strategy and cluster sample.  In \S~\ref{sec:redux} we describe our real time image reduction pipeline for discovering and characterizing transient candidates, along with our final products derived from data taken over the entire survey.  Section 4 details our spectroscopic observations to confirm cluster SNe Ia.  In \S~\ref{sec:clusterlum} we calculate the luminosity and stellar masses of the MENeaCS cluster sample and \S~\ref{sec:SNhost} details our host galaxy photometry.  Next, \S~\ref{sec:deteffs} details our SN detection efficiency calculations.  Finally, in \S~\ref{sec:rates} we put all of the elements together and calculate the MENeaCS cluster SN Ia rate, along with our SN Ia rate associated with red sequence galaxies.  In \S~9 we compare our results with both cluster and field SN Ia rates in the literature.  We calculate the late time DTD from our own cluster red sequence SN Ia rate, and similar results in the literature, in \S~\ref{sec:DTD} in order to constrain the SN Ia progenitor.
We conclude and discuss future directions in \S~\ref{sec:future}.  Unless otherwise noted, we assume a flat $\Lambda$CDM cosmology with $\Omega_{m}$=0.3, $\Omega_{\Lambda}$=0.7 and a Hubble constant of $H_{0}$ = 70 km s$^{-1}$ Mpc$^{-1}$.

\section{MENeaCS Survey Design \& Goals}\label{sec:techdesign}

The MENeaCS survey was designed with two primary scientific goals in
mind: measuring the SN~Ia rate in $z\sim0.1$ galaxy clusters and
utilizing galaxy-galaxy lensing to measure the dark matter radius of
early-type galaxies as a function of clustercentric distance.   We detail our cluster sample and observing strategy here, with an emphasis on those aspects of the program most relevant for our SN science goals.

\subsection{The Cluster Sample}\label{sec:clussample}

To accomplish both the SN and galaxy-galaxy lensing goals of
MENeaCS, we chose the brightest X-ray clusters accessible from the CFHT
at $0.05 < z < 0.15$.  The logic behind this choice is simple: the most X-ray luminous clusters tend to have the largest stellar mass, and will thus have the most SNe and provide a large sample of galaxy lenses.  The cluster sample is detailed in
Table~\ref{table:clustertable} and consists of 57 clusters.  Redshifts are those reported in the
NASA/IPAC Extragalactic Database (NED), and the X-ray luminosities are
those reported directly from the BAX online cluster database, which
attempts to homogenize luminosities in the literature and uses a
$H_{0}$=50 \kms Mpc$^{-1}$ and $\Omega_{m}$=1 cosmology.  We refer the
reader to the BAX web site for
details\footnote{http://bax.ast.obs-mip.fr/}.

Since it will be useful to know the approximate mass and radius of our cluster sample when calculating SN rates and trends, we also list $M_{200}$ and $R_{200}$ in Table~\ref{table:clustertable} using the bisector regression $L_{X}-M_{200}$ scaling relation of \citet{Reiprich02}. While there are more up to date measurements of the slope and normalization of the $L_{X}-M $ relation \citep[e.g.][]{Stanek06,Vikhlinin09,Mantz10,Hoekstra11}, our chosen relation is consistent with most literature values at the $\sim$1$\sigma$ level, and we seek consistency with our previous results on the IC SNe in the MENeaCS survey \citep{Sand11}.  All but six of the MENeaCS clusters can be probed out to $R_{200}$ given the CFHT/Megacam field of view, and even for these clusters, we were able to cover out to $\sim 0.9 R_{200}$.  Note that a factor of two systematic uncertainty in our cluster masses will result in a $\sim$25\% systematic in our value of $R_{200}$.  We discuss the effect that such systematic uncertainties in our derived $R_{200}$ value have on our SN Ia rate measurements in \S~\ref{sec:rates}.

\subsection{Observing Strategy}

\begin{figure}
\begin{center}
\mbox{\epsfysize=9.0cm \epsfbox{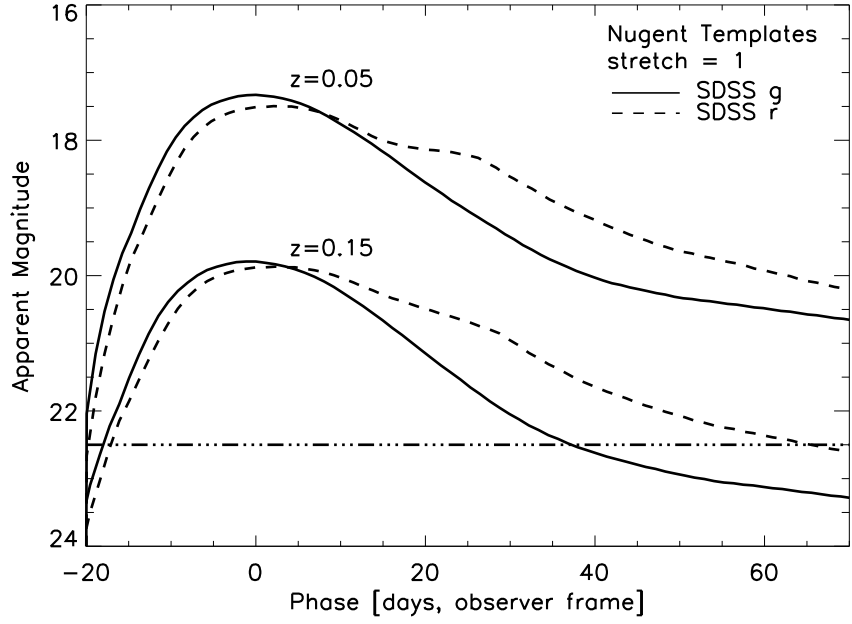} 
}
\caption{ Template SN ia light curves at $z=0.05$ and $z=0.15$ \citep{Nugent02}, the redshift range of the MENeaCS survey.  We have a self-imposed detection limit for SNe of $g\sim22.5$ mag to accommodate efficient spectroscopic followup, as indicated by the dot-dashed line in the figure.  We are capable of spectroscopically identifying our cluster SNe Ia roughly two or more months after explosion.  
\label{fig:lc_example}}
\end{center}
\end{figure}

Our goal was to obtain $g'$ and $r'$ images (with $2\times120$s per band) of each of our clusters once every month during the
two-week queue run of CFHT/Megacam, typically with an image quality of $\lesssim$1\farcs0, in order to simultaneously fulfill our weak lensing requirements.  These individual epoch images were used for SN discovery (see \S~\ref{sec:imagesub}), while the ultimate deep stacks over all epochs are also being used for programs on cluster physics and gravitational lensing.  This strategy allows us to discover and follow SN
Ia at the redshifts of our clusters for several months (see e.g. the
template light curves in Figure~\ref{fig:lc_example}, spanning our redshift range), and it also allows multiple opportunities for spectroscopic confirmation.

We had a dedicated program for SN followup and confirmation.  When possible, follow-up
photometry of our cluster fields was obtained with the 90Prime imager \citep{90prime}
on the Steward Observatory 2.3m Bok telescope, for which we were generally allocated
several nights per month.  These observations were primarily for rejecting slow-moving objects identified as possible intracluster SN candidates, and for filling in SN light curves to better inform our spectroscopic priorities, using techniques similar to those described in \citet{SandSN08}.  Since these observations were used solely for these purposes, they are not discussed further in this work.  Spectroscopic follow up of SN candidates
 was primarily done at the MMT, either with the Blue Channel
Spectrograph or Hectospec (which afforded a limited target of opportunity mode, since Hectospec is queue scheduled).  We typically were allocated $\sim$1 night per month for spectroscopic follow up during MENeaCS.  Additionally, target of opportunity observations with
Gemini North were used for intracluster supernova candidates; those are presented in \citet{Sand11}.

\section{Imaging Reduction \& Analysis}\label{sec:redux}

Our primary data reduction goals were geared toward the nearly real time detection and calibration of transients, to be discussed in \S~3.1--3.3 below, the best of which could then be followed up spectroscopically.  Additionally, we maintained a running deep stack image during the course of the survey of each cluster field specifically for identifying host galaxies and our hostless, intracluster events (\S~3.4).  Ultimately, we made a set of final deep stack images for each cluster field based on the calibrated data from the CFHT (\S~3.5) for our final host galaxy photometry.

\subsection{Real Time Imaging Data Reduction}\label{RTdata}

As mentioned, our primary SN discovery images came from the square-degree CFHT/Megacam camera \citep{cfhtmegacam}. Megacam has one $\sim2$ week queue run per month, and we aimed for one epoch every queue run. Very occasionally, two epochs were taken in one queue run to make up for epochs missed due to weather, i.e. when the target field was setting. Each epoch consists of $2\times 120$s images in $g'$ and $r'$, dithered by $\sim 5\arcmin$ to fill in the chip gaps while ensuring the brightest cluster galaxy (BCG) is centrally positioned on one of Megacam's central chips. During the first epoch of observation of a cluster, we took $4\times 120$ s $g'$ and $r'$ images in order to have sufficient depth to determine the initial astrometric solution. This first, slightly deeper epoch was used as the reference template for SN detection throughout the survey.  No attempt was made, after the survey was complete, to use an optimal template or final deep stack image to recover SNe not found during the course of the survey.

Bias subtracted and flat fielded CFHT Megacam images
were provided by the CFHT-developed Elixir data reduction system
\citep{elixir} within $\sim$24 hours of acquisition.
The final Elixir-processed data for a given queue run used
master flat fields, but these are only available {\it after} the queue
run -- too slow for our needs.
We thus use the real-time data for SN detection, preliminary photometry,
and to build interim ``deep stacks'' during the survey, and the
Elixir-processed data for our final deepstacks and photometric calibrations.

For the real-time data, we generated object catalogs with {\sc
SExtractor} \citep{sexbib}
and determined the astrometric solution with {\sc SCAMP}
\citep{scampbib}, matching objects with the 2MASS catalog. The solution
was constantly refined, as every new epoch used all previous images.
We combined each epoch's data with {\sc SWarp}\footnote{version 2.15.7; http://terapix.iap.fr/soft/swarp}, using the {\sc lanczos3}
interpolation function for image resampling.

\subsection{Image Subtraction and Transient Detection}\label{sec:imagesub}

As described above, each epoch results in one $g'$ and one $r'$
image, and our slightly deeper first epoch images are used as the
reference template. Briefly, the image with a sharper point spread
function (PSF) was aligned, PSF-matched and flux-scaled to that of the
poorer quality image, and the reference is subtracted from the new
epoch. Sources on the difference image are detected using the IRAF task
{\sc daofind}. These sources are automatically culled to reject those
which have declined/disappeared since the reference;
sources which are not round; sources in close proximity to a
negative source, indicating a moving object; and sources in groups or in
rows/columns with multiple sources, indicating association with
e.g. diffraction spikes or bad pixels. To make further automatic culls, we
then measure source fluxes in four apertures of radius 2, 4, 6, and 8 pixels,
an integrated flux from the best fit PSF, and its $\chi^{2}$ representing
goodness-of-fit. Sources are rejected if they are not detected
simultaneously in both $g'$ and $r'$; have $g' \geq 23.5$; do not have a
constantly increasing flux in successively larger apertures; or have a $\chi^2>50$. 

After these automatic culls, triplet images are created for
the remaining transient candidates, showing the target image, template
image and the difference. Visual review of the $g'$ and $r'$ triplets,
and rejection only of those objects which are clearly 
spurious artifacts, leads to the final candidate list.
These transient candidates were then roughly prioritized according to their likelihood of being cluster SN Ia to facilitate spectroscopic followup.  We discuss this process in \S~\ref{sec:spectarg}.

\subsection{Real Time Photometric Calibration}\label{sec:photcal}

Calibrated photometry for SN candidates in real time was essential
information for our photometric and spectroscopic follow-up campaigns.
Queue run zeropoints were derived using all
cluster fields covered by SDSS which were imaged in photometric
conditions during that queue run. For real-time analysis, photometric
epochs were simply identified by visual inspection of the CFHT SkyProbe
plots of attenuation over time for the night.
For epochs taken during non-photometric conditions, a correction for
attenuation was derived by comparing star magnitudes to those from a frame taken under photometric conditions.  Such real-time
photometric analysis was only moderately precise, good to within $\sim$0.1--0.2
magnitudes, but this was sufficient for comparison with our imaging
follow-up at the Bok telescope, and vetting and exposure time estimation
for spectroscopic follow-up.

 \subsection{Real Time Host Identification}\label{sec:hostid}

As mentioned in \S~\ref{RTdata},  we maintained interim deep-stack images of
each cluster and source catalogs to facilitate quick identification of
SN host galaxies and likely cluster members.  
An interim host galaxy was assigned to each SN candidate using the method of the SNLS \citep{Sullivan06}, which uses a host-SN elliptical separation normalized by the candidate host galaxy size to assign hosts -- the dimensionless $R$ parameter described in Equation (1) of \citet{Sullivan06}.  For details, we refer the reader to the original work of \citet{Sullivan06}, and \citet{Sand11}, who corrected a typographical error in the original version.  Four hostless, intracluster SNe, defined as those cluster SN Ia with no host within $R$=5, were also discovered and presented in \citet{Sand11}.

\subsection{Final deep stack images}\label{sec:deepstacks}

Beyond the running deep stack images constructed during the survey for on-the-fly host identification, we have also made a uniform set of deep stack images from our entire CFHT/Megacam data set.  These stacks will be used for host photometry and colors,  identification of the red sequence in each of our clusters, and for overall measurement of our cluster luminosities.  Each of these measurements are key for determining and understanding our SN Ia rate.

For these final deep stacks we utilize ``Elixir" processed Megacam images
\citep{elixir} from the Canadian Astronomy Data Center\footnote{See
http://cadcwww.dao.nrc.ca/}.  The seeing of each input image was
determined automatically with an initial run of the automated source
extraction package {\sc SExtractor}.  Any images with
$>1\farcs0$ seeing were not included, although this amounted to only
$\sim$10\% of the total.  The images were resampled to a common grid with {\sc SCAMP} and median-combined with SWarp, as discussed in \S~3.1.  These final deep stack images were calibrated directly to the SDSS filter system, including both a zeropoint and color term.  Cluster fields with no SDSS overlap were calibrated from photometric nights on which an SDSS cluster was observed.

 For those fields that hosted cluster SNe, we have also made similar deep stack images including only epochs {\it before} supernova discovery, specifically to measure uncontaminated host colors.  These SN-free stacks were generated and calibrated as described above, with our SN host photometry presented in \S~\ref{sec:SNhost}.

\section{Supernova spectroscopy}

In this section, we detail our methodology for obtaining SN spectroscopy, and ultimately identifying cluster SN Ia -- a critical ingredient for calculating the cluster SN Ia rate.

\subsection{Spectroscopic prioritization}\label{sec:spectarg}

For two weeks in any given month during the MENeaCS program, we would obtain queue observations of our cluster fields from the CFHT and quickly identify viable cluster SN candidates.   Nearly all of our spectroscopy was scheduled classically, with roughly one night allocated per month.  Often this night would be allocated during the middle of the Megacam queue run, so that a potentially good cluster SN Ia target might have to wait a number of weeks before being spectroscopically confirmed. Weather or instrument problems could also bump the spectroscopic confirmation of a SN to a subsequent month.

Nonetheless, our spectroscopic campaign was relatively robust to these scheduling and weather difficulties due to the relative brightness of our SNe (Figure~\ref{fig:lc_example}), and the relatively small number of viable cluster SN Ia targets.  As discussed in \citet{Sand11}, we focused our spectroscopic follow up on likely cluster SN Ia, e.g. a newly discovered SN candidate had to be consistent with a SN Ia within $\lesssim$30 days of explosion at the cluster redshift.  This rough magnitude and color cut changed depending on the cluster redshift but ranged from $-0.5\lesssim g-r \lesssim 0.8$ and $17.5 \lesssim g \lesssim 22.5$.  A good spectroscopic target which was not observed for weather or other reasons would be pursued in subsequent spectroscopy runs.  Preference was given for good cluster SN candidates with a clustercentric distance of $R\lesssim1.2R_{200}$, although this was not a hard rule, especially if no other viable candidates were available.  We do note that any SN candidate that appeared to be hostless was given the highest spectroscopic priority regardless of color or brightness, and was generally sent to Gemini as a target of opportunity \citep{Sand11}.

We have also utilized the NASA/IPAC Extragalactic Database (NED) to
identify known variable sources and objects with redshifts in the
literature.  This allows us, for example, to cull known variable
objects (e.g.~active galactic nuclei and variable stars) and to
identify redshifts of host galaxies.  For our
purposes, we ignore SN candidates that occur in host galaxies with
redshifts inconsistent with cluster membership. 

During a typical CFHT queue run, we would identify $\sim$3-4 high priority cluster SN Ia candidates, which could be easily followed up with a night of 6m-class spectroscopy, with time to spare for other transient spectroscopy.  There was a small fraction of time during the MENeaCS survey ($\sim$9\%) where we had successive weather and instrument related problems and were not able to perform spectroscopy on viable cluster SN Ia candidates (see \S~4.1 of Sand et al. 2011 for a discussion on this).  We factor this into our SN rate calculations in \S~8.

\subsection{Observations}

Spectroscopic identification of our viable cluster SN Ia candidates is critical for confirming their nature.  We focus on our cluster SN Ia here, while our cluster core collapse SNe are presented in a companion paper (Graham et al. in prep).    

Spectroscopic
observations were obtained with one of three telescope/instrument
combinations: Gemini North and the Gemini Multi Object Spectrograph  \citep[GMOS;][]{Hook04}, the MMT and its Blue Channel
Spectrograph \citep[BCS;][]{BCS}, and the MMT with Hectospec \citep{Hectospec}.  For our GMOS observations (obtained solely for our IC SNe candidates) we used the R400 grating centered at 680 nm, along with the GG455 order blocking filter; see Sand et al. (2011) for the details of these observations.  With BCS, the 300 line grating with a central wavelength of 580 nm was always used.  Hectospec was used with the 270 line grating, giving wavelength coverage between $\sim$360-800 nm, with one exception.  For Abell2670\_9\_08\_1 (Figure~\ref{fig:SN_weirdhecto}), the Hectospec 600 line grating was used due to scheduling constraints, giving wavelength coverage between $\sim$550-750 nm.  We discuss the spectroscopic identification of this object in more detail in the next subsection.
Standard CCD processing, one dimensional spectrum extraction and
calibration were accomplished with {\sc IRAF} for our GMOS and BCS spectra.  The Hectospec spectra were pipeline processed at the Harvard-Smithsonian Center for Astrophysics.  

\subsection{Supernova Identification}\label{sec:SNID}

\begin{figure}
\begin{center}
\mbox{ 
\epsfysize=6.0cm \epsfbox{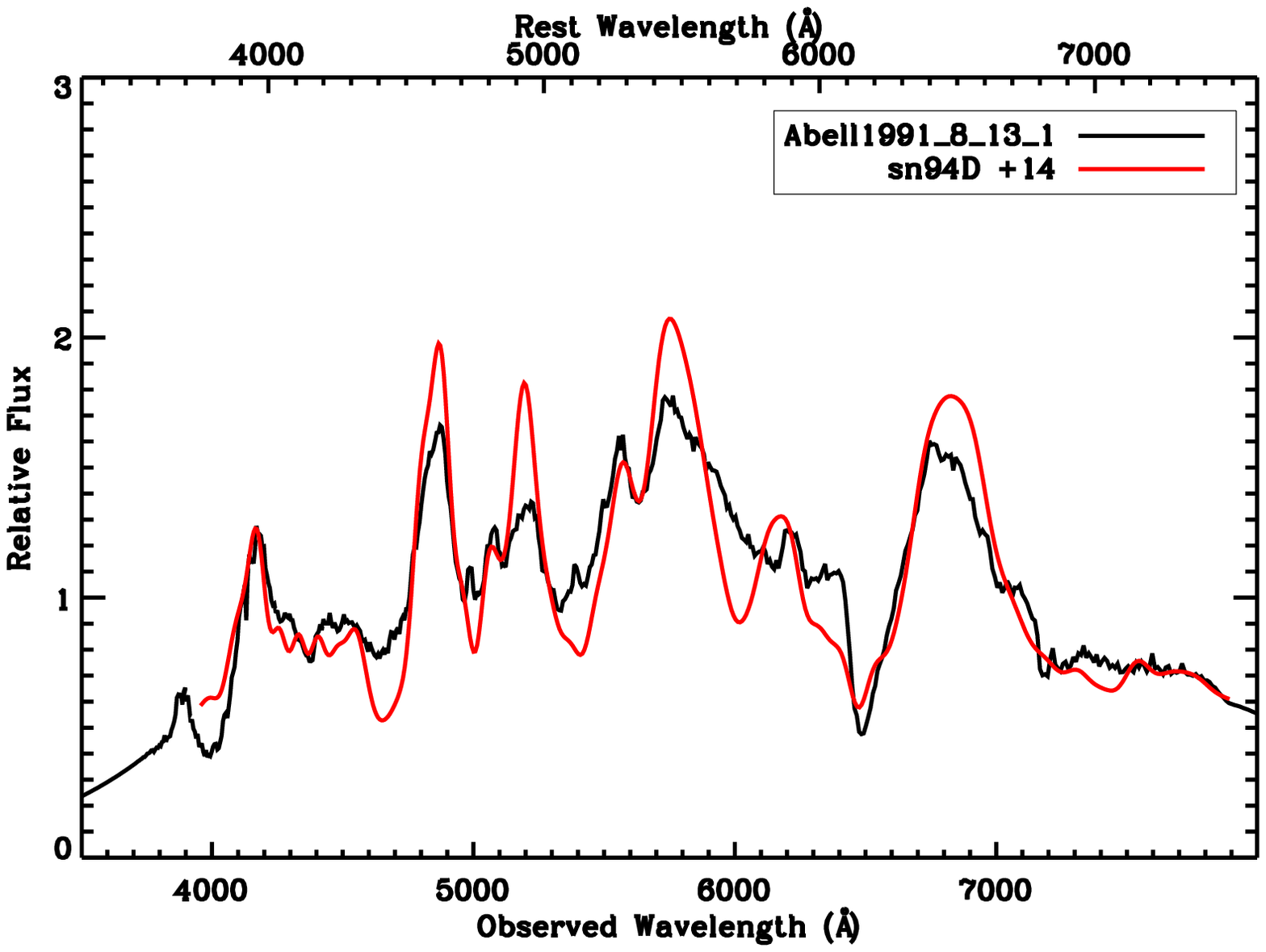} 
\epsfysize=6.0cm \epsfbox{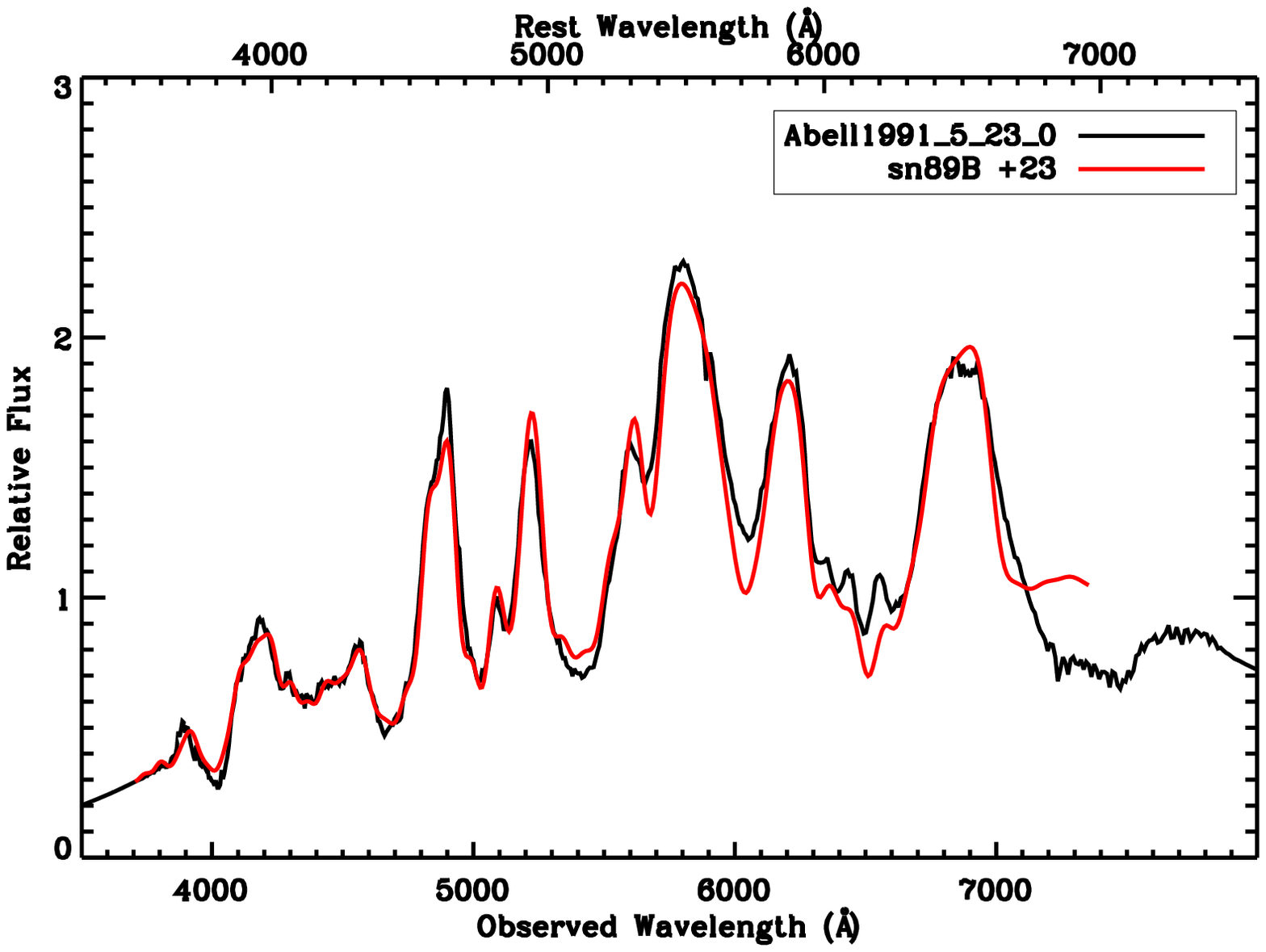} 
}
\mbox{ 
\epsfysize=6.0cm \epsfbox{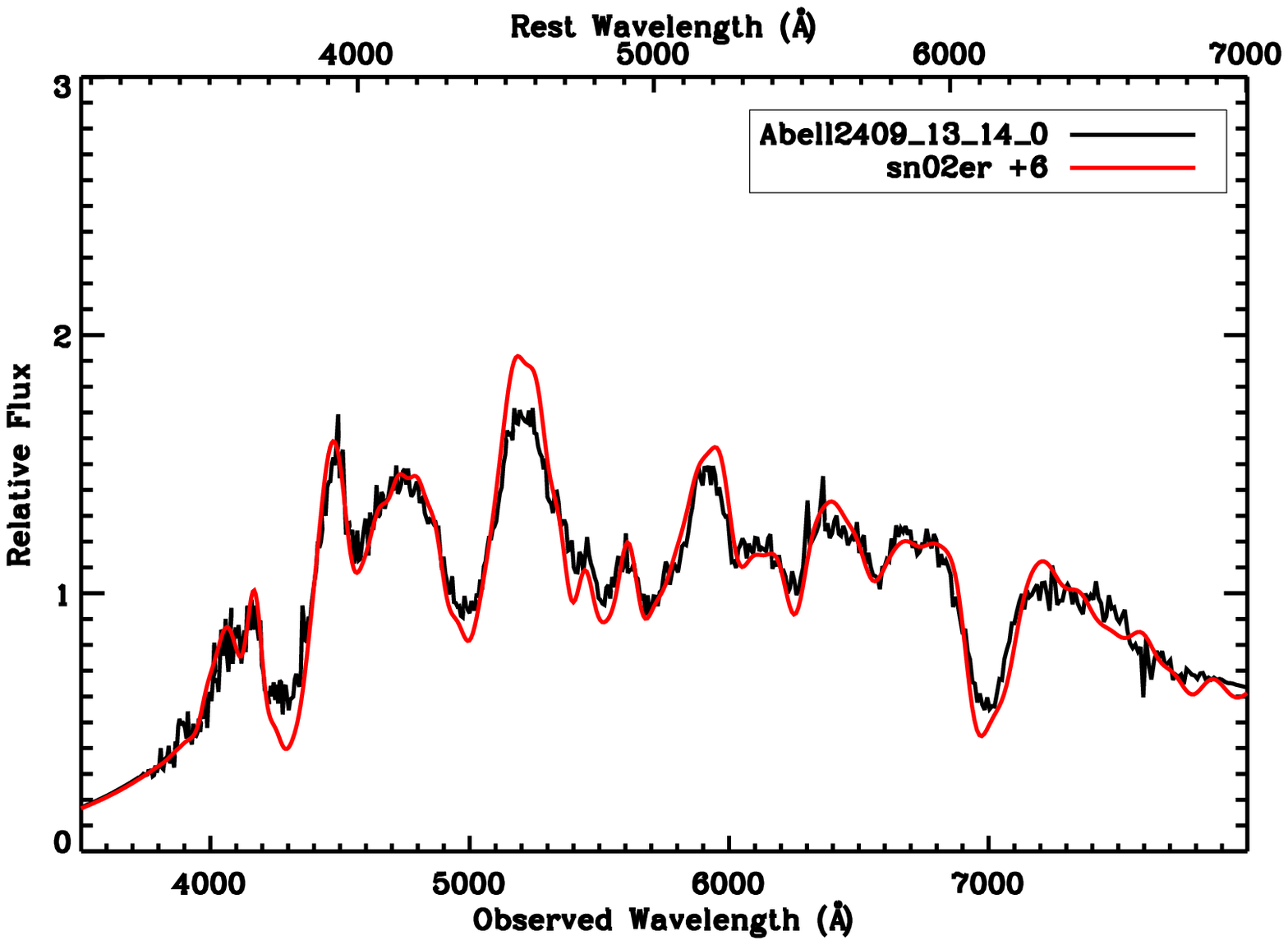} 
\epsfysize=6.0cm \epsfbox{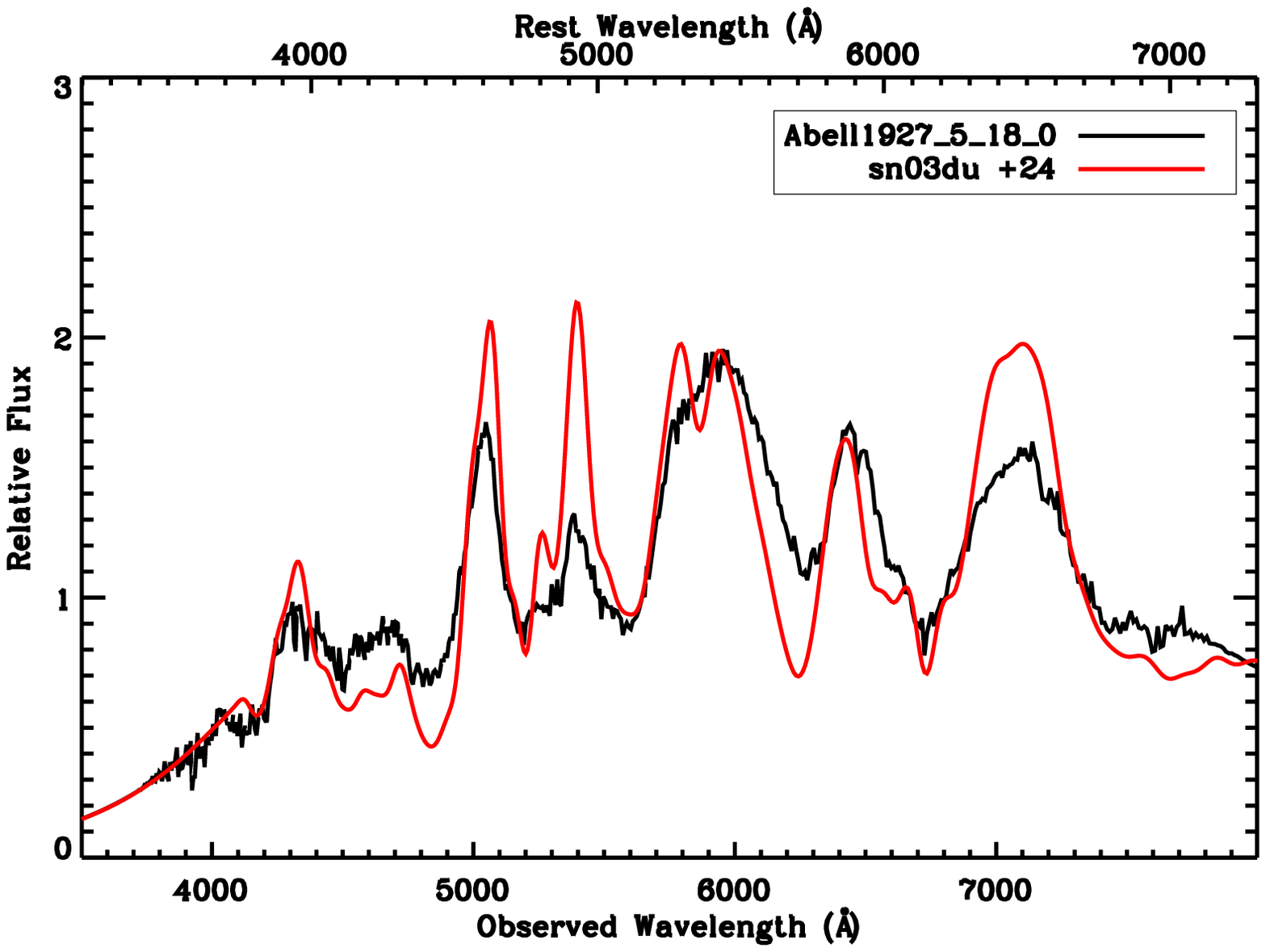} 
}
\caption{Spectra of cluster SN Ia, along with the best-fitting template SN as determined by SNID.  See \S~\ref{sec:SNID} for details of the spectroscopic typing and Table~\ref{table:SNItable} for results.  SNID removes a pseudo-continuum from each spectrum before fitting to minimize host galaxy contamination.  Top left -- Abell1991\_8\_13\_1, a cluster normal SN Ia at
$z=0.0526$. Top right -- Abell1991\_5\_23\_0, a cluster normal SN Ia at $z=0.0580$.
Bottom left -- Abell2409\_13\_14\_0 is a normal cluster SN Ia at $z=0.1458$.
Bottom right -- Abell1927\_5\_18\_0 is a normal cluster SN Ia at $z=0.0945$. 
\label{fig:SNset1}}
\end{center}
\end{figure}

\begin{figure}
\begin{center}
\mbox{ 
\epsfysize=6.0cm \epsfbox{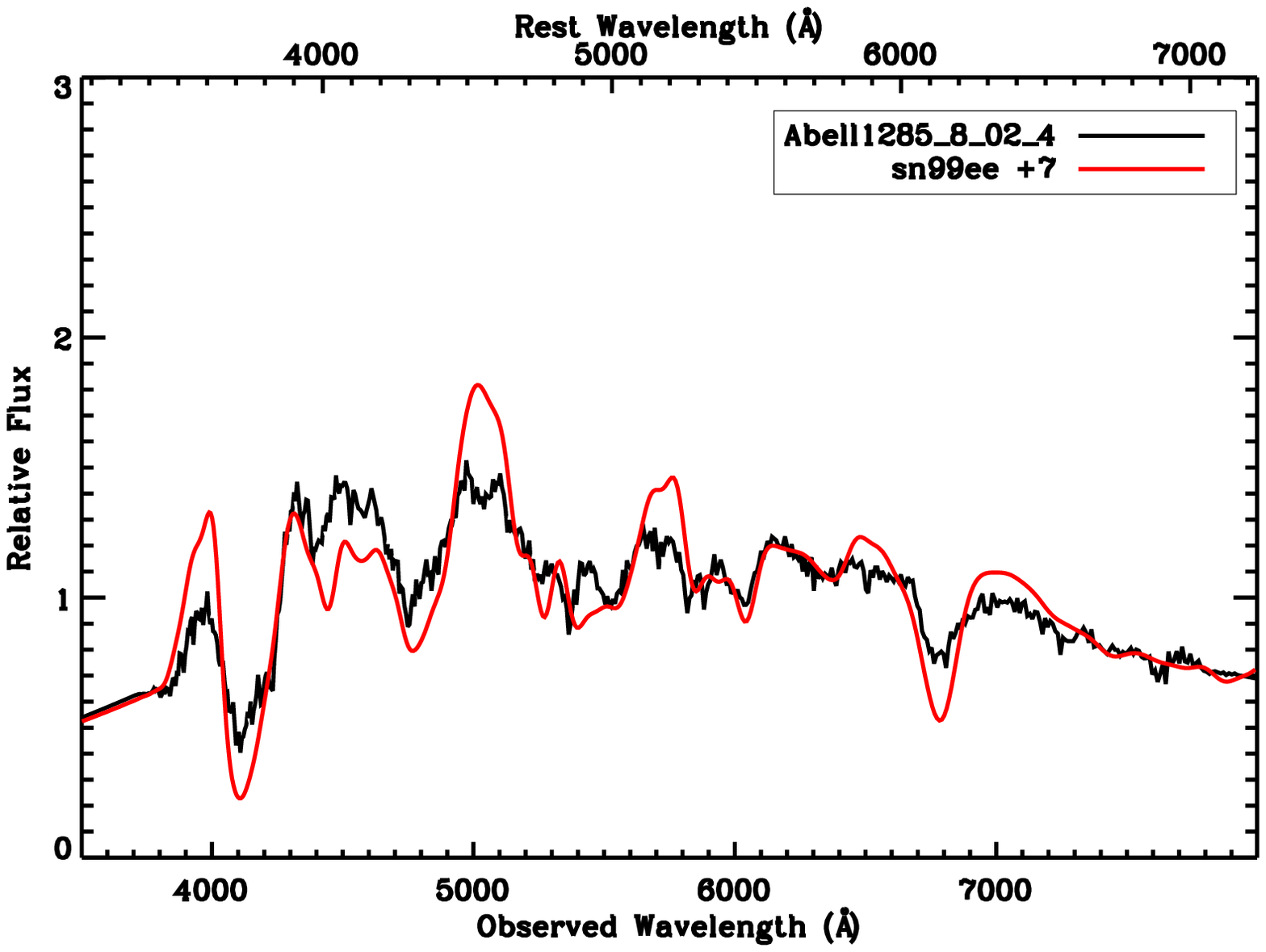} 
\epsfysize=6.0cm \epsfbox{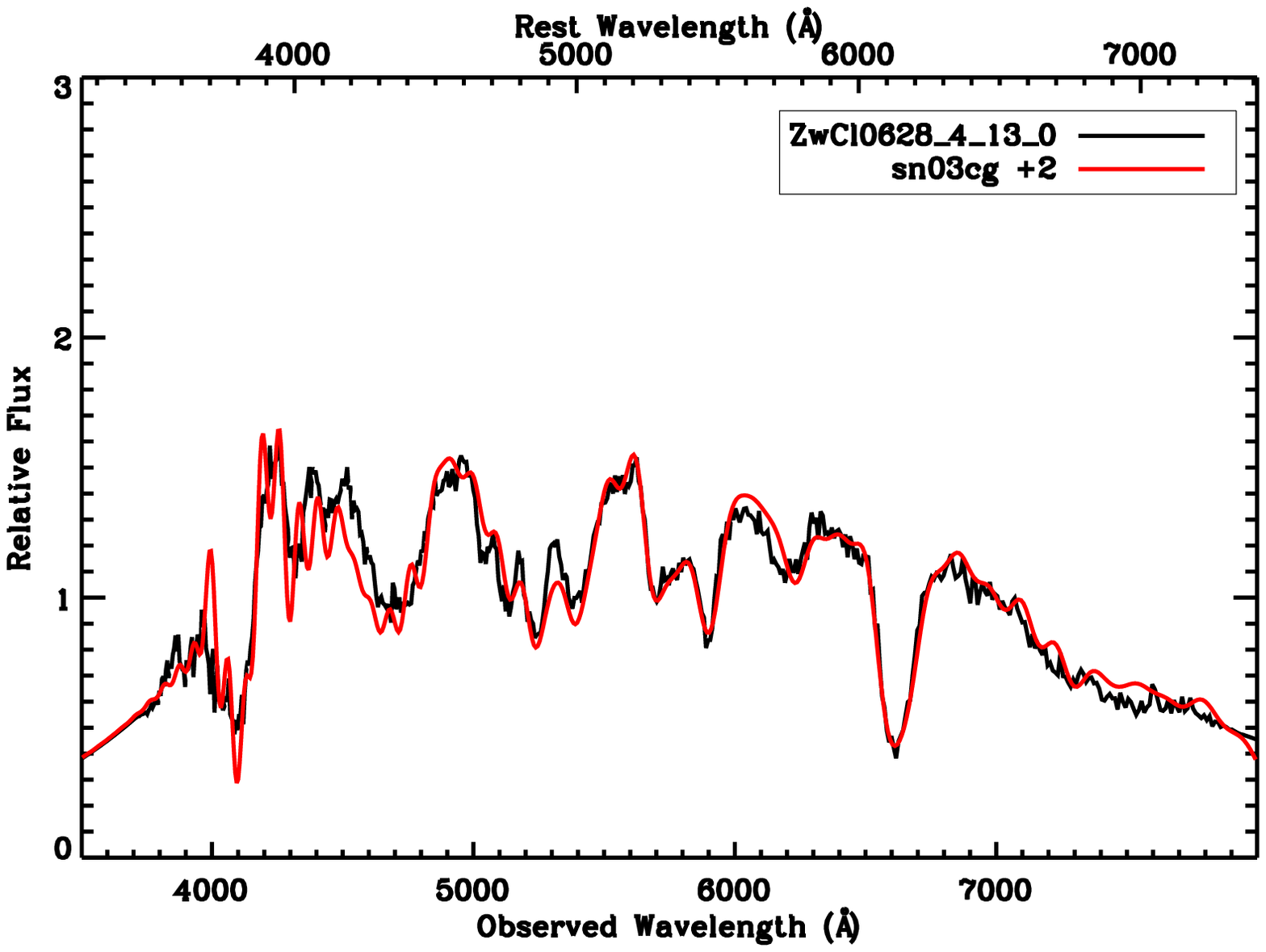} 
}
\mbox{ 
\epsfysize=6.0cm \epsfbox{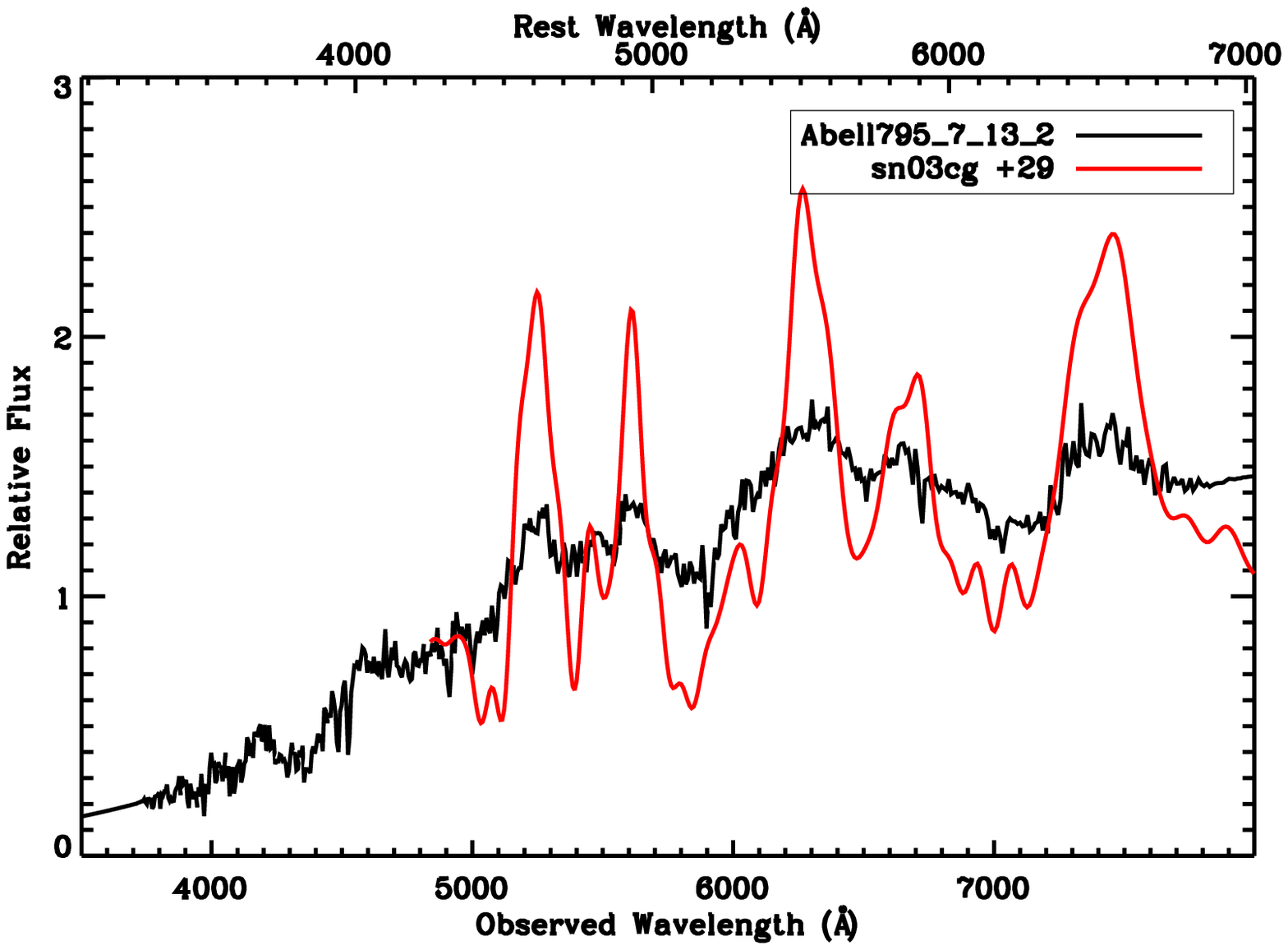} 
\epsfysize=6.0cm \epsfbox{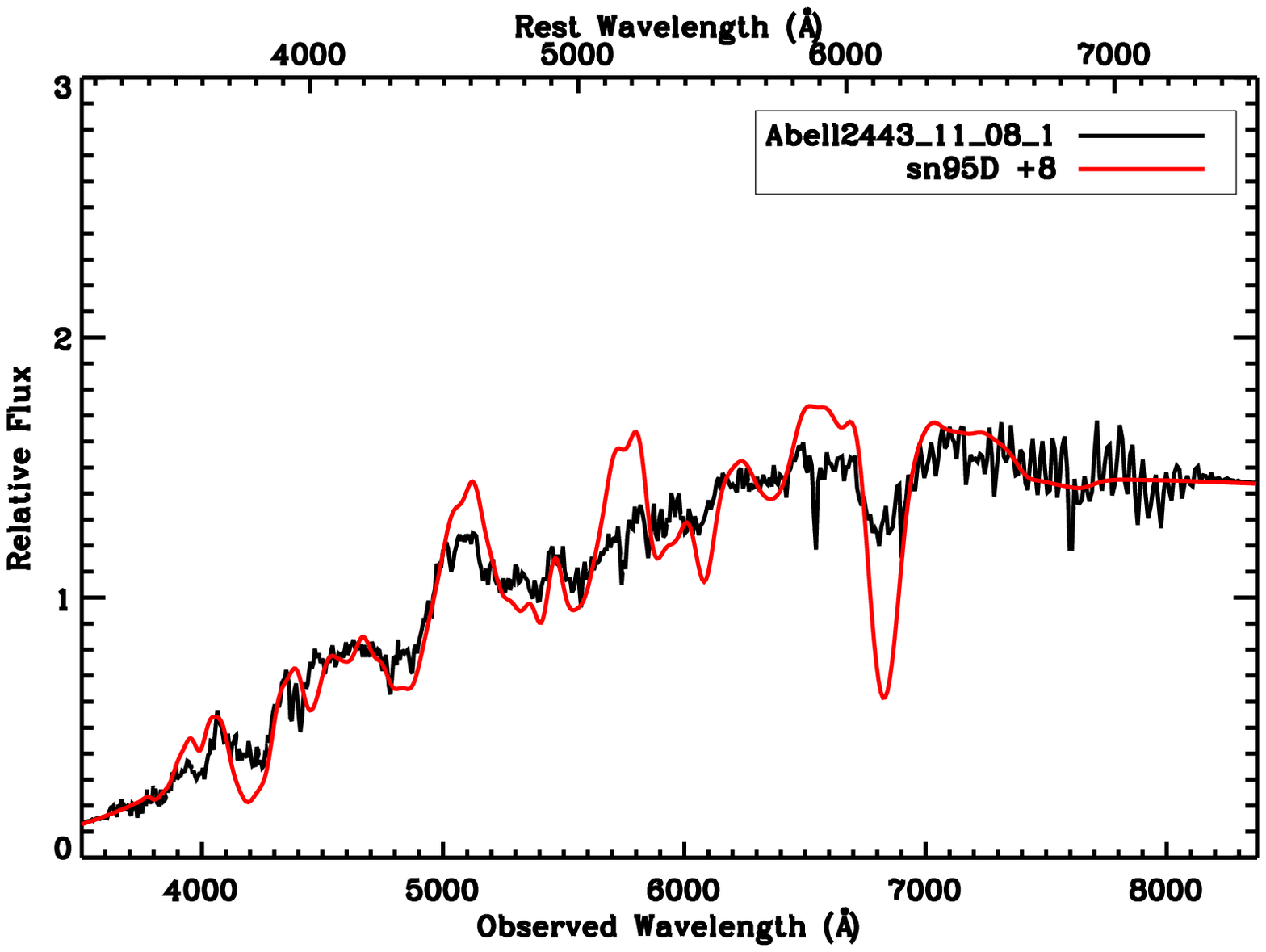} 
}

\caption{ Spectra of cluster SN Ia, along with the best-fitting template SN as determined by SNID.  See \S~\ref{sec:SNID} for details of the spectroscopic typing and Table~\ref{table:SNItable} for results.  SNID removes a pseudo-continuum from each spectrum before fitting to minimize host galaxy contamination.  Top left -- Abell1285\_8\_02\_4, a cluster
normal SN Ia at $z=0.1071$. Top right -- ZwCl0628\_4\_13\_0 is a normal cluster SN
Ia at $z=0.0822$.  Bottom left -- Abell795\_7\_13\_2 is a cluster SN Ia at $z=0.1414$
Bottom right -- Abell2443\_11\_08\_1 is a cluster SN
Ia at $z=0.111$.  
\label{fig:SNset2}}
\end{center}
\end{figure}

\begin{figure}
\begin{center}
\mbox{ 
\epsfysize=6.0cm \epsfbox{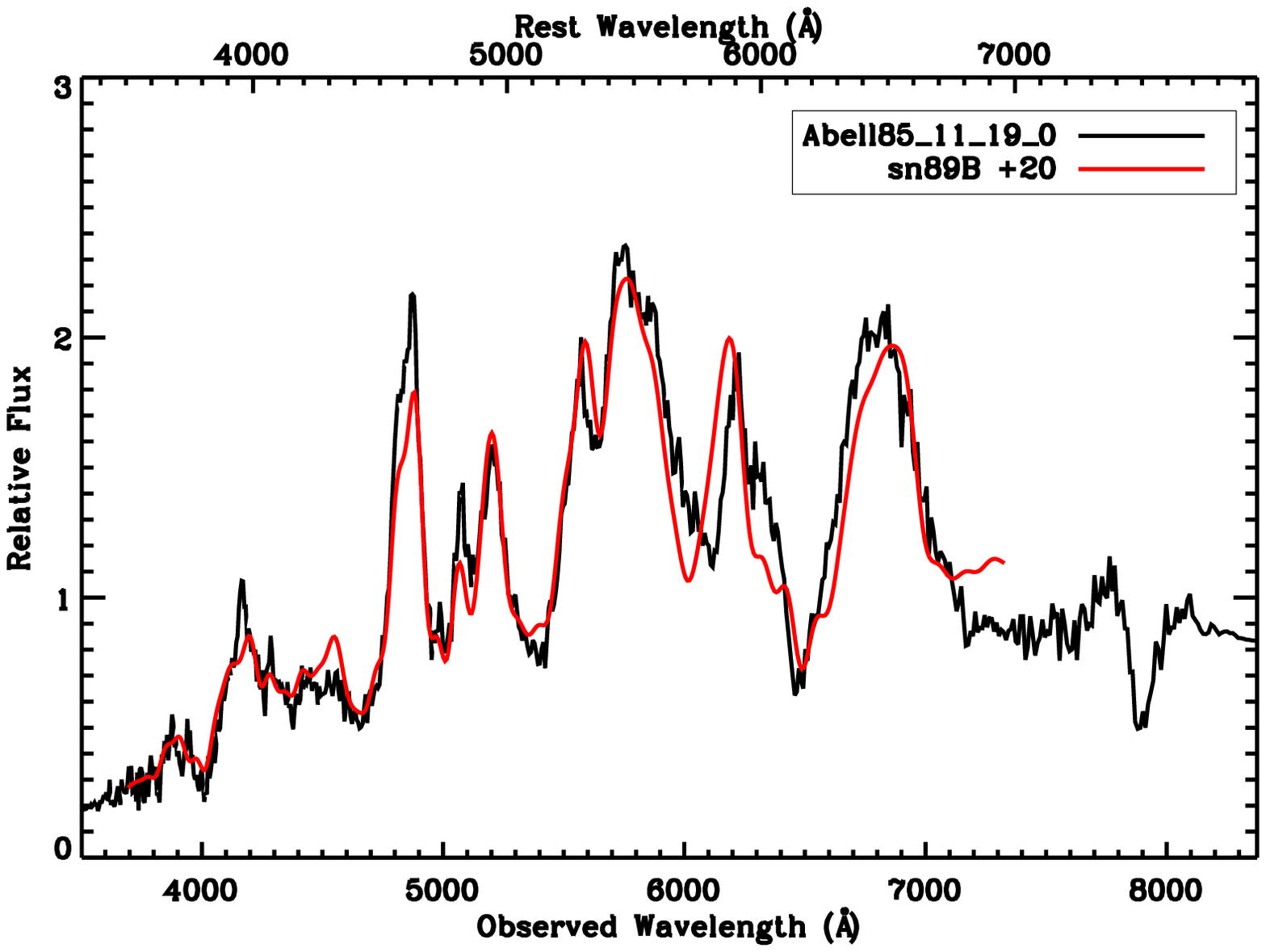} 
\epsfysize=6.0cm \epsfbox{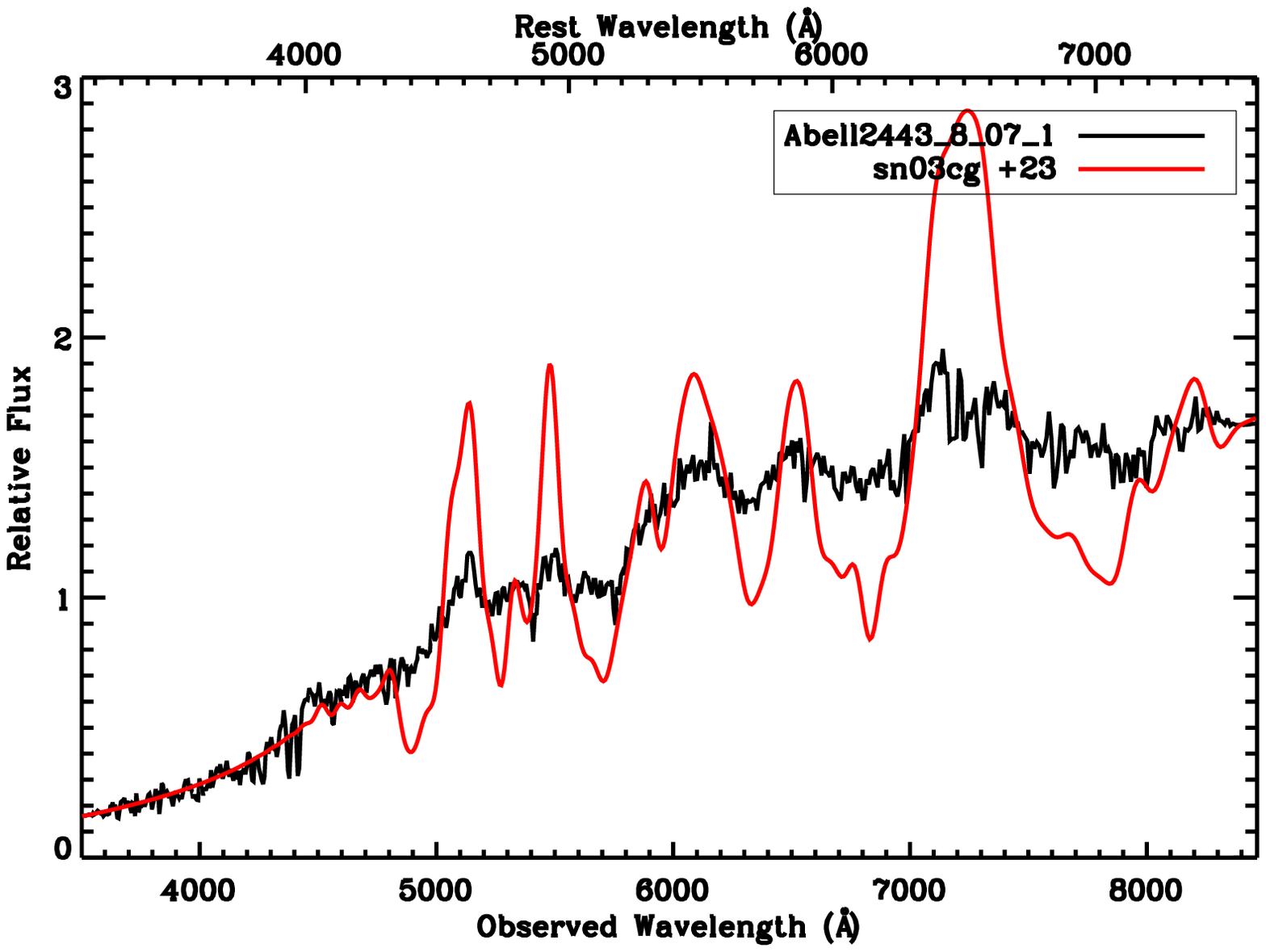} 
}
\mbox{ 
\epsfysize=6.0cm \epsfbox{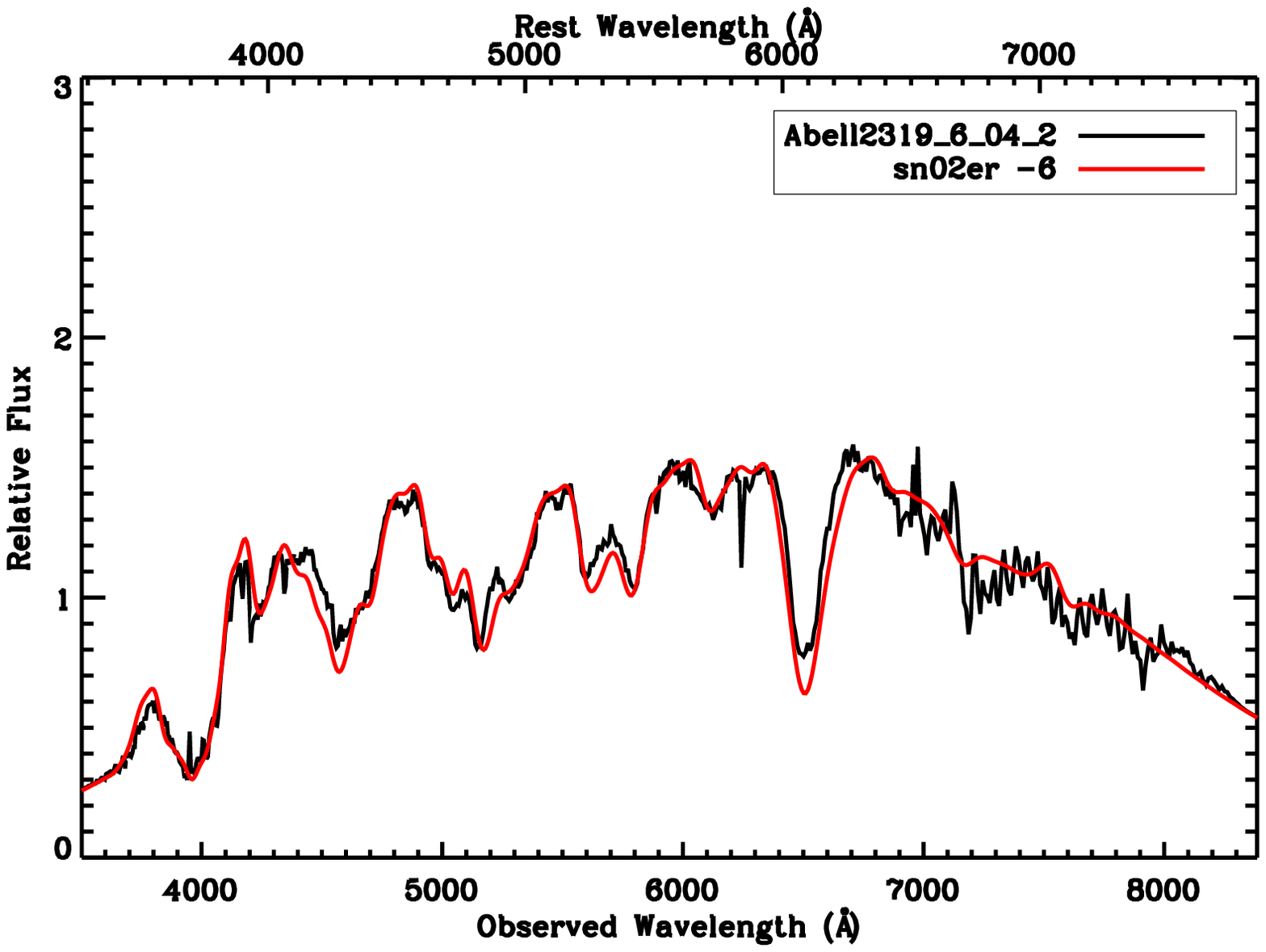} 
\epsfysize=6.0cm \epsfbox{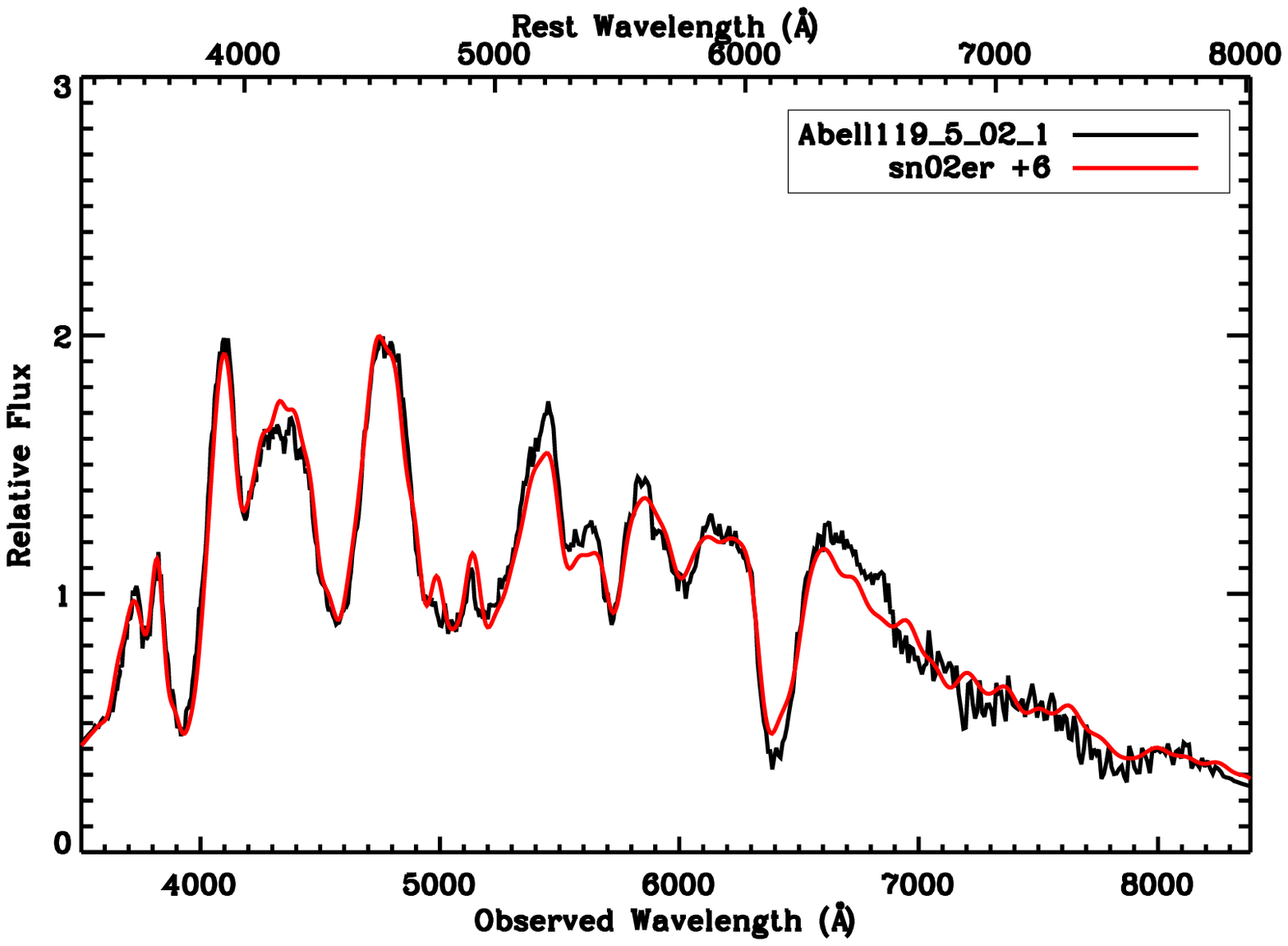} 
}

\caption{Spectra of cluster SN Ia, along with the best-fitting template SN as determined by SNID.  See \S~\ref{sec:SNID} for details of the spectroscopic typing and Table~\ref{table:SNItable} for results.  SNID removes a pseudo-continuum from each spectrum before fitting to minimize host galaxy contamination.  Top left -- Abell85\_11\_19\_, a normal
cluster SN Ia at $z=0.0534$. Top right -- Abell2443\_8\_07\_1 is a cluster SN Ia at
$z=0.1135$.  Bottom left -- Abell2319\_6\_04\_2 is a normal cluster SN Ia at
$z=0.0606$.  Bottom right -- Abell119\_5\_02\_1 is a normal cluster SN Ia at $z=0.0444$.
\label{fig:SNset3}}
\end{center}
\end{figure}

\begin{figure}
\begin{center}
\mbox{ 
\epsfysize=6.0cm \epsfbox{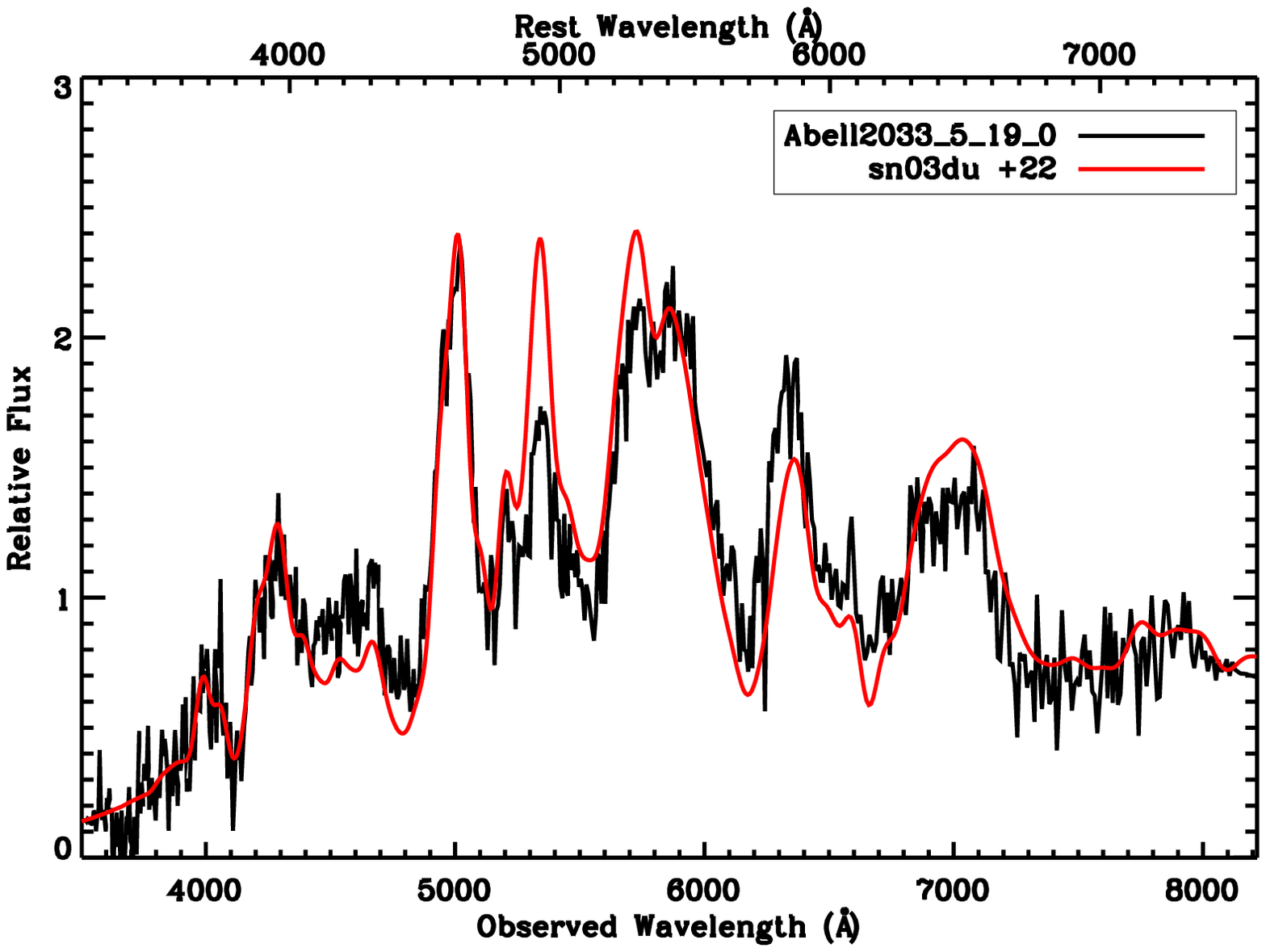} 
\epsfysize=6.0cm \epsfbox{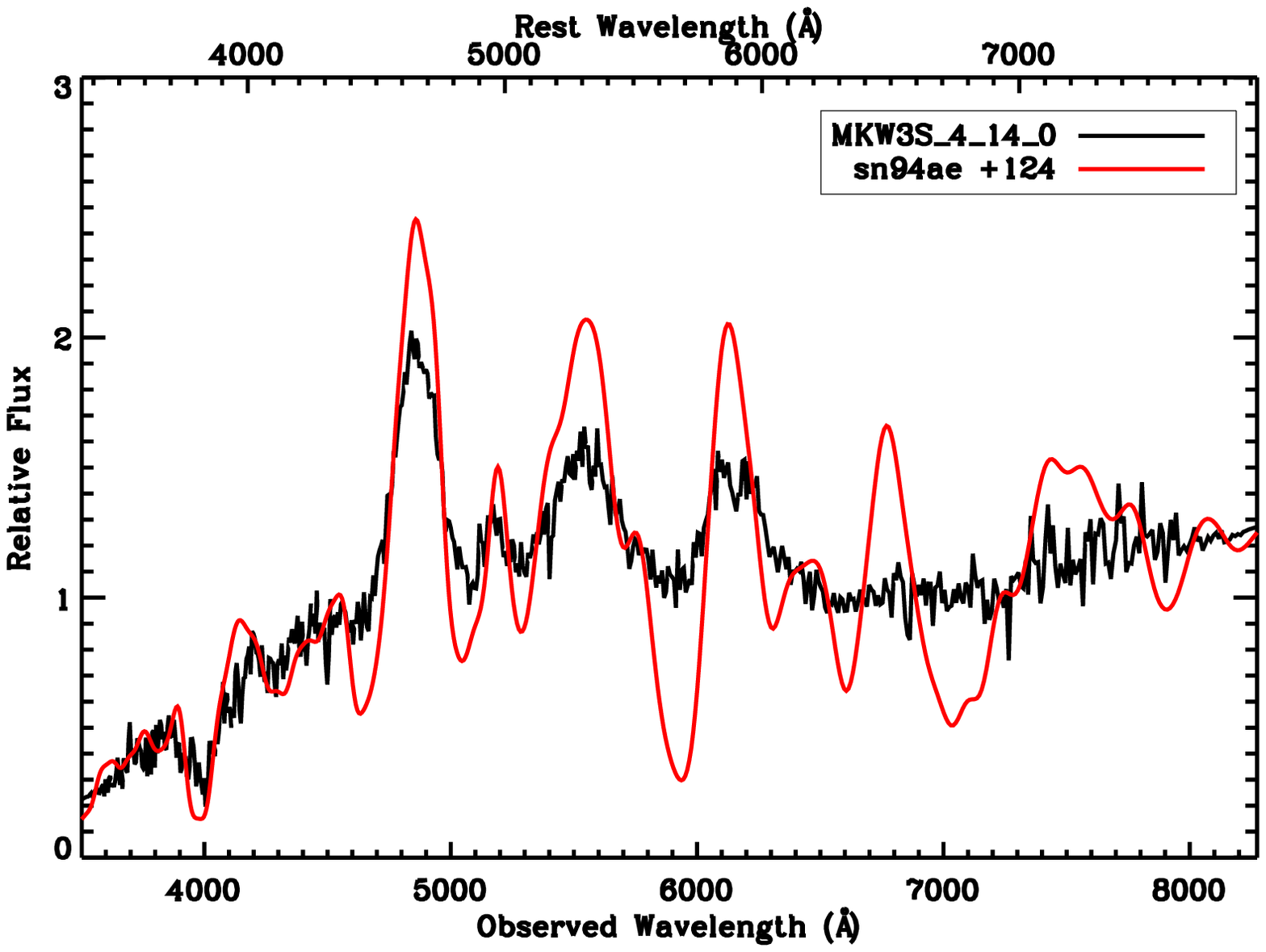} 
}
\mbox{ 
\epsfysize=6.0cm \epsfbox{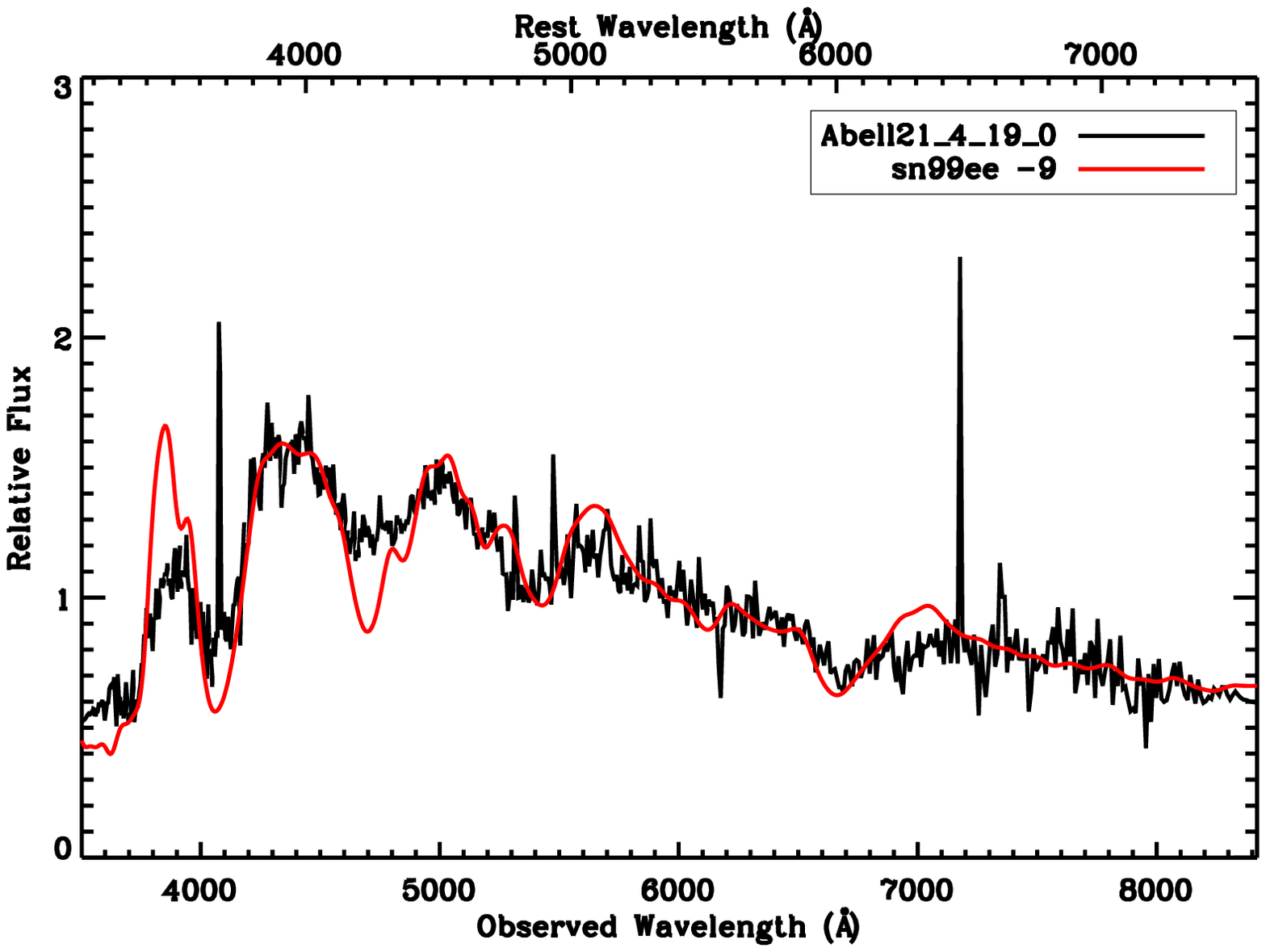} 
\epsfysize=6.0cm \epsfbox{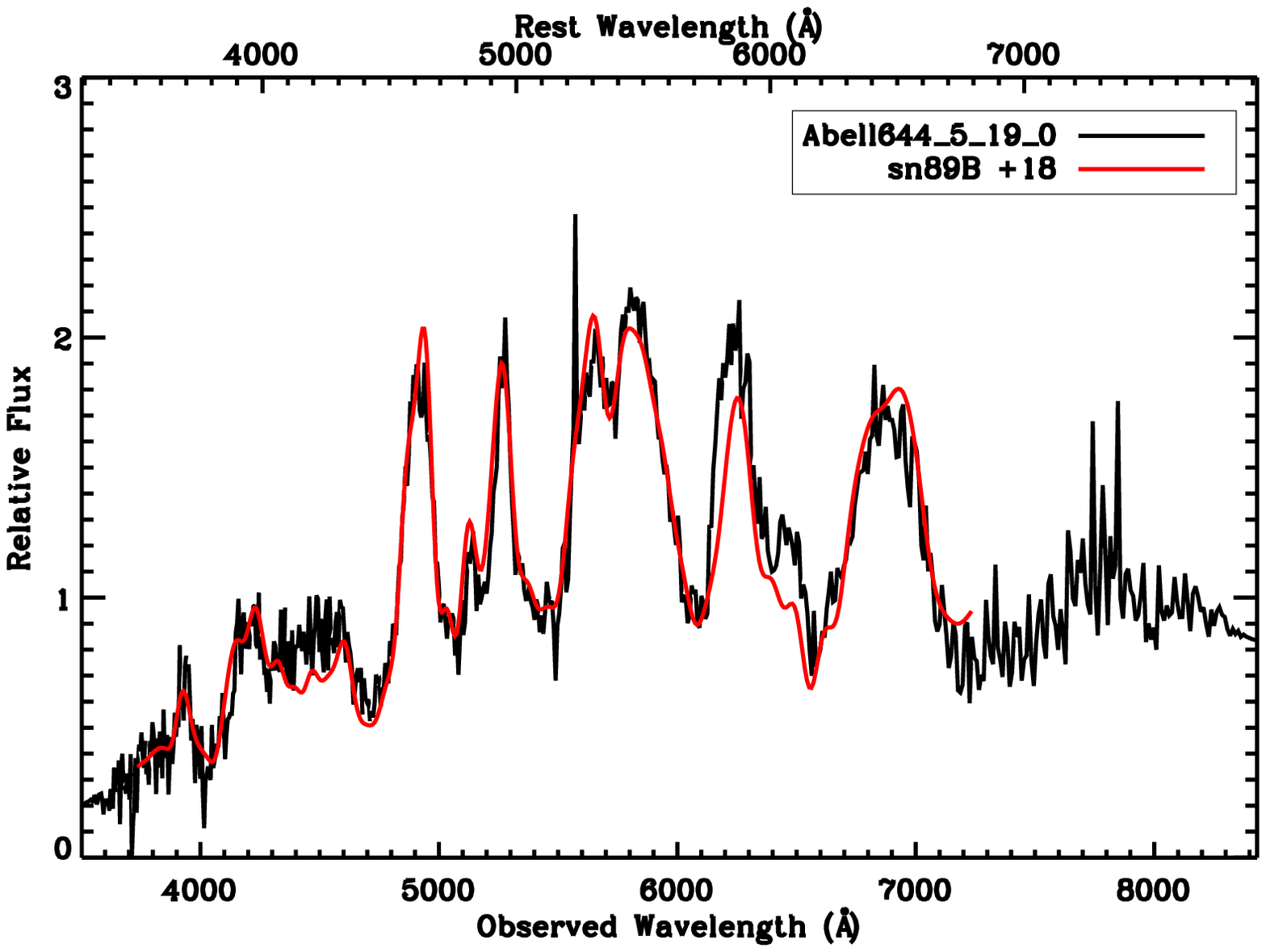} 
}

\caption{Spectra of cluster SN Ia, along with the best-fitting template SN as determined by SNID.  See \S~\ref{sec:SNID} for details of the spectroscopic typing and Table~\ref{table:SNItable} for results.  SNID removes a pseudo-continuum from each spectrum before fitting to minimize host galaxy contamination.  Top left -- Abell2033\_5\_19\_0, a normal
cluster SN Ia at $z=0.0837$. MKW3S\_4\_14\_0 is a cluster SN Ia at
$z=0.0526$.  Abell21\_4\_19\_0 is a normal cluster SN Ia at
$z=0.0947$.  Abell644\_5\_19\_0 is a normal cluster SN Ia at $z=0.0656$.
\label{fig:SNset4}}
\end{center}
\end{figure}

\begin{figure}
\begin{center}
\mbox{ 
\epsfysize=6.0cm \epsfbox{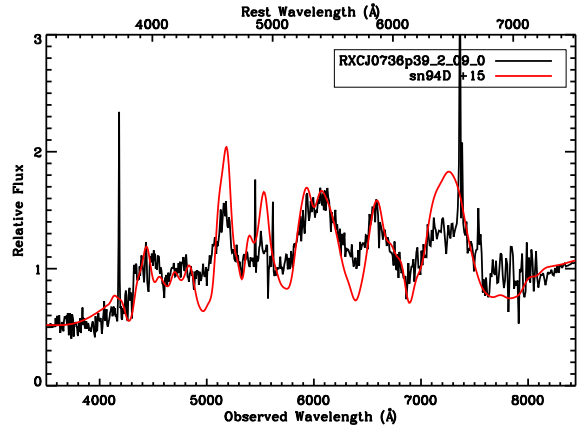} 
}
\mbox{ 
}
\caption{A spectrum of a cluster SN Ia, along with the best-fitting template SN as determined by SNID.  See \S~\ref{sec:SNID} for details of the spectroscopic typing and Table~\ref{table:SNItable} for results.  SNID removes a pseudo-continuum from each spectrum before fitting to minimize host galaxy contamination.  RXCJ0736p39\_2\_09\_0 is a normal
cluster SN Ia at $z=0.1225$.
\label{fig:SNset5}}
\end{center}
\end{figure}

\begin{figure}
\begin{center}
\mbox{ 
\epsfysize=6.0cm \epsfbox{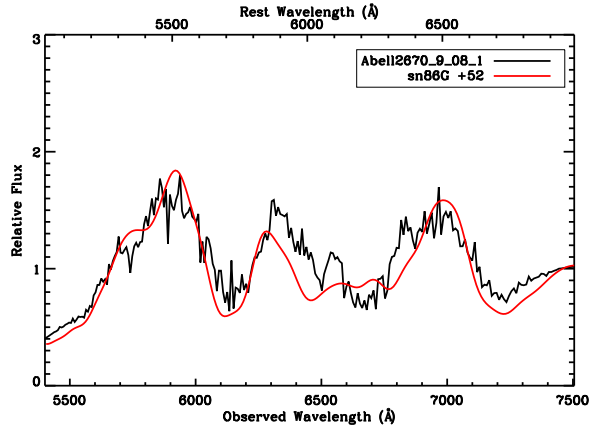} 
}

\caption{A spectrum of a cluster SN Ia, along with the best-fitting template SN as determined by SNID.  See \S~\ref{sec:SNID} for details of the spectroscopic typing and Table~\ref{table:SNItable} for results.  SNID removes a pseudo-continuum from each spectrum before fitting to minimize host galaxy contamination.  Due to Hectospec scheduling, it was necessary
to get this spectrum in 600 line mode, only allowing for wavelength
coverage between $\sim$5500 to 7500 \AA.  Nonetheless, it is a cluster SN Ia, although its subtype was not obtainable due to the
wavelength coverage.  Abell2670\_9\_08\_1 is a cluster SN Ia at
$z=0.076$.  
\label{fig:SN_weirdhecto}}
\end{center}
\end{figure}

\begin{figure}
\begin{center}
\mbox{ 
\epsfysize=6.0cm \epsfbox{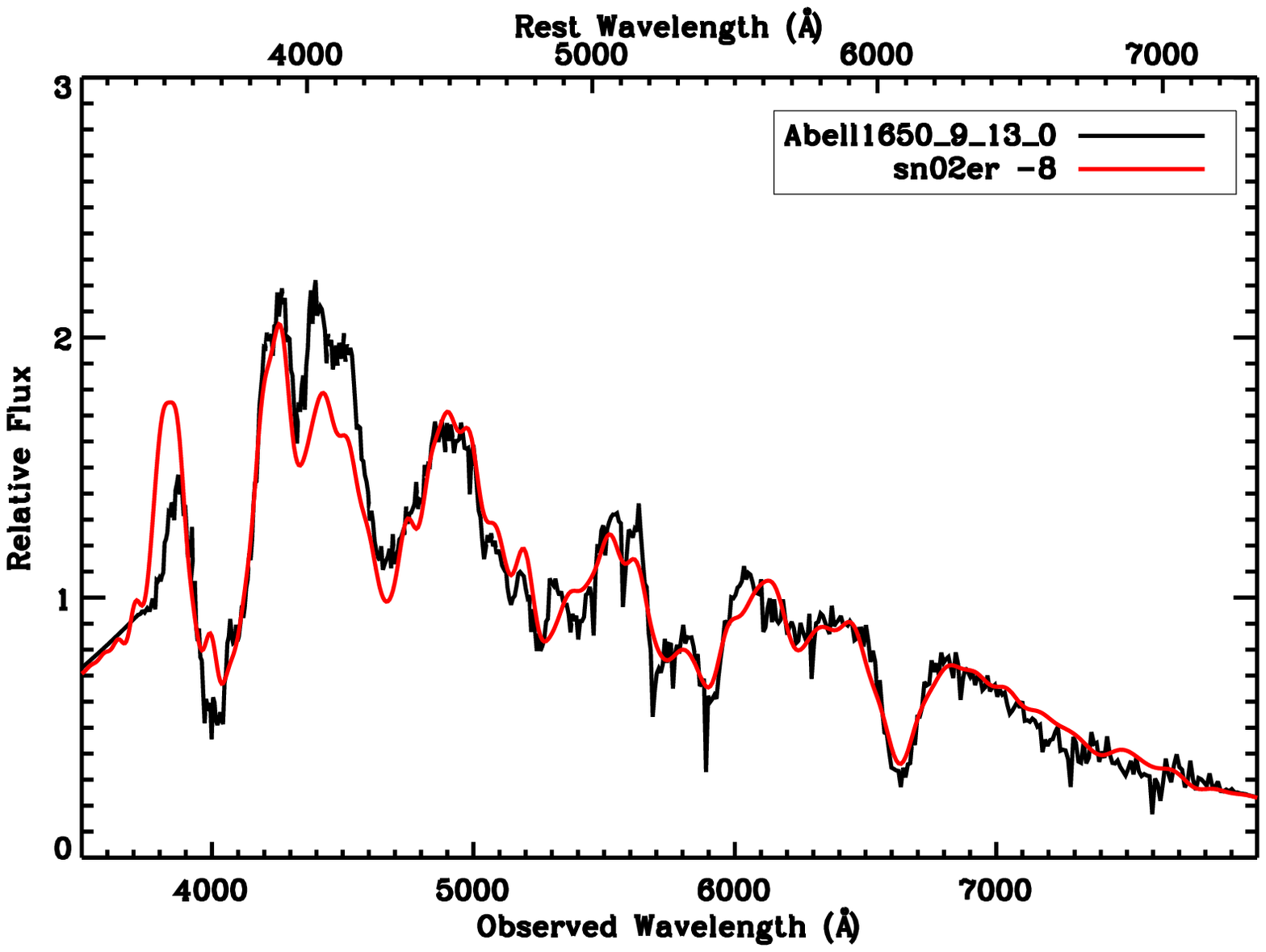} 
\epsfysize=6.0cm \epsfbox{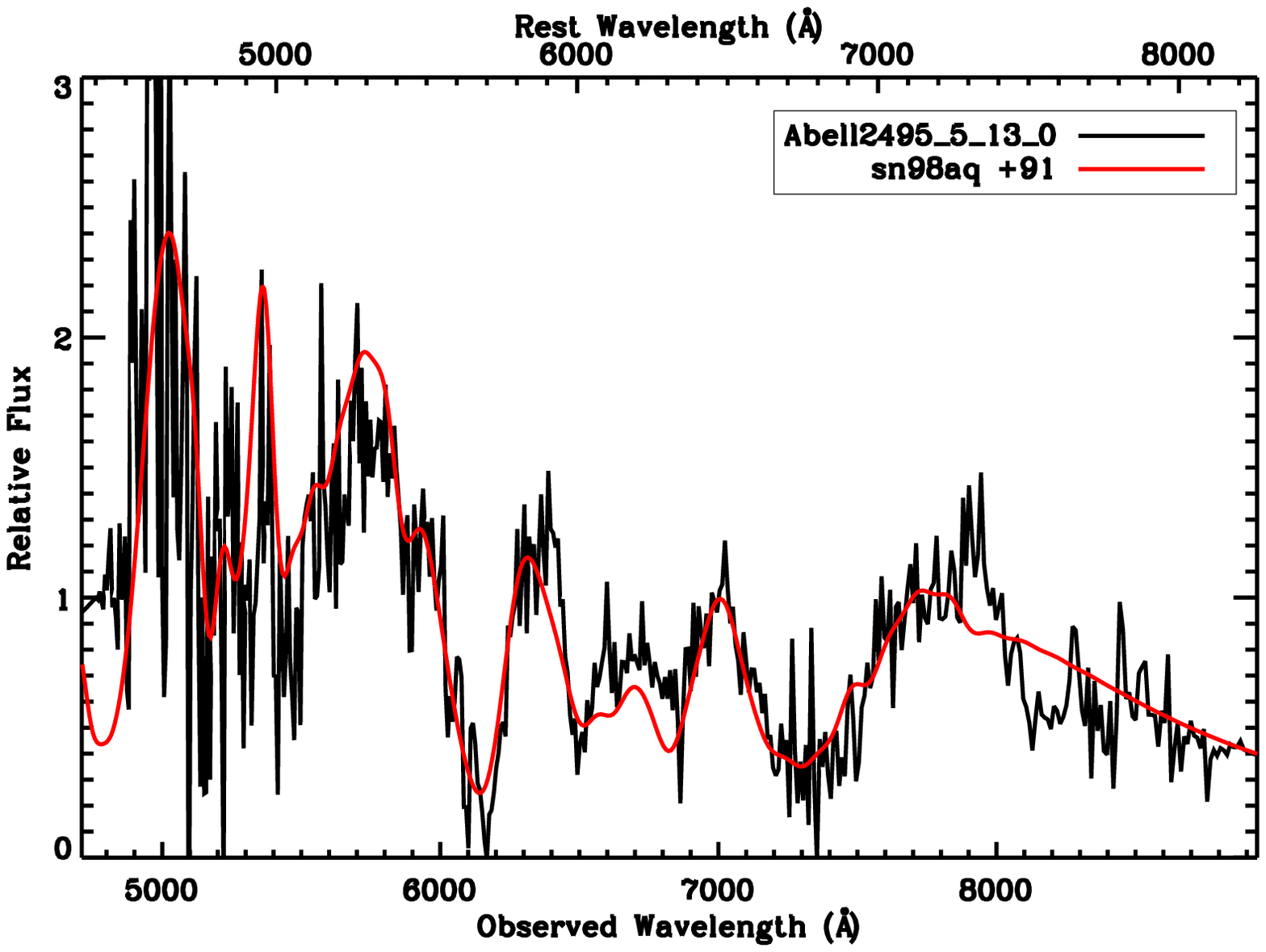} 
}
\mbox{ 
\epsfysize=6.0cm \epsfbox{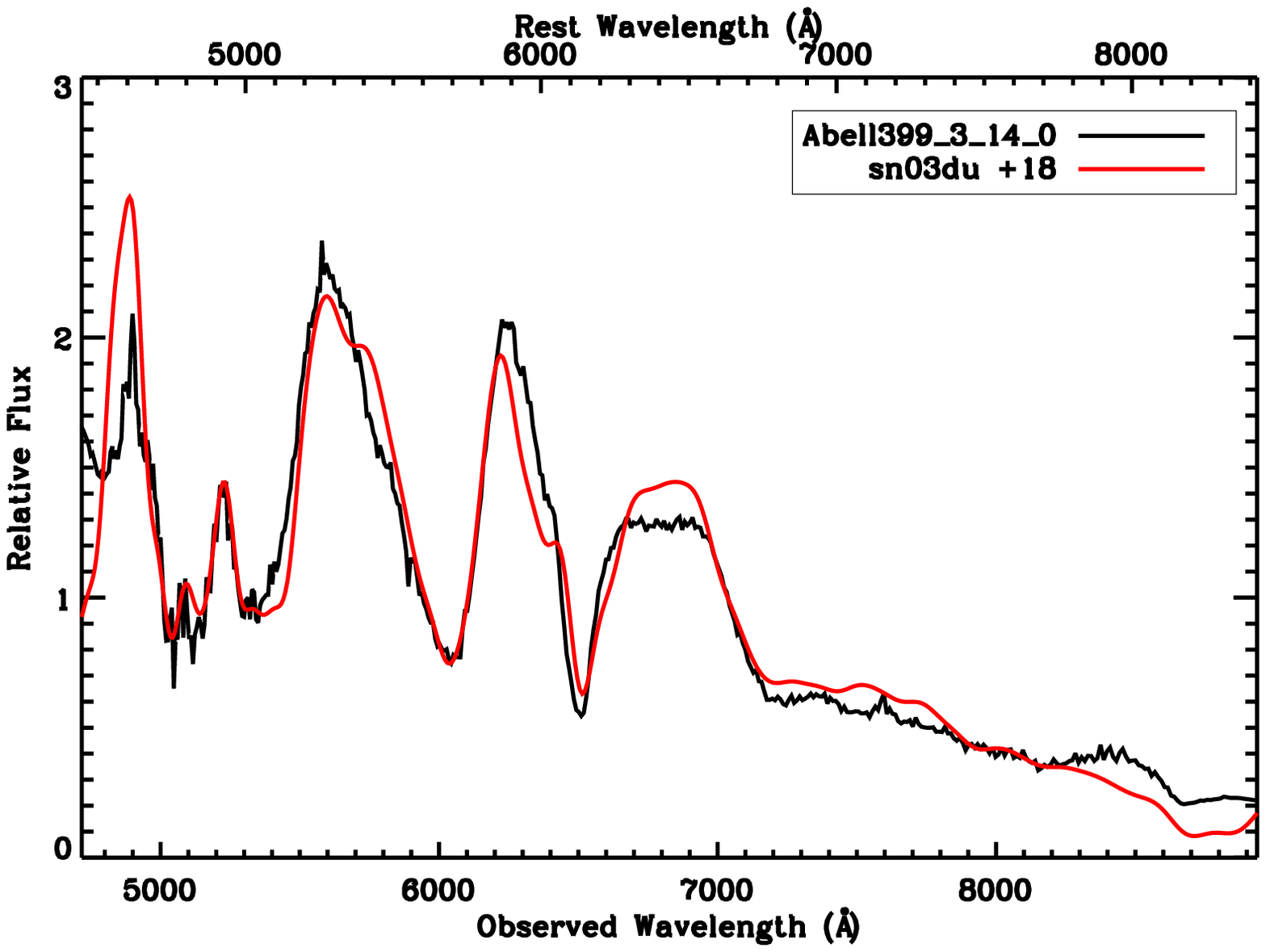} 
\epsfysize=6.0cm \epsfbox{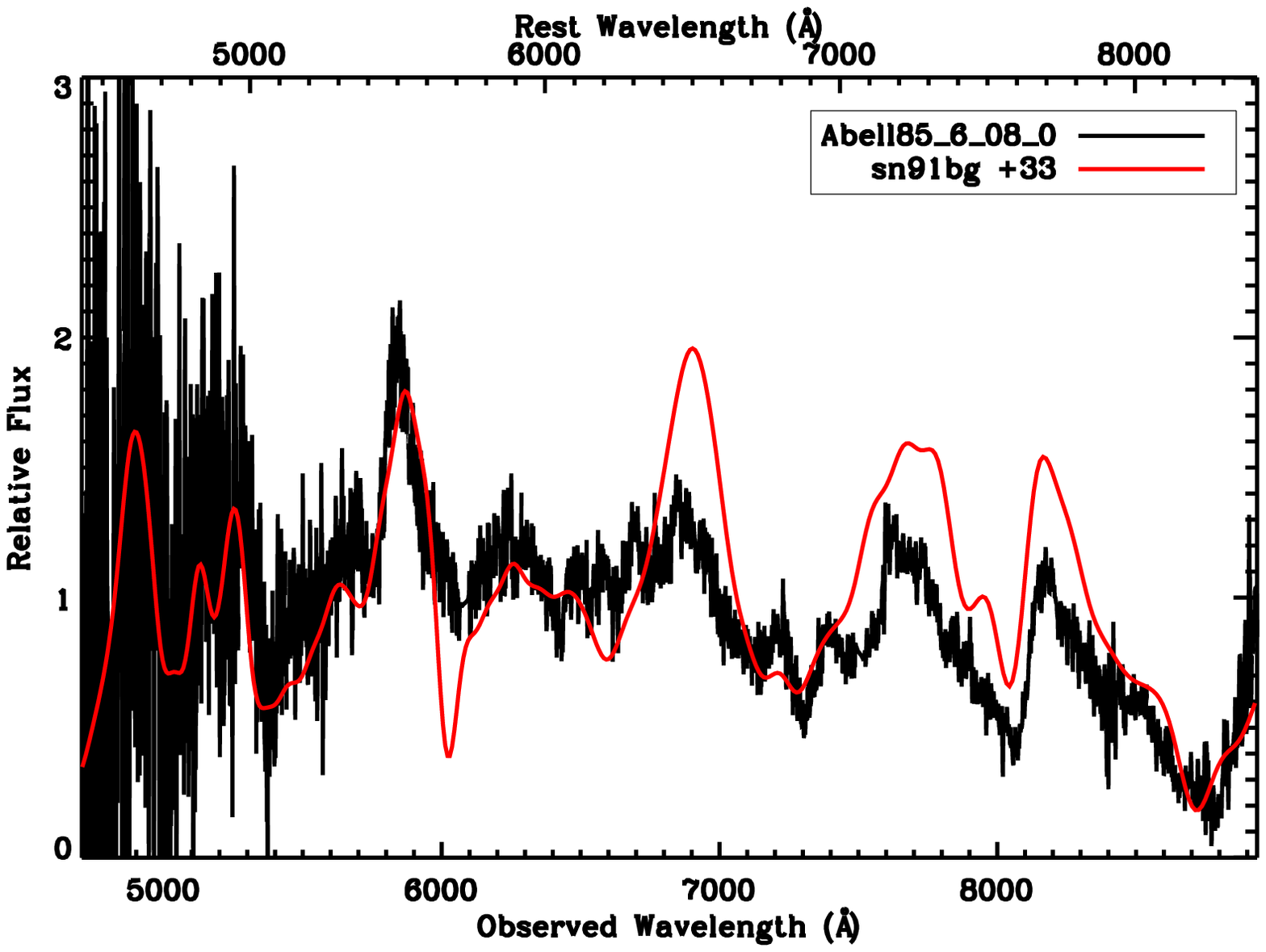} 
}

\caption{The four intracluster SN Ia found by MENeaCS, and first reported in \citet{Sand11}.  See that work for details. Top left -- Abell1650\_9\_13\_0, a cluster
normal SN Ia at $z=0.0.0836$. Top right -- Abell2495\_5\_13\_0, a cluster normal SN
Ia at $z=0.0796$.  Bottom left -- Abell399\_3\_14\_0 is a normal cluster SN Ia at
$z=0.0613$.  Bottom right -- Abell85\_6\_08\_0 is a normal cluster SN Ia at
$z=0.0617$.  
\label{fig:SNset6}}
\end{center}
\end{figure}

To determine supernova
type from our spectra, we use the publicly available Supernova Identification (SNID)
software \citep{SNID}.  SNID correlates an input spectrum with a large library of template spectra, including SNe of all types, variable stars, luminous blue variables, active galactic nuclei (AGN) and galaxies.  We use the same definition of a `good' correlation as previous works \citep{Miknaitis07,Foley09}, which we summarize briefly here while describing the basic SNID procedure.  First, SNID places input and template spectra on the same logarithmic wavelength scale and divides out a pseudo-continuum so that correlations are not dependent on flux calibration errors or reddening.  Also, since many of our SN spectra have significant host galaxy contamination, the removal of a continuum shape minimizes any false correlations that this may cause.  A correlation redshift is calculated based on the highest peak in the correlation function, and a correlation parameter, $r$, is determined and defined as the ratio of the height of the correlation peak to the root mean square of the antisymmetric component of the correlation function around the correlation redshift \citep[see][for details]{SNID}.  To ensure that correlations are meaningful, they are weighted by the overlap in ln($\lambda$) (the `lap' parameter) between input and template spectrum.  We consider, as in previous work, a `good' correlation to have $r$(lap) = $r \times$ lap $\geq$ 5 and lap = ln ($\lambda_{1}/\lambda_{0}$) $> 0.40$.  The quality of our best matches can be seen in Figures~\ref{fig:SNset1}--\ref{fig:SNset6}.

For each candidate SN spectrum, we attempt to determine the type, subtype, redshift and age by executing four SNID runs, as in \citet{Foley09}, who we refer the reader to for details.  A confident SN type is reported only if 50\% of the `good' SNID correlations were of that same type, along with the best-matched template.  Likewise, it was often impossible to determine a reliable SN subtype even if its type was clear, since we formally require that 50\% of the 'good' correlations (along with the best match) must be of that subtype.  Even so, our subtype determinations should be taken with a grain of salt given that they are based on a single spectrum, and we have only sparse light curves.  A summary of our cluster SNe Ia are in
Table~\ref{table:SNItable} -- note that our four IC SN Ia have already been presented in \citet{Sand11}, but we include them here as well for completeness.  In Figures~\ref{fig:SNset1}--\ref{fig:SNset6} we present all of our cluster SN Ia spectra, twenty-three in all.  We also overplot the best-matched SNID template, which are listed in Table~\ref{table:SNItable} as well.

One exception for our standard supernova typing scheme was made for
Abell2670\_8\_08\_1.  This object was observed with Hectospec with the
600 line grating (with a wavelength range of $\sim$550-750 nm).  Due
to the limited wavelength range, SNID had difficulty in converging to
a solution.  To narrow the search range, we input an initial guess redshift corresponding to that of the cluster ($z=0.072$),
and restricted the type of supernova searched to SNe Ia.  This led to
excellent fits to the available data, as can be seen in Figure~\ref{fig:SN_weirdhecto}.
Given this, we take this object to be a SN Ia at $z=0.076\pm0.0036$, in
accordance with the median and dispersion of redshifts resulting in
good fits.  No subtype determination was attempted.

We also inspect each SN spectrum for host galaxy absorption and/or emission lines to pinpoint its redshift precisely, and we report this redshift in Table~\ref{table:SNItable} when available.  When these features are found, they agree with the SNID-derived redshift in all cases.

\subsection{Cluster Membership}

\begin{figure}
\begin{center}
\mbox{ 
\epsfysize=9.0cm \epsfbox{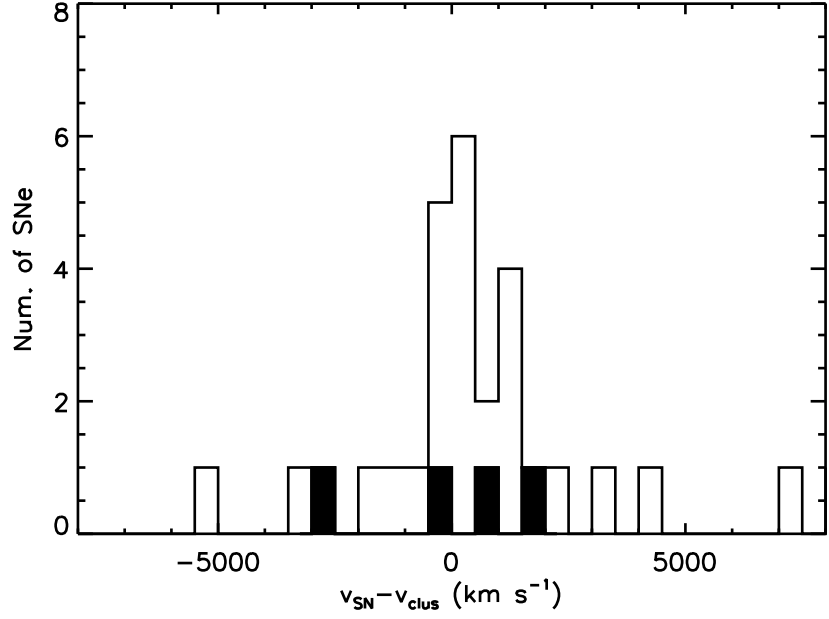} 
}
\caption{ The histogram of the velocity difference between the cluster redshifts  and our SNe Ia.  Our hostless SNe Ia are shown in the filled histogram.  Only those SN Ia with $\Delta v < 8000$ km s$^{-1}$ are shown.  We choose a $\Delta v < 3000$ km s$^{-1}$ as our definition for a cluster SN Ia.  Larger values of $\Delta v$ only add SNe beyond $R_{200}$, beyond which MENeaCS becomes incomplete due to the Megacam field of view in any case.
\label{fig:SN_vdiff}}
\end{center}
\end{figure}

MENeaCS spectroscopically confirmed $\sim$70 SNe, and we must separate foreground/background objects from those associated with the target clusters.  We start by plotting the histogram of the velocity difference between the cluster redshift and our SNe Ia ($v_{SN}-v_{clus}$) in Figure~\ref{fig:SN_vdiff}.  A cutoff value for cluster membership of $\Delta v < 3000$ km s$^{-1}$ is a natural fit for SN membership in a survey like MENeaCS.  MENeaCS clusters have a velocity dispersion of $\sim$600-1000 km s$^{-1}$ (Sand et al. 2011), while SN velocities can only be measured to $\sim$1000-2000 km s$^{-1}$ due to their broad spectral features and intrinsic diversity (see Table~\ref{table:SNItable}).  As previously noted in \S~\ref{sec:techdesign}, MENeaCS becomes incomplete at $\sim R_{200}$ due to the field of view of Megacam.  Thus, while we report all of our SN Ia with $\Delta v < 3000$ km s$^{-1}$ in Table~\ref{table:SNItable}, we focus on those within either 1 Mpc or $R_{200}$ to calculate our SN rate in \S~\ref{sec:rates}.  If we were instead to place our $\Delta v$ cutoff value at 5000 km s$^{-1}$, we would only add three SN Ia into contention, all of which are at $R>1.2R_{200}$, and so no new objects would be added into our rate calculations.  In the end, we have identified eleven cluster SN Ia within 1 Mpc of the cluster center, and sixteen within $R_{200}$.

We note that our cut of $\Delta v < 3000$ km s$^{-1}$ and $R_{SN} < R_{200}$ is comparable with other studies in the literature.  The intragroup SN study of \citet{McGee10} only included SNe with $R_{SN}<R_{vir}$ and $\Delta v < 3000$ km s$^{-1}$, while the SDSS cluster SN results of \citet{Dilday10} used $R_{SN} < $1.4 Mpc and $\Delta v < 4500$km s$^{-1}$.  The $z\sim0.5-0.9$ cluster SN study of \citet{Sharon10} used a velocity cut of $\Delta v \lesssim2600$km s$^{-1}$.

\section{Cluster Stellar Luminosity and Mass}\label{sec:clusterlum}

In this section, we calculate the luminosity of each MENeaCS cluster and infer stellar masses -- a crucial ingredient for measuring the SN rate.  Given that we do not have a complete redshift survey to identify all cluster members, we measure our luminosities in a statistical sense, as has been done in previous cluster SN rate analyses \citep[e.g.][]{Sharon07,Sharon10,Barbary10}.  

The basic approach is straightforward.  Galaxy luminosities are measured in each cluster deep stack image out to various clustercentric radii.  Since these fields contain both cluster galaxies and foreground/background contamination, several background fields are used to gauge the typical contamination.  K-corrections are determined for each cluster's redshift to determine rest frame properties and the statistical background is subtracted.  From there, further corrections can be made for photometric completeness (utilizing a standard cluster luminosity function) and the  inclusion of the intracluster light component.  Luminosities of specific cluster galaxy populations can be determined (i.e. red sequence galaxies), so that their SN rate can be measured directly.

We used the stacked images of each cluster, described in \S~\ref{sec:deepstacks}, to create galaxy catalogs.  We run {\sc SExtractor} \citep{sexbib} in dual image mode, carefully setting the parameters so that faint galaxies were deblended from bright nearby galaxies, as is common near the BCG.  We use {\sc SExtractor} {\sc MAG\_AUTO} photometry for galaxy magnitudes, and a fixed 3 arcsecond circular aperture for galaxy colors.  Stars are removed from the catalogs by flagging objects that lie on the stellar locus in half-light radius ($r_{1/2}$) versus magnitude.
All of our photometry is corrected for foreground Galactic extinction using E(B-V) values from \citet{Schlegel98}.  As in previous cluster SN work, we discard galaxies in both the cluster frame and the background fields which are brighter than the identified BCG in each cluster.

Once the catalogs are cleaned of stellar objects, a color magnitude diagram is made, and a red sequence is fit.  This process will be discussed in detail by Bildfell et al. (2012, in prep), and is a similar procedure to that presented in \citet{Pimbblet02}.  We find tight red sequences in all of our clusters, with typical scatter about the linear fit of $\sim$0.03-0.04 mag for any given cluster.  These measurements are important for two reasons.  First, we do not include galaxies redder than the cluster red sequence (accounting for its scatter) into our cluster luminosity measurements (either in the background field or in the cluster field), as these objects, aside from the possibility of an extremely reddened cluster member, are likely background galaxies \citep[e.g.][]{Gladdersred}.  Second, we want to measure the red sequence SN Ia rate (\S~\ref{sec:rates}) directly, as these galaxies have a relatively pure stellar population of old, $\sim$10 Gyr-old stars and are ideal for constraining the SN Ia delay time distribution (\S~\ref{sec:DTD}).

For our `background' fields, we use the four CFHT Legacy Survey Deep Fields, which cover a total of four square degrees.  We downloaded the images from the MegaPipe CFHTLS web page\footnote{http://www2.cadc-ccda.hia-iha.nrc-cnrc.gc.ca/community/CFHTLS-SG/docs/cfhtls.html}\citep{Gwyn08}, using the ``full" version containing all usable data.  Galaxy catalogs are created using {\sc SExtractor} in a fashion identical to that of our clusters fields.  These deep images are complete to $g\sim25.5$ and $r\sim25.5$ as indicated by their galaxy number counts, which is deeper than our cluster deep stack fields.  We only count background galaxies down to our chosen faint magnitude level for each cluster.  We take the average luminosity per arcmin$^{2}$ over the four background fields as our background value, and we take the standard deviation in luminosity in the four different fields as our uncertainty, since this is an estimate of cosmic variations in large scale structure.

Since only frames with PSF FWHM $<1.0$ arcsec were included in our deep stack images, each cluster and each band have slightly different depths.  To gauge the completeness of each field, we plotted histograms of galaxy number counts versus magnitude, and did a direct comparison with our deeper CFHTLS background fields.  For the sake of simplicity, and because all of our cluster fields were complete at least down to $g,r$=23.5 mag, we only included galaxies brighter than this in our cluster luminosity calculation.   

We use K-corrections derived from  the {\sc kcorrect} software package, as described in \citet{Blanton07}, which we refer the reader to for details.  Briefly, {\sc kcorrect} fits a spectral energy distribution (SED) to galaxy fluxes based on a linear combination of a small number of known galaxy templates, which are derived from the Bruzual-Charlot stellar evolution synthesis code \citep{Bruzual03}.  This SED is then used to find the K-correction necessary to measure absolute g and r band magnitudes.  Additionally, since the templates can be interpreted physically, the SED fit also provides an estimate of the stellar mass-to-light ratio.  We use these mass-to-light ratios to convert from stellar luminosity to stellar mass.  We investigate the limitations of using only two photometric bands ($g$ and $r$) for our K-corrections, by comparing with 5-band SDSS photometry when there is overlap with the MENeaCS sample in \S~\ref{sec:sdsscompare}.

For each cluster, we sum the K-corrected $g$ and $r$-band luminosity of all galaxies brighter than our chosen detection limit of $g$, $r$=23.5 mag, fainter then the cluster's BCG, and within our red-color cut to exclude objects redder than our measured cluster red sequence.  We apply an identical cut to the four K-corrected CFHTLS Deep Field catalogs, and remove this statistical background.  We then make a small correction to account for the total luminosity of all faint cluster galaxies below our detection limit, mimicking the procedure described in \citet{Sand11}.  We adopt the same faint-end slope ($\alpha$=$-$1.03) and normalization as that found for the Virgo Cluster in \citet{Trentham02} in our cluster luminosity function.  The derived correction factors -- at most $\sim$1 to 2\% of the clusters' overall luminosity -- are insignificant in comparison to our overall luminosity measurement uncertainties, and so we do not discuss any further possible ambiguities associated with our choice of the faint end slope or normalization here.  An additional factor is necessary to account for the intracluster light component of the clusters' luminosity, as we did discover four intracluster SN Ia during the course of MENeaCS \citep{Sand11}.  We directly adopt the intracluster stellar mass fraction found for the MENeaCS sample, $f_{ICL}$=0.16$^{+0.13}_{-0.09}$, and applied this factor to obtain our final cluster luminosities.

As mentioned, {\sc Kcorrect} also outputs a stellar mass to light (M/L) ratio corresponding to the best-fit galaxy SED.  The {\sc Kcorrect} code assumes a Chabrier initial mass function (IMF), while the cluster SN rate results in the literature have all been normalized to a `diet' Salpeter IMF \citep{Maoz10clus}.  We therefore convert our M/L to a 'diet' Salpeter IMF by using the conversions of \citet{Bernardi10}.  We use these M/L ratios to convert our cluster luminosities to a stellar mass.  The average stellar M/L in our clusters (dividing the total stellar mass by luminosity) is $M/L_{r}=2.4$, $M/L_{g}=3.5$ for the cluster luminosity as a whole, and $M/L_{r}=2.5$, $M/L_{g}=3.8$ for our red sequence galaxies.

We list our cluster luminosities and stellar masses within $R_{200}$ in Table~\ref{table:clusterlum}.  We include values for the cluster as a whole (including galaxies below our detection limit and the intracluster light) and for our designated red sequence galaxies only. 
Similar values were calculated at radii of 1 Mpc, 1.25$R_{200}$ and 0.75$R_{200}$ so that we could gauge the sensitivity of our SN rate measurements to our $R_{200}$ values and out to a fixed cluster radius (\S~\ref{sec:rates}).  Finally, we also list $B$-band luminosities in Table~\ref{table:clusterlum}, as many SN rates have traditionally been reported per unit $B$-band luminosity.  To convert our $g$ and $r$ band luminosities to $B$-band, we use the filter transformations presented in \citet{Blanton07}.

\subsection{Comparison to SDSS luminosities}\label{sec:sdsscompare}

To test our luminosity measurement procedure, we have also derived cluster luminosities for the 43 clusters in the MENeaCS sample which are in the SDSS footprint.  We downloaded $u,g,r,i,z$ photometry from SDSS Data Release 8\footnote{http://www.sdss3.org/dr8/}, choosing the {\it model} magnitudes of extended objects, and excluding point sources.  We have also downloaded SDSS data covering three of the four CFHTLS Deep Fields (the fourth is not in the SDSS footprint), and use these as background fields.  We adopt a similar procedure for measuring the clusters' luminosity as described above, with two exceptions.  First, we assume faint detection limits of $g$,$r$=22.2 mag, in line with the SDSS depth.  Second, we derive K-corrections using the full 5-band photometry, rather than the single color provided by the MENeaCS $g,r$ photometry.

For a fair comparison, we have recalculated our MENeaCS cluster luminosities with  faint detection limits of $g$,$r$=22.2 mag, and use    only the three of the four CFHTLS Deep Fields in the SDSS footprint as a background.  Then, in principle, the only differences in our derived cluster luminosities between the MENeaCS and SDSS data will arise from differences in our K-corrections derived in 5 bands versus 2, and from differences in photometry between the SDSS pipeline and our own {\sc SExtractor} based approach.

The comparison between the SDSS and MENeaCS data sets were relatively good, with the final cluster luminosities agreeing at the $\lesssim$10\% level, with the MENeaCS data systematically brighter in $r$-band and dimmer in $g$-band.  We are satisfied with this measurement agreement at the $\sim$10\% level given the differences in the two data sets.   We consider systematic uncertainties of $\pm10\%$ in our luminosities and explore the consequences for our SN rate measurement in \S~\ref{sec:rates}.


\section{Supernova host galaxies}\label{sec:SNhost}

While a detailed study of the host galaxy properties of the MENeaCs sample is beyond the scope of the current work, it is of interest to measure our hosts' basic photometric properties to divide the sample between red sequence versus bluer galaxies, and derive SN rates for each respective population.  A full study of the MENeaCS SN host galaxy properties, including ultraviolet, optical and near infrared photometry, along with optical spectroscopy, will be presented in a future contribution (Graham et al. in preparation).

First, we create SN-free deep stack images for each cluster field which hosted a SN, analogous to our deep stack combination procedure outlined in \S~\ref{sec:deepstacks}, including only images taken before SN discovery.  Some experimentation has shown that SN light is still visible in our differenced images $\sim$12 months after SN discovery, making our approach conservative and appropriate.  We then perform photometry identically to that described in \S~\ref{sec:clusterlum}, and we present our hosts' $r$-band magnitudes and $g-r$ color (within a 3" aperture) in Table~\ref{hosttab}.  We also list the clustercentric radius of the host galaxy, both in kpc and scaled by $R_{200}$.

\begin{figure}
\begin{center}
\mbox{ 
\epsfysize=4.5cm \epsfbox{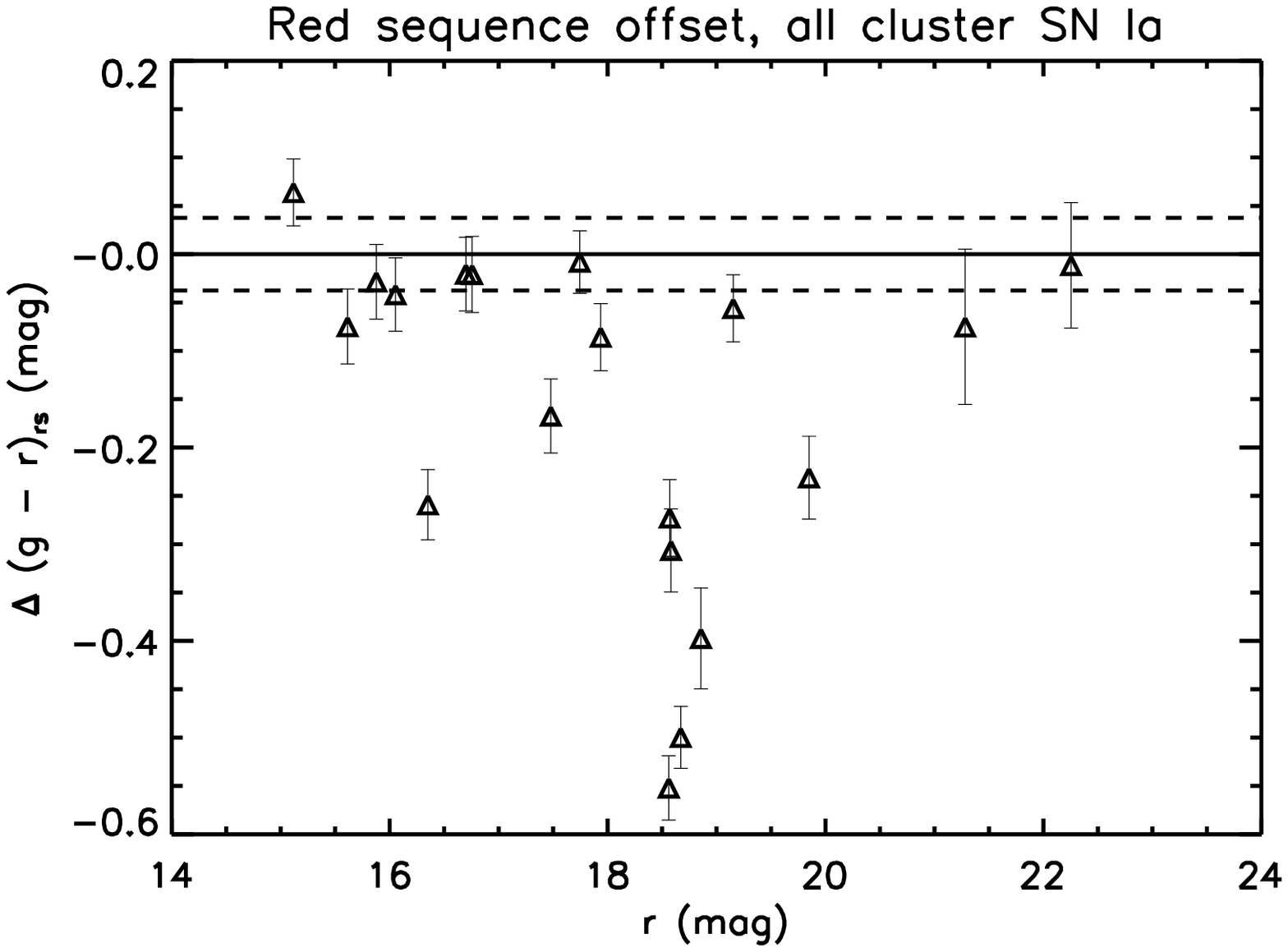} 
\epsfysize=4.5cm \epsfbox{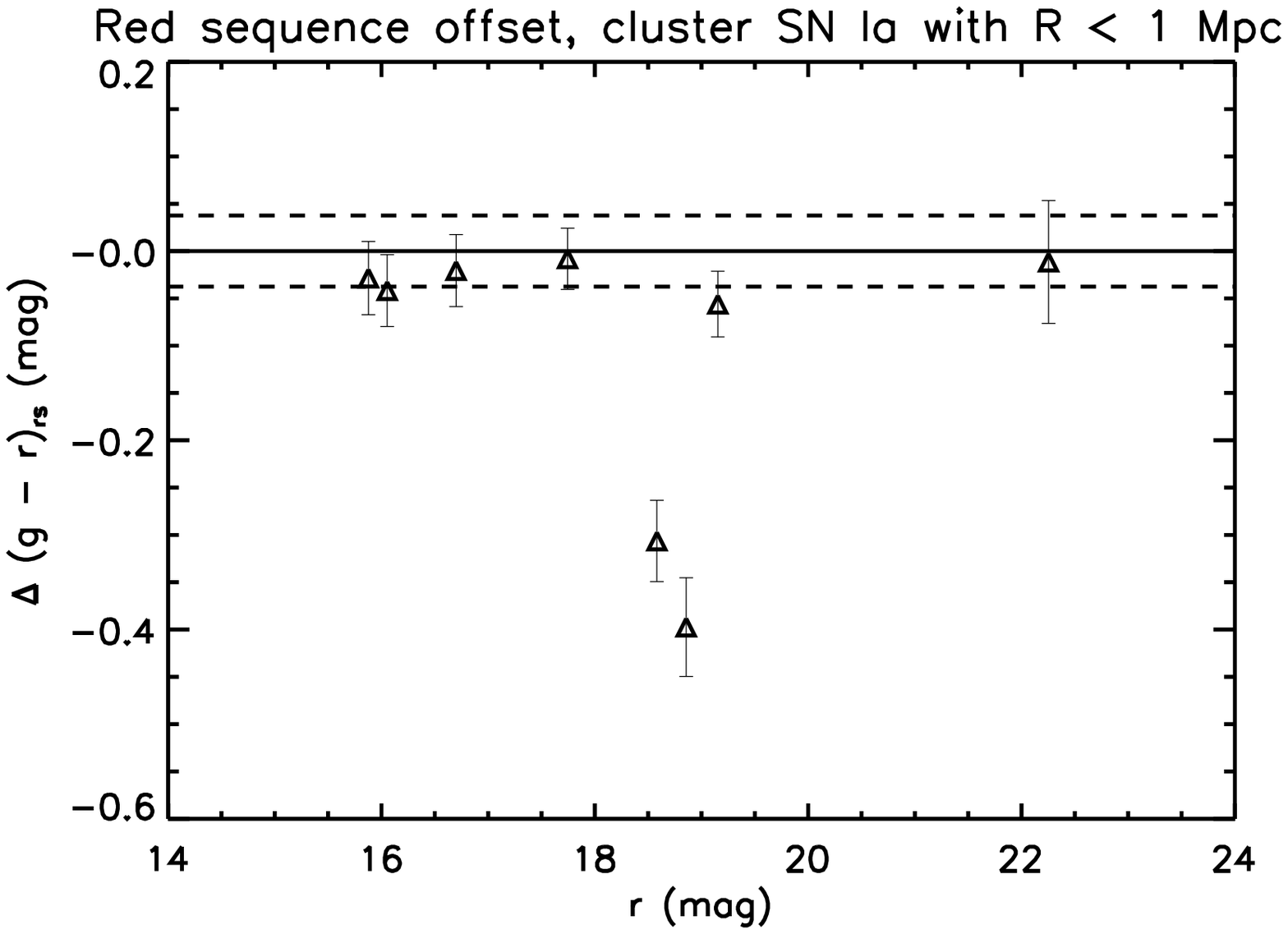} 
\epsfysize=4.5cm \epsfbox{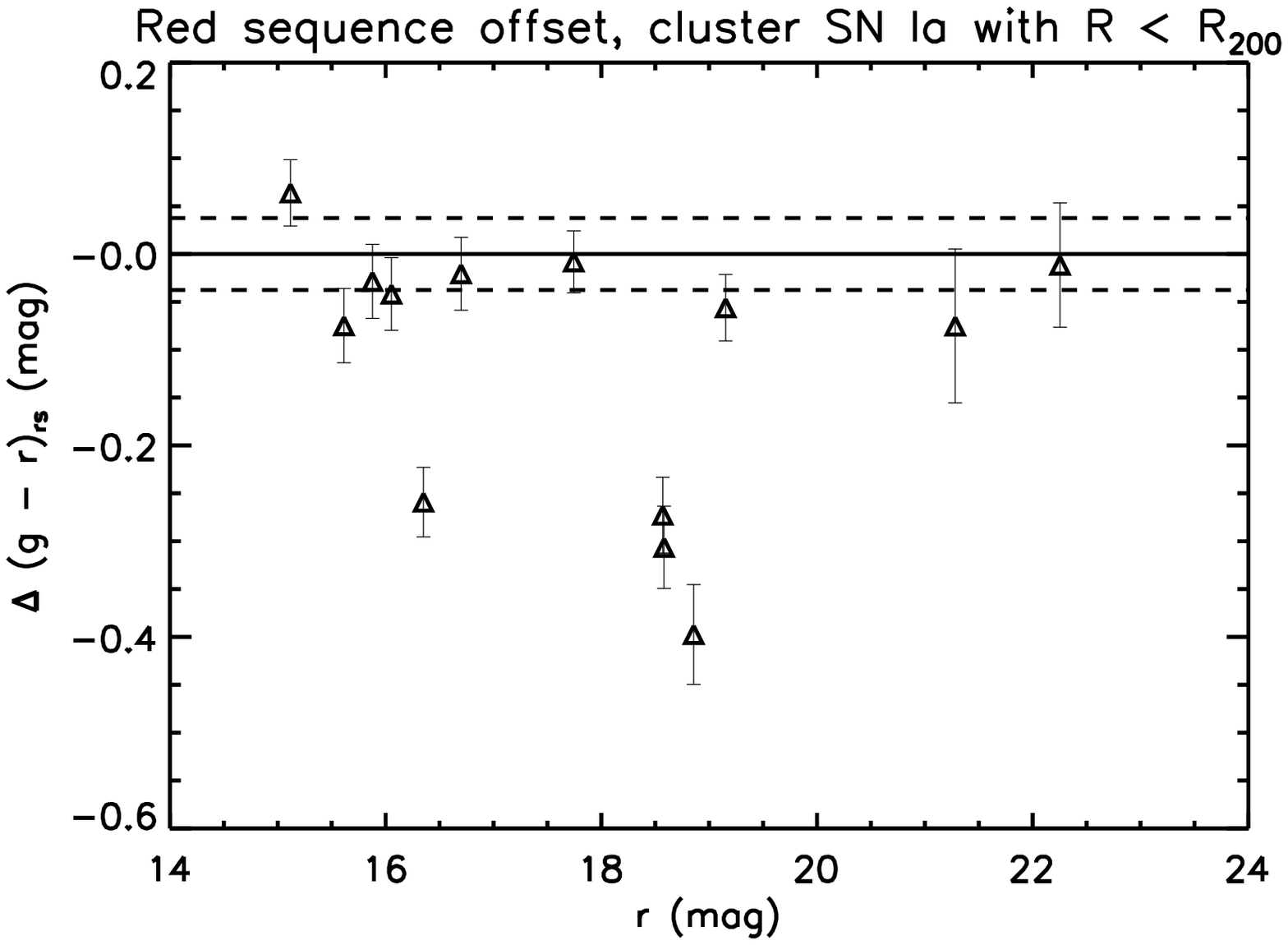} 
}
\caption{Red sequence offsets of SN host galaxies in MENeaCS.  The dashed lines indicate the median scatter in the red sequence over the enitre MENeaCS sample.  {\it Left} SN host galaxy offsets for the entire MENeaCS cluster SN Ia sample.  Ten out of the 19 SN Ia with hosts were consistent with the cluster red sequence, with bluer hosts becoming more prominent in the cluster outskirts.  {\it Middle} SN host galaxies which were within 1 Mpc of the cluster center, of which 6 out of the 8 were consistent with the red sequence.  {\it Right} SN host galaxies within $R_{200}$, of which 9 out of 13 were consistent with the cluster red sequence.  We use these SN numbers for determining out red sequence SN Ia rate in \S~\ref{sec:rates}. \label{fig:hostprops} }
\end{center}

\end{figure}

\begin{figure}
\begin{center}
\mbox{ 
\epsfysize=5.0cm \epsfbox{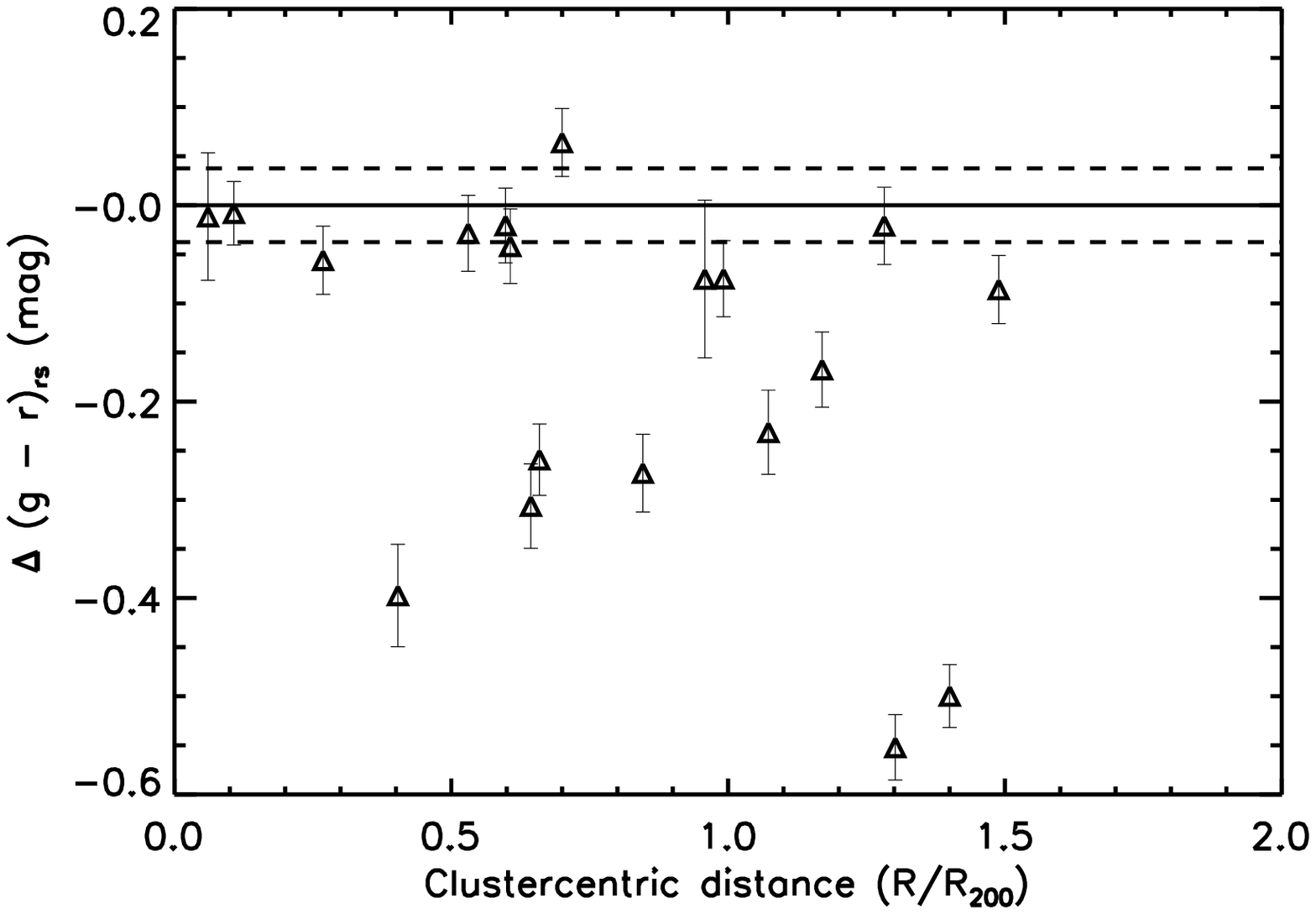} 
\epsfysize=5.0cm \epsfbox{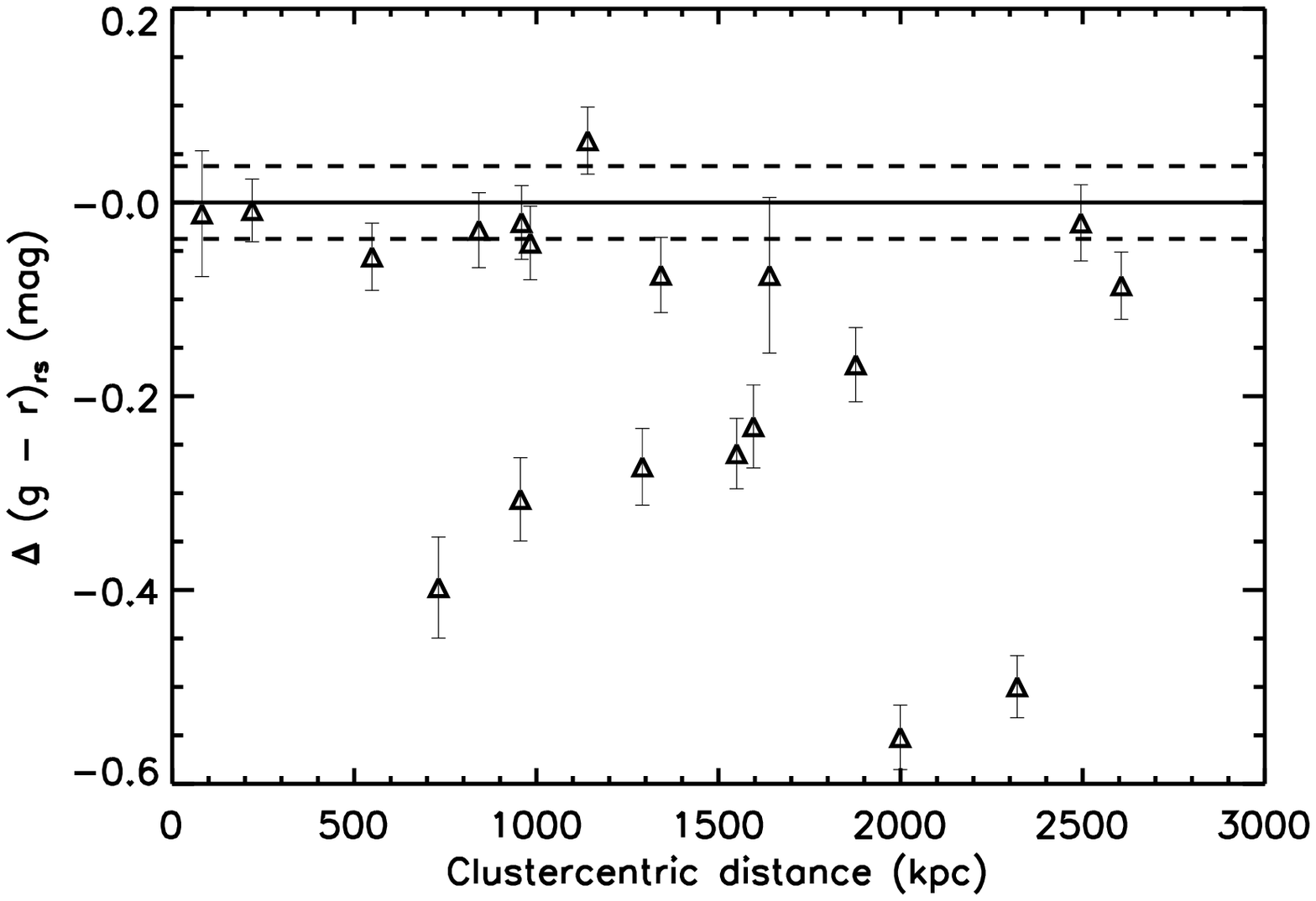} 
}
\caption{Red sequence offsets of SN host galaxies in MENeaCS as a function of clustercentric distance in units of $R_{200}$ (Left) and kpc (Right).    The dashed lines indicate the median scatter in the red sequence over the enitre MENeaCS sample.  Note that the most central host galaxies are all hosted by red sequence hosts, but the sample size is too small to say anything definitive.  \label{fig:hostprops_rad} }
\end{center}
\end{figure}

We utilize our well-defined cluster red sequences to measure a color offset between the red sequence and each of our host galaxies.  This serves as an indicator of the relative color of the host with respect to the cluster galaxy population as a whole, and allows us to pick out those cluster SN Ia which were hosted by red sequence galaxies -- we also list these color offsets in Table~\ref{hosttab}.  A plot of our hosts' red sequence offsets as a function of host magnitude can be seen in Figure~\ref{fig:hostprops}, with the horizontal dashed lines indicating the median scatter in the red sequence over the entire MENeaCS sample.  We simply define a host to be consistent with the cluster red sequence if its 1$\sigma$ color uncertainty overlaps with the median scatter in the red sequence.  In Figure~\ref{fig:hostprops_rad} we also plot our hosts' red sequence offsets as a function of clustercentric radius.  For our SN hosts within 1 Mpc of the cluster center, 6 out of 8 are consistent with the red sequence, while 9 out of 13 are red sequence hosts within $R_{200}$.  In the overall cluster SN Ia sample, 10 out of the 19 cluster SN Ia which were hosted (four others were hostless intracluster SNe) were consistent with the red sequence.  We use the above SN numbers when deriving our SN rates at different radii and in different galaxy populations in \S~8.  Although the most central host galaxies are all on the red sequence, the sample size is too small to say anything definitive about the radial distribution of host properties in MENeaCS.

Eleven out of the 19 SNe that were hosted were within one half-light radius, as reported by {\sc SExtractor}.  A `by eye' $r^{1/4}$ fit to the radial distribution of SNe from their host (see Figure 1 of Sand et al. 2011) is excellent.   This suggests that there is no SN detection deficiency in our galaxy cores  (as verified by our detection efficiency simulations in \S~7) and that the SNe ``follow the light" of the galaxy population, at least in the sense of their radial distribution.  More detailed work on the structure and morphology of our hosts will be presented in the future.

We finally note that the bulk of our SN host galaxies consistent with the cluster red sequence are on the blue side of the sequence.  We will defer a thorough discussion as to whether this might be an indication  of some residual star formation in these hosts for a future work, where we will gather much more detailed information about each.

\section{Supernova Detection Efficiency}\label{sec:deteffs}
In order to determine SN\,Ia rates from the MENeaCS survey, we must have a measure of our detection efficiency. To do this we generated a population of artificial SNe with realistic intrinsic properties, added them to our data, and ran them through the SN detection pipeline described in Section \ref{sec:redux}. The fraction of planted SN recovered was then used to derive our efficiency, described below, and included in our rates calculation, as described later in Section \ref{sec:rates}.

\subsection{The Population of Fake SNe\,Ia}

For a robust evaluation of our detection efficiencies, we planted artificial SNe\,Ia in clusters and epochs with a variety of properties. Since a calculation of our detection efficiency on every single image would be prohibitive, we used a representative sample of our data, constructed a look-up table of detection efficiencies from these, and then applied it to our sample as a whole for our SN rate calculation (\S~\ref{sec:rates}).  We used 18 clusters distributed evenly in redshift, and for each cluster chose 2-6 epochs which covered a range of PSF FWHMs, sky backgrounds, and dates. This included 74 of the 580 total epochs in MENeaCS (13\%).

We plant many artificial SNe\,Ia per field, but avoid overcrowding, so the total number of artificial coordinates for each cluster are set by the number of galaxies in the field. We used three hosting situations for the artificial SNe: red-sequence (RS) cluster hosts, non-RS hosts, and hostless. We plant a SN in half of the RS galaxies and 1\% of the other galaxies in the field, along with dozens of hostless SNe.  Without redshifts we cannot know which non-RS galaxies are cluster members, but for the purposes of detection efficiencies we can treat them as such. All artificial SNe are given the redshift of the cluster.

To generate artificial SNe coordinates which ``follow the light" in galaxies (i.e. occur more often in more luminous galaxies, and are distributed proportionally to their radial luminosity profiles) we used a custom routine which, given the reference image, the positions of potential hosts, and the number of SNe, returned a set of coordinates distributed proportional to galaxy light.  This resulted in realistic distributions of host luminosity and SNe\,Ia host offsets for the sample of artificial SNe.  Hostless artificial SNe were distributed randomly, but we did not allow a `hostless' artificial SN to be within 10 effective radii of any catalog galaxy.

Our population of artificial SNe\,Ia were given a very wide distribution of apparent magnitudes, bracketing our realistic detection sensitivities. A total of 68687 individual transient sources were simulated. The raw instrumental flux of a artificial SNe\,Ia was calculated from its given apparent magnitude and our real-time photometric calibrations (described in Section \ref{sec:photcal}).  A custom routine was used to plant artificial SNe\,Ia as a Moffat profile source with a PSF FWHM of that image. These altered images were run through the transient detection and automatic culling procedure described in Section \ref{sec:redux}. 

For recovery of artificial SNe, the visual review of detection triplets was unnecessary because only obviously spurious artifacts, not potential SN, are rejected at this stage, especially within our magnitude range of interest.  Indeed, our faint end detection limit, as described in the next section, is dominated by our spectroscopic constraints, and not our ability to detect transients down to even fainter magnitudes.

\subsection{Recovery Statistics of Fake SNe\,Ia}

\begin{figure}
\begin{center}
\mbox{ 
\epsfysize=6.0cm \epsfbox{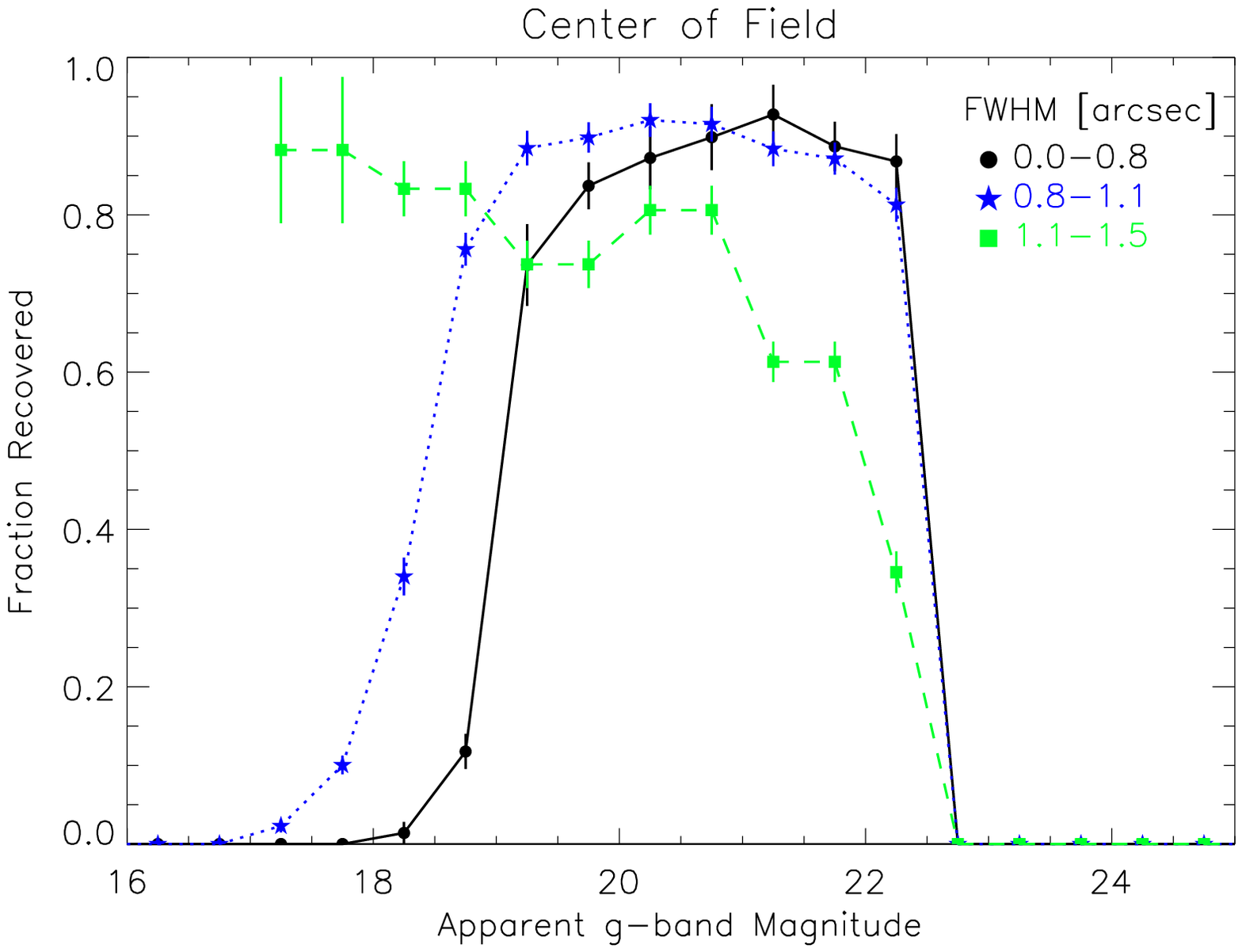} 
\epsfysize=6.0cm \epsfbox{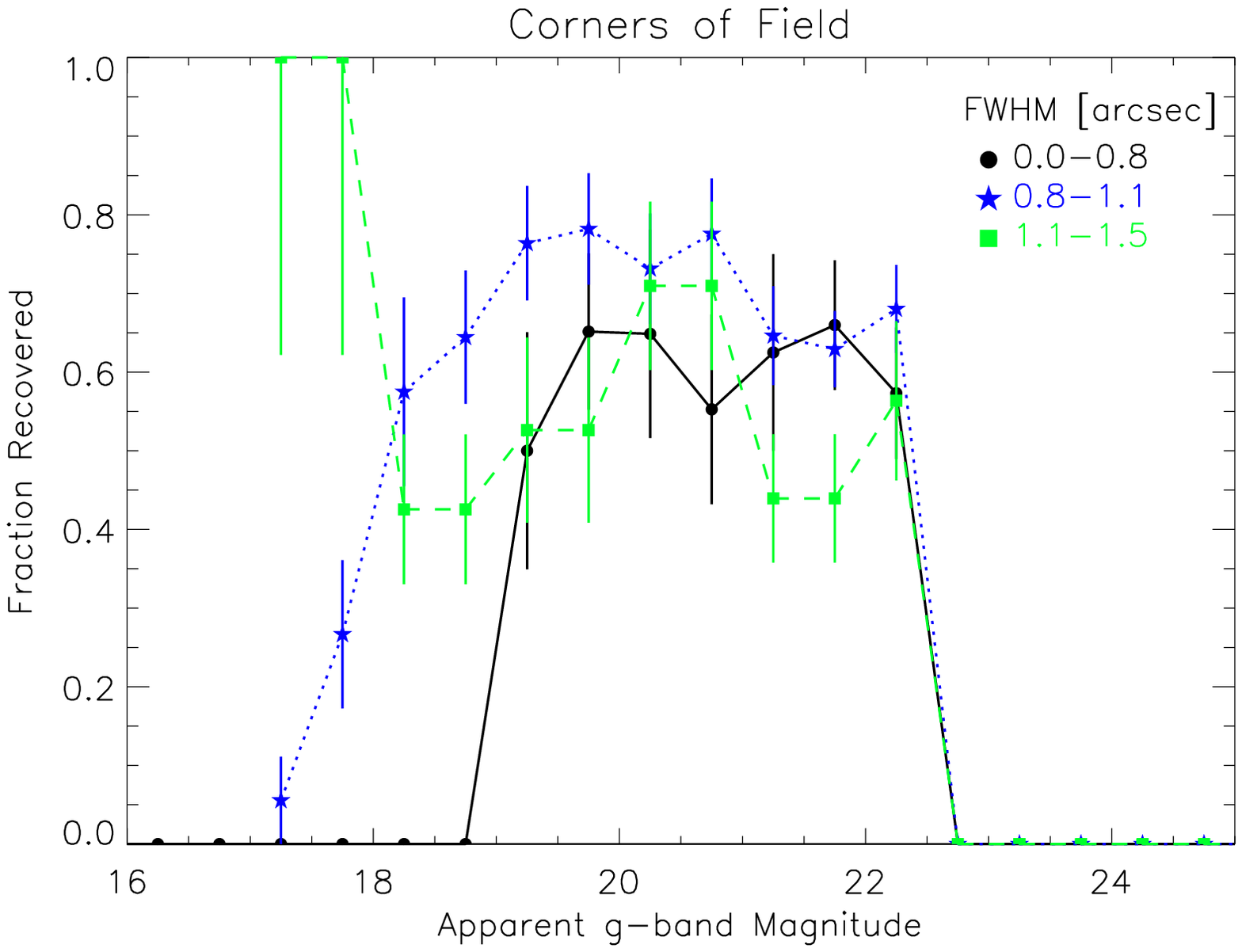} 
}
\caption{Sample supernova detection efficiencies ($\rm \eta(m)$) in the MENeaCS survey. Detection efficiencies for the central regions of our images sections (left) and  corner image sections (right) are shown.  Efficiencies for good seeing (black), acceptable seeing (blue), and bad seeing (green) are shown in each.  See \S~\ref{sec:deteffs} for details, along with an explanation for the detection efficiency fall off at faint and bright magnitudes. \label{fig:SNdeteff}}
\end{center}
\end{figure}

For the rates calculation described in Section~\ref{sec:rates}, we require the detection efficiency as a function magnitude: $\rm \eta(m)$. This is the fraction of planted objects which were recovered, in each apparent magnitude bin. We found the primary two factors that affected the shape of $\rm \eta(m)$ are position on the MegaCam field of view and the image PSF FWHM, as shown in Figure \ref{fig:SNdeteff}. Other surveys find that epoch zeropoint or sky background affect their efficiency, but CFHT's queue-scheduling resulted in nearly all our data being taken in very good conditions. In the few cases where our sky backgrounds or zeropoints are worse, so was the seeing, and thus accounting for efficiency changes with image PSF FWHM is the only necessary factor.
 
In processing we divide every image into 25 sections (5x5), and found the detection efficiencies are lowest in the corners, and lower along the edges, than in the center 3x3 sections of the field.  There is a simple reason for this. We {\sc SWarp} all exposures of an epoch into 20x20k images centered on the BCG. Of our multi-point dither pattern, only 2 pointings are done per epoch, which results in shallower, single-exposure regions along some edges and in some corners (the initial reference image has 4 pointings for complete coverage). In these single-exposure regions, bad columns and chip gaps are not filled, and the PSF convolution process creates larger artifacts around these areas, degrading our detection efficiencies at the field edges. The effect of this reduced $\rm \eta(m)$ on our rates is minimal as the field edges contain only a small amount of the cluster stellar mass. The shape of $\rm \eta(m)$ is also altered in bad seeing. When the PSF FWHM was large, bright transients had better recovery statistics. Although transients bright enough to benefit from this are rare, we use the appropriate detection efficiency for image PSF FWHM, as described in Section \ref{sec:rates}.  For reference, 84\% of our data had a PSF FWHM better than 1\farcs1, while 38\% was better than 0\farcs8.

The detection efficiency cutoff at faint and bright magnitudes, as seen in Figure~\ref{fig:SNdeteff}, are easily explained.   First, we imposed our spectroscopic followup criteria of $-0.5 \lesssim g-r \lesssim 0.8$ and $17.5 \lesssim g \lesssim 22.5$ into our artificial SN recovery statistics, leading to the faint-end cutoff in detection efficiency seen in Figure~\ref{fig:SNdeteff}.  The bright-end cutoff in detection efficiency is due in part to image saturation (thus our detection efficiency for bright objects is higher in poor seeing) and a small bug in our SN detection software algorithm.   An important and automated step in the SN detection pipeline fit a PSF to any SN candidate in our difference images.  A candidate with a high $\chi^{2}>50$ was deemed an artifact and rejected as a legitimate SN candidate.  However,  real, bright SN candidates were being rejected with high $\chi^{2}$ values, due to their high signal to noise ratio and the inadequacy of the PSF model being applied to them.  Although a variable $\chi^{2}$ PSF fitting cut would have allowed us to routinely detect SNe $\sim1$ magnitude brighter, our final detection efficiencies allowed us to discover SNe over several CFHT epochs.  The essential task is to identify our true detection efficiencies so that  an accurate SN rate can be calculated.

Many surveys find a degraded efficiency in galaxy cores. We do see this, but only in the very centers of the brightest ($\rm r \lesssim16$ mag) galaxies. Thus, it affects only a very small fraction of our surveyed mass, and is accounted for by the fact that our artificial SNe were realistically planted proportional to luminosity; we need no additional correction for this.  As verification, we found no deficiency in the number of actual SNe discovered in the central regions of our hosts, as discussed in \S~\ref{sec:SNhost}.

\section{The Cluster SN rate}\label{sec:rates}

In this Section we present our final cluster SN rate measurements, along with a thorough discussion of our sources of statistical and systematic uncertainty.  We present our SN Ia rate within 1 Mpc of the cluster center, as well as within $R_{200}$.  Additionally, we specifically measure the SN~Ia rate in red sequence galaxies at both radii, as well as the cluster galaxy population as a whole.  The red sequence SN rate will be used in \S~\ref{sec:DTD} for constraining the late-time delay time distribution of SNe Ia.

\subsection{The SN\,Ia Rate Calculation} \label{sec:ratecalc}

Our calculation of the SN\,Ia rate is similar in spirit to that presented in Sharon et al. (2007) and Barbary et al. (2011).  The rate of Type Ia supernovae, $\rm R_{Ia}$, is: 

\begin{equation}
\label{eq:rate}
R_{Ia} = \frac{ N_{Ia}/C_{Spec} }{ \sum_{j=1}^{j=N_{im}} \Delta t_j M_j  }
\end{equation}

\noindent
where $\rm N_{Ia}$ is the number of SNe\,Ia discovered and $\rm C_{Spec}$ is our spectroscopic incompleteness due to weather and scheduling difficulties, which we have calculated to be $0.91$ (see \S~\ref{sec:spectarg} and Sand et al. 2011 for details). The cluster luminosity or stellar mass, depending on which you calculate the SN rate with respect to, surveyed in the $\rm j^{th}$ observation is represented by $M_j$. The control time, $\rm \Delta t_j$, for the $\rm j^{th}$ of $\rm N_{im}$ images is expressed as:

\begin{equation}
\Delta t = \int_{t_1}^{t_2} \eta(m(t)) dt
\label{eq:deltat}
\end{equation}

\noindent
where $\rm m(t)$ is the supernova light curve (magnitude as a function of time), and $\eta$ is the SN detection efficiency, which varies with field position and image FWHM, as described in Section \ref{sec:deteffs}. The integration limits, $\rm t_1$ and $\rm t_2$, are set to practical boundaries defined by our limit on color at time of discovery, $\rm g-r\lesssim0.8$. 

Since we are able to observe a given cluster SN for several months, it is likely for it to be re-discovered in subsequent epochs, or for it to be discovered by our pipeline after a second observation if the object was missed in the first (for example, if it occurred near a chip gap or the observing conditions were less favorable in the first epoch).  To account for this, when multiple images of a cluster are taken within $\rm <50$ days of each other, we subtract the probability that it was also detected previously:

\begin{equation}
\eta = \eta_j - \eta_j \eta_{(j-1)} - \eta_j \eta_{(j-2)} - \eta_j \eta_{(j-1)} \eta_{(j-2)}
\label{eq:eta}
\end{equation}

\noindent
where $\rm \eta_j$ is for the $\rm j^{th}$ epoch, and $\rm \eta_{j-1}$ and $\rm \eta_{j-2}$ are the detection efficiencies from the two previous epochs \citep[see][for a similar methodology]{Sharon07}.  

As described in Section \ref{sec:deteffs}, the shape of $\rm \eta(m)$ varies across the field. The stellar luminosity/mass, $M_j$ in the denominator of Eqn~\ref{eq:rate}, also varies across the field.  To account for the spatial variation of $\rm \eta(m)$ we determine three control time values for the center, edge, and corner regions of the image.  These are each multiplied by the fraction of the total $M_j$ contained in each field area and summed.  We have assumed that the cluster luminosity (and stellar mass) profile is a projected NFW profile \citep{NFW96,NFW97} with concentration parameter, $c=3$, as has been observed for the galaxy profile in galaxy clusters \citep{Lin04b}.

\subsection{Distributions of Intrinsic SN\,Ia Parameters}

Although SNe\,Ia are used as standard candles, there is diversity among their intrinsic light curves \citep{Phillips93} which we must include to properly calculate $\rm R_{Ia}$.  We use a Monte Carlo method in which the rate is calculated over many SN\,Ia light curves, with sub-types and peak magnitude drawn randomly from observed distributions (described below). 

Previous cluster SN rates have not explicitly considered the 91bg-like (sub-luminous) and 91T-like (over-luminous) SN\,Ia subtypes in their rate calculations.  Distributions of SN\,Ia subtype  depend on the galaxy sample observed; 91bg-like SNe\,Ia happen more often in early-type galaxies, and  91T-like SNe\,Ia prefer late-type galaxies \citep{Howell09}.  Recently, \citet{Li11} have shown that $\sim$16\% and $\sim$10\% of all SNe Ia were 91bg- and 91T-like, respectively, while these ratios change to $\sim$34\% and $\sim$3\% in early type galaxies.  As discussed in \S~\ref{sec:SNID}, MENeaCS did not have the well-sampled light curves or spectroscopic sequences necessary to definitively determine SN Ia subtype, and so our reported SN rates technically include all SN Ia.  We will present our SNe rates both with and without consideration for sub-type, allowing for simple comparison with values in the literature.  These rates always agree within 1$\sigma$.

Thus, we use the template light curves for normal, 91bg-like (sub-luminous), and 91T-like (over-luminous) SNe\,Ia from \citet{Nugent02}, and apply the following to obtain a light curve in apparent magnitudes, $\rm m(t)$: the randomly chosen peak $B$-band magnitude; a stretch value based on the established stretch-brightness correlation \citep[e.g.][]{Phillips93,Perlmutter97}; the cluster's redshift; the appropriate $B$- to $g$ (and $r$)-band SN\,Ia k-correction \citep{Hsiao07}; and the inverse of the photometric corrections for the epoch.

For the Monte Carlo procedure, SN Ia subtype is drawn randomly from the relative rates of \citet{Li11} and the peak B-band magnitude is drawn from the volume-limited luminosity function of the appropriate subtype, which are derived from Table 3 of \citet{Li11}.  Note that the \citet{Li11} SN Ia luminosity function does not include a correction for host galaxy extinction, and by drawing from it we implicitly include the effects of host extinction into our SN rate measurement.  For normal SNe\,Ia we use a stretch value correlated with its peak magnitude, as in $\rm M_{B,peak}=-19.3+\alpha(1-s)$ where $\rm \alpha\sim1.5$ \citep{Astier06}. Altogether, this yields a realistic distribution of SN\,Ia light curves for input to the Monte Carlo rates calculation. The 91bg- and 91T-like SNe Ia have their own Nugent template light curves and require no additional stretching.

The cluster rates measurements of many past surveys (e.g. Mannucci et al. 2008) used a Gaussian distribution of SN\,Ia peak magnitudes, as described in \citet{Cappellaro97} --   we also present SN rates using that luminosity function for comparison purposes, which did not explicitly include the 91bg/91T subtypes.  Note that we do not include an extra host galaxy extinction term when using this SN luminosity function for the rate calculation, but find results consistent with the SN Ia rate derived from the \citet{Li11} luminosity function -- see the next section for details.

\subsection{Results}

Putting together the elements in the previous subsections, we proceed to calculate the cluster SN rate within 1 Mpc and $R_{200}$, both for the red sequence population and for the cluster as a whole.  We ran 1000 trials of our SN rate calculation code for each scenario aforementioned, varying the SN Ia light curve, detection efficiency, cluster luminosity/stellar mass and SN number, $\rm N_{Ia}$, according to their uncertainties or assigned distributions. In this way, our errors for detection efficiency (see error bars in Figure \ref{fig:SNdeteff}), for cluster luminosity/stellar mass (discussed in Section \ref{sec:clusterlum}), and for Poisson statistics on $\rm N_{Ia}$ are included in the final statistical uncertainties.  Our reported SN rate corresponds to the median of the 1000 Monte Carlo trials, with the quoted statistical uncertainties corresponding to the 16th and 84th percentiles (the 68\% confidence interval).

In Table \ref{table:SNrates} we present the MENeaCS SN rate measurements for the three discussed distributions of SN\,Ia luminosity functions: 1) the \citet{Li11} SN~Ia luminosity function without SN91bg/91T subtypes, 2) the \citet{Li11} luminosity function with SN91bg/91T subtypes and 3) the Gaussian SN Ia luminosity function of \citet{Cappellaro97}, which included no SN91bg/91T subtypes.  We note that our results for these three SN\,Ia luminosity functions are consistent to within our 1$\sigma$ statistical uncertainties.  We compare our rates with those in the literature in \S~\ref{sec:compare}, and will use our \citet{Li11} SN Ia luminosity function results, without subtype, for that purpose.

Our quoted statistical uncertainties come from several factors: the intrinsic distribution of light curve properties,  Poisson statistics on the SN number, the uncertainty on the cluster luminosity, and uncertainties in our detection efficiencies.
To determine the relative contribution of each factor, we also run the Monte Carlo with the random draw deactivated for each factor in turn, and instead use only its fiducial value. From this analysis it is very clear that we are still dominated by the small number of discovered SN Ia, with  $\gtrsim$90\% of the error budget originating from this.

We are aware of two possible sources of systematic uncertainty.  The first was identified in \S~\ref{sec:sdsscompare}, where we found a offset of up to $\sim$10\% in our cluster luminosities depending on if we used SDSS versus MENeaCS photometry.  To investigate, we rescaled our SN rates, adjusting our cluster luminosities/stellar masses by $\pm$10\%.  Since the SN rate is inversely proportional to cluster luminosity/stellar mass this systematic is straightforward to compute, and can be seen to be smaller than our statistical uncertainties in all cases.  Our second systematic is due to a possible misestimation of $R_{200}$ in our cluster sample, based on the $L_{X}-M_{200}$ scaling relation of \citet{Reiprich02} -- see \S~\ref{sec:clussample}.  If this scaling relation is  systematically off from the true $L_{X}-M_{200}$ relation, than this would systematically effect our $R_{200}$ values.  By comparing our SN rates at 1 Mpc and $R_{200}$ in Table \ref{table:SNrates}, one can see that the SN rate is not a strong function of cluster radius. Pressing the point, we assume a factor of two systematic offset in our cluster masses, which results in a $\pm$25\% systematic in our value of $R_{200}$.  Recalculating our SN rates at these radii yields SN rates which are offset by $\lesssim$10\%, and we include this systematic uncertainty into our reported results.

Our final systematic uncertainties in Table~\ref{table:SNrates} consist of the two factors discussed above, and the two values are summed in the case of our rates at $R_{200}$.
Assuming we have identified all of the appropriate sources of systematic uncertainty, then we can safely say that our SN Ia rate measurements are dominated by our statistical uncertainty, given that our systematics are always subdominant.

\section{Comparison to previous work}\label{sec:compare}

Cluster SN rates, at several redshifts, have been reported by multiple groups -- we present a summary of the literature results in Table~\ref{table:SNsumm}.   We have omitted the high redshift measurement of \citet{Galyam02}, given that they found only one likely SN Ia candidate at $z\sim0.8$.  When necessary, we correct SN rate values to that expected for a ``diet" Salpeter IMF since this is the dominant IMF utilized in the cluster SN rate literature. We note in the Table at which clustercentric radius the SN Ia rate was taken, especially if rates at multiple radii were reported.  Since most cluster SN rate measurements are not tied to $R_{200}$, but a physical radius, we report our own rates within 1 Mpc in this Table.  Likewise, we present our rate derived from the \citet{Li11} SN~Ia luminosity function, with no 91bg/91T subtypes.
   
\begin{figure}
\begin{center}
\mbox{ 
\epsfysize=6.0cm \epsfbox{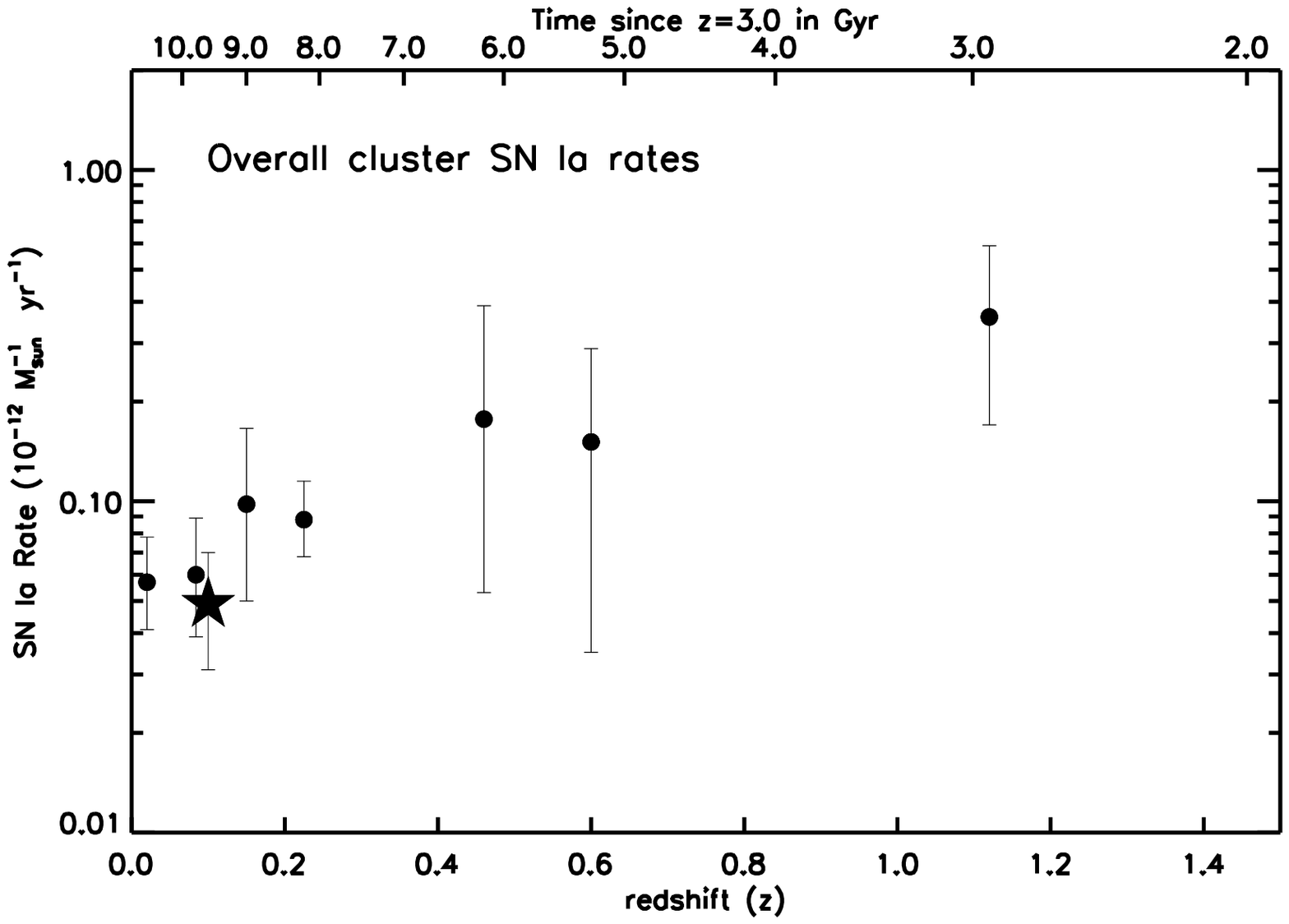} 
\epsfysize=6.0cm \epsfbox{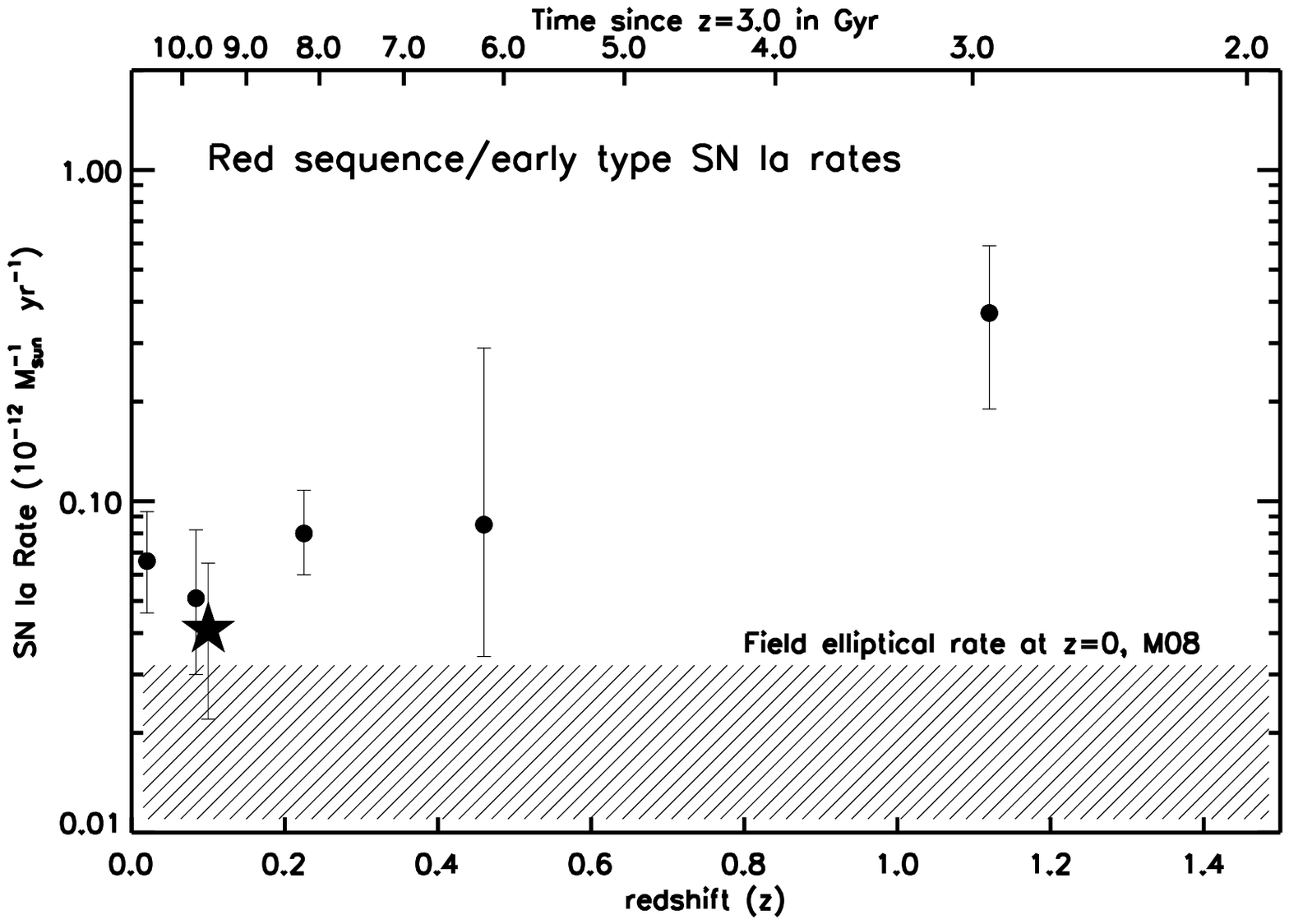} 
}
\caption{A compilation of cluster SN Ia rates as a function of redshift over all galaxy types (left), and in early type/red sequence galaxies (right).  Exact values, and their sources, are in Table~\ref{table:SNsumm}.  The star-point represents data measured in the current work.  In the right panel, we also indicate the field elliptical SN Ia rate at $z=0$, as determined by \citet{Mannucci08} as the thatched box.  The MENeaCS red sequence rate is consistent with the field elliptical rate at $z=0$. \label{fig:SNrates}}
\end{center}

\end{figure}

A plot of the extant cluster SN Ia measurements as a function of redshift can be seen in the left panel of Figure~\ref{fig:SNrates}.  The MENeaCS cluster rate, plotted as the star, is consistent with the literature rates at $0 < z < 0.2$ to within 1$\sigma$, although they are systematically higher.

In Table~\ref{table:SNsumm} and the right panel of Figure~\ref{fig:SNrates}, we present SN rates in cluster red sequence/early type galaxies, when those values have been calculated.  Again, the MENeaCS red sequence rate is consistent with other low and moderate redshift measurements.  In fact, the only evidence that the red sequence/early type SN rate evolves with redshift is via the \citet{Barbary10} $z\sim$1.1 measurement, at the $\sim2\sigma$ level.  A more precise measurement of the $z\sim0.5$ rate would confirm the evolution, and aid in constraining the DTD (see next Section).

If we assume that red sequence galaxies are synonymous with early type galaxies, there is no evidence for a cluster rate enhancement in the early-type SN Ia rate, as has been reported in the literature previously \citep[][but see, Cooper et al. 2009]{Sharon07,Mannucci08,Dilday10}, albeit at low statistical significance.  The early-type galaxy field rate, as reported by the SDSS-II Supernova Survey \citep{Dilday08,Dilday10} at $z\sim0.1$, was $0.078^{+0.03}_{-0.02}$ SNur --  statistically identical to our own $r$-band SN rate among red sequence galaxies.  Likewise, the `average' SN Ia rate in  elliptical galaxies in the LOSS survey was $0.051^{+0.018}_{-0.017}$ SNuM \citep[see Table~6 of ][]{Li11}, in agreement with our own red sequence rates, although this rate may include both field and cluster elliptical galaxies.  Mannucci et al. 2008 did disentangle field and cluster elliptical rates from the \citet{Cappellaro99} rate study, and found $0.019^{+0.013}_{-0.008}$ SNuM in the field, which is $\lesssim$1$\sigma$ lower than our own measurement, which we illustrate in the right panel of Figure~\ref{fig:SNrates}.  Given that our own red sequence rates agree with field measurements from SDSS-II and \citet{Cappellaro99}, along with the combined LOSS rate, to the levels of our measurement uncertainties, we must await higher precision field and cluster measurements before any differences will be detectable.

A simple model for the SN~Ia rate is the `A+B' representation, with $R_{Ia} = A \times M + B \times \dot{M} $, where $M$ is the stellar mass surveyed and $\dot{M}$ is the current star formation rate \citep{Scannapieco05,Mannucci06}.  Both $A$ and $B$ are constants determined from the data.  While this model is too simplistic if the true SN~Ia DTD has a broad distribution, it is still useful for comparison with coefficient values in the literature.  Recent updates to the `A+B' model have suggested that the rate component associated with the $A$ value may go like $M^{0.67}$ \citep{Smith11}, although we do not consider this further for our simple comparison.  If our red sequence galaxy sample is composed solely of old stars, with no recent star formation, than their SN~Ia rate is a clean measurement of the `A' coefficient.  In that case, we find an $A$  value of $4.1\pm1.7 \times 10^{-14}$ SNe yr$^{-1}$ M$_{\odot}^{-1}$.  For comparison, the SDSS-II SN survey recently found $A = 2.75^{+0.57}_{-0.47} \times 10^{-14}$ SNe yr$^{-1}$ M$_{\odot}^{-1}$ \citep{Smith11}, while the SNLS found $5.3\pm1.1 \times 10^{-14}$ SNe yr$^{-1}$ M$_{\odot}^{-1}$ \citep{Sullivan06} -- the MENeaCS value is in agreement with both of these measurements.

We also checked for signs of a SN rate change as a function of cluster mass within the MENeaCS survey itself by splitting our clusters into a `high mass' and a `low mass' sample, dividing them at the median MENeaCS cluster mass of $6.61\times10^{14} M_{\odot}$.  When we calculate our red sequence SN Ia rate for each of these samples separately, using the \citet{Li11} luminosity function with no 91bg/91T subtypes and a limiting clustercentric radius of 1 Mpc, we get a SN rate of $\rm 0.0372_{-0.0239}^{+0.0249} (stat) \ _{-0.0034}^{+0.0041} (sys)$ SNuM for our high mass sample and $\rm 0.0450_{-0.0293}^{+0.0310} (stat) \ _{-0.0041}^{+0.0050} (sys)$ SNuM for our low mass sample.  There is no discernible change in red sequence SN Ia rate as a function of cluster mass in the MENeaCS sample, although we are hindered by the small number of SNe in the high and low mass cluster bins.

Finally, we note that even though we do not detect a cluster SN rate enhancement in red sequence/early type galaxies versus that seen in the field, it is still possible that some or all of these SNe belonged to a young stellar population, as has been suggested to be the cause of the cluster elliptical rate enhancement.  Even galaxies that lie on the optical red sequence can have some recent star formation activity \citep[e.g.][among others]{Kaviraj07,Trager08}, and the fact that nearly all of our red sequence hosts lie on the blue side of the red sequence indicates that more detailed study is warranted.  However, as we constrain the late-time DTD of SN Ia in the next section, we make the simplest assumption consistent with our current data -- red sequence/early type galaxies are made of stars formed at high redshift ($z\sim3$) in a single burst, and have evolved passively until the present epoch.

\section{Constraints on the Delay Time Distribution}\label{sec:DTD}

In this section we combine the MENeaCS cluster SN Ia rate measurement at $z\sim0.1$ with others from the literature at $0 < z < 1.4$ to constrain the late time DTD of SN Ia.  In contrast to previous work, we focus on cluster rates in red sequence (or early type) galaxies which we assume are composed of stars that formed at high redshift in a single burst, and have evolved passively to this day.  Utilizing the cluster SN rates over all galaxy types risks including SN Ia from younger stellar populations, weakening the power of deriving the late-time DTD from cluster SN measurements.  Indeed, MENeaCS has spectroscopically confirmed seven cluster core collapse SNe (Graham et al., in preparation), even though they were not the primary target of the survey, suggesting that star formation continues in our cluster fields.

Under the assumption that all cluster red sequence/early type galaxy stars formed in a single burst at high redshift \citep[e.g.][among many others]{Stanford98}, the recovery of the DTD is straightforward -- it is the SN rate renormalized to the stellar mass at the formation epoch rather than at the epoch of measurement.  To account for mass lost during stellar evolution, we take the tabulated values of \citet{Bruzual03} for a Salpeter IMF, and convert to a ``diet" Salpeter by dividing the tabulated mass lost by a factor of 0.7 \citep[see e.g. ][]{Maoz10clus}.  Thus, the DTD is $\Psi(t) = m(t) R_{SNIa}(t)$, where $m(t)$ is the remaining fraction of stellar mass at time $t$ after the star formation burst, which for our chosen IMF is $m(t)=1-m_{loss}(t)/0.7$.

\begin{figure}
\begin{center}
\mbox{ 
\epsfysize=6.0cm \epsfbox{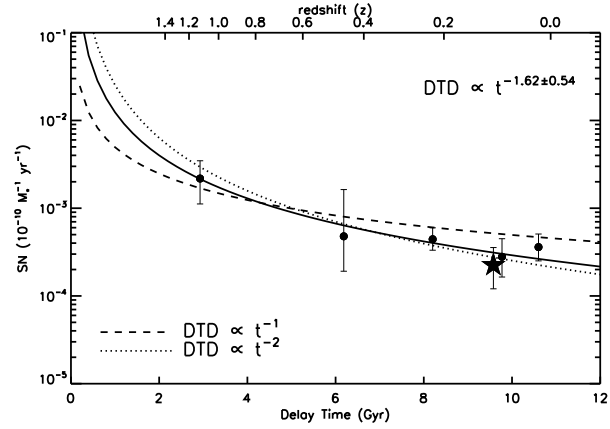}
}
\caption{Recovered delay time distribution based on early type/red sequence SN Ia rates in clusters, taken from Table~\ref{table:SNsumm}, and assuming that the constituent stellar population formed at $z=3$ in a single burst.  The star-point represents data measured in the current work.  The solid DTD curve is the best-fitting power law, with $\Psi \propto t^{s}$, and an $s$ value of $-1.62$.  The $1\sigma$ error region for the power law is $-1.08 < s < -2.16$.  We also plot the best-fitting $t^{-1}$ and $t^{-2}$ power laws for illustrative purposes, as these are the expected power law forms for the double degenerate and double detonation scenarios, respectively.  \label{fig:DTD}}
\end{center}

\end{figure}

The DTDs derived from the red sequence/early type cluster SN Ia rates in the literature are plotted in Figure~\ref{fig:DTD}.  A simple, and theoretically motivated, way to parameterize the late-time DTD is to adopt a power law, $\Psi (t) \propto t^s$, where $t$ is the time since the stellar system's formation.  We choose to leave the normalization as a free parameter in our fits since we have no leverage on the DTD at early times using the cluster SN technique.
For our assumed star formation epoch of $z_{f}=3$, we find a best fit value for the late time DTD scaling exponent of $s=-1.62\pm0.54$, using the {\sc MPFIT} package in IDL \citet{Markwardt09}.  If we change the formation epoch to $z_{f}=2.5$, we find $s=-1.50\pm0.49$.  Likewise, a formation epoch of $z_{f}=4.0$ yields a scaling exponent of $s=-1.78\pm0.60$.  

These values are largely consistent with similar late-time DTD measurements in the literature, although most find values closer to $s\sim-1$.  Using SN Ia in elliptical galaxies from the  Subaru/XMM-Newton Deep Survey, \citet{Totani08} found $s=-1.08\pm0.15$, although their SNe were not spectroscopically confirmed.  The most recent derivation of the DTD from cluster SN rate measurements have come from \citet{Maoz10clus} and \citet{Barbary10}, finding power law scaling exponents of $s=-1.2\pm0.3$ and $s=-1.31^{+0.55}_{-0.40}$, respectively, for an assumed star formation burst at $z_{f}=3$. Both of these analyses used the cluster SN rate over all galaxy types, with possible contamination from SNe Ia from young stellar populations.

\subsection{Comparison with theoretical predictions}

As touched on in \S~1, there are currently three major progenitor/explosion paths to SNe Ia;  the SD scenario, the DD scenario and the double detonation scenario.  We will compare DTD predictions for each scenario with our measurement of $\Psi \propto t^{s}$ where $s=-1.62\pm0.54$ to constrain the possible progenitor.  At present, DTD predictions involving binary population synthesis have trouble matching the observed SN Ia rates \citep[e.g.][]{Yungelson00,Ruiter09,Mennekens10}, and so we will focus on the shape of the DTD rather than on its normalization, although this caveat should be kept in mind.

A generic prediction of the DD scenario is a late time DTD with a power law form which goes like $\Psi \propto t^{-1}$ \citep[e.g.][among others]{Greggio05,Ruiter11}, although the specific exponential slope is dependent on the initial binary star separation distribution.  Our power law scaling exponent of $s=-1.62\pm0.54$ can be considered consistent with the DD scenario given this slight ambiguity in theoretical predictions.

In general, DTD predictions for the SD scenario predict late-time power law slopes which are steeper than for the DD scenario \citep{Greggio05}. In the binary population synthesis calculations of \citet{Ruiter09}, the SD DTD looks more like a step function with the SN Ia rate quickly dropping by roughly an order magnitude $\sim$6-7 Gyr after star formation.  The SD DTD of \citet{Mennekens10} has a similar shape, with the drop off age perhaps occurring a couple of Gyr earlier.  This is not inconsistent with our observed DTD, especially given the current uncertainty in the cluster early type SN rate at $z\sim0.5$, and an improved measurement here would be a promising path forward.

Recently, \citet{Ruiter11} determined the DTD for the double detonation scenario, and found a power law falloff with $\Psi \propto t^{-2}$.  This is in general agreement with our own late time DTD measurement to within the uncertainties.

Finally, \citet{Pritchet08} have proposed a SN Ia DTD which is proportional to the formation rate of white dwarfs, and should have a power law form which goes like $\Psi \propto t^{-1/2}$.  While this DTD is compatible with the trend of SN Ia rate versus specific star formation rate among the hosts in the Supernova Legacy Survey, it is incompatible with our results at the $>2\sigma$ level.  Indeed, \citet{Pritchet08} themselves suggest that an additional SN Ia channel is necessary even in the presence of their model.

To conclude, our late-time DTD measurement is in broad agreement with recent binary population synthesis calculations for the double detonation scenario, and is only $\sim1\sigma$ discrepant from expectations for the DD scenario.  An improved measurement of the cluster SN rate at $z\sim0.5$ would help to definitively rule out DTDs with shapes consistent with the SD scenario.

\section{Conclusions and Future Work}\label{sec:future}

MENeaCS surveyed 57 X-ray selected galaxy clusters ($0.05 < z < 0.15$) over a two year span and discovered 23 cluster SN Ia.  After carefully calculating our SN detection efficiency, cluster luminosities and stellar masses, and taking into account the known dispersion in SN Ia properties, we implemented a Monte Carlo technique for calculating our SN Ia rate.  Our final cluster SN Ia rate within $R_{200}$ (1 Mpc) is $0.042^{+0.012}_{-0.010}$$^{+0.010}_{-0.008}$ SNuM ($0.049^{+0.016}_{-0.014}$$^{+0.005}_{-0.004}$ SNuM), while our SN Ia rate in red sequence galaxies is $0.041^{+0.015}_{-0.015}$$^{+0.005}_{-0.010}$ SNuM ($0.041^{+0.019}_{-0.015}$$^{+0.005}_{-0.004}$ SNuM), assuming the SN luminosity function of \citet{Li11} with no SN~Ia subtypes.   First we note that the cluster SN Ia rates among red sequence galaxies and among the cluster galaxy population as a whole are statistically consistent with each other, naively suggesting that the cluster galaxy SN Ia rate is not strongly contaminated by SN Ia associated with recent star formation.  A comparison between our red sequence SN Ia rates and those reported for red/early type galaxies in the literature are consistent at the $\sim1\sigma$ level, a finding predicted by \citet{Cooper09}.  Although we do not see a cluster red sequence SN Ia rate enhancement versus the field, our results are also consistent, within the uncertainties, with the slight enhancement seen in previous work\citep{Sharon07,Mannucci08,Dilday10}.  We can not conclusively rule out the slight early type SN Ia rate enhancement seen in these earlier works.


We gathered cluster red sequence/early type SN Ia rates from the literature, along with our own measurements, to calculate the late time ($>2$ Gyr) DTD.  We fit a power law to the extant data  --  $\Psi(t) \propto t^s$ -- and we found $s=-1.62\pm0.54$, assuming a brief star formation epoch of $z_f=3$ for the red sequence stellar population, followed by passive evolution.  This DTD is consistent with both the DD and double detonation scenarios for the SN Ia progenitor.  An improved measurement of the red sequence SN Ia rate at $z\sim0.5$ would also constrain the SD progenitor model, since it is predicted to have an order of magnitude drop-off at roughly that epoch.

There are several future avenues of research.  A companion paper to the current work will present our cluster core collapse SN rate, and discuss the implications for current star formation in the cluster environment.  Additionally, we are gathering detailed data on the host properties of our SNe Ia, and of the cluster galaxies generally.  It is intriguing that nearly all of our red sequence hosts lie on the {\it blue} side of the sequence, and we are gathering deep optical spectroscopy to search for recent star formation in these systems.
Is it possible that the mismatch in normalization between theoretical late time DTDs and the measured SN Ia rates in supposedly old stellar populations is because of small amounts of star formation in cluster early type galaxies \citep[e.g.][]{Kaviraj07,Trager08}?

Finally, a new cluster SN search at $z\sim0.5$ of sufficient size to beat down the current SN rate uncertainties would allow for finer discernment of the late time DTD shape.   Additionally, the discovery of  intracluster SNe would provide the first definitive measurement of the ICL fraction at that redshift.

\acknowledgments

We are extremely grateful to the operators of the CFHT queue without whom this project would not be possible.  We thank Nelson Caldwell for managing the MMT/Hectospec queue.  We are grateful to Stephenson Yang for his patience and diligence regarding computer and network maintenance.  HH acknowledges support from a Marie Curie International Reintegration Grant.
This research has made use of the VizieR catalogue access
 tool, CDS, Strasbourg, France.  This research has made use of the
 NASA/IPAC Extragalactic Database (NED) which is operated by the Jet
 Propulsion Laboratory, California Institute of Technology, under
 contract with the National Aeronautics and Space Administration.
 
 This work is based in part on data products produced at the Canadian Astronomy Data Centre as part of the Canada-France-Hawaii Telescope Legacy Survey, a collaborative project of NRC and CNRS.
 
Observations reported here were obtained at the MMT Observatory, a
 joint facility of the Smithsonian Institution and the University of
 Arizona.  
 
Based on observations obtained with MegaPrime/MegaCam, a joint project of CFHT and CEA/DAPNIA, at the Canada-France-Hawaii Telescope (CFHT) which is operated by the National Research Council (NRC) of Canada, the Institut National des Science de l'Univers of the Centre National de la Recherche Scientifique (CNRS) of France, and the University of Hawaii. 

Based on observations obtained at the Gemini
Observatory (under program IDs GN-2009A-Q-10 and GN-2008B-Q-3), which is
operated by the Association of Universities for Research in Astronomy,
Inc., under a cooperative agreement with the NSF on behalf of the Gemini
partnership: the National Science Foundation (United States), the
Science and Technology Facilities Council (United Kingdom), the National
Research Council (Canada), CONICYT (Chile), the Australian Research
Council (Australia), Ministerio da Ciencia e Tecnologia (Brazil) and
Ministerio de Ciencia, Tecnologia e Innovacion Productiva (Argentina).

\bibliographystyle{apj}
\bibliography{mybib}

\clearpage

\begin{deluxetable}{lcccccc}
\tabletypesize{\small}
\tablecolumns{7}
\tablecaption{MENEACS Cluster Fields \label{table:clustertable}}
\tablehead{
\colhead{Cluster} & \colhead{$z_{clus}$} & \colhead{$\alpha$} & \colhead{$\delta$} & \colhead{$L_{X}$} & \colhead{M$_{200}$} & \colhead{$R_{200}$}\\
\colhead{} & \colhead{} & \colhead{(J2000.0)} & \colhead{(J2000.0)} & \colhead{$10^{44}$ ergs $s^{-1}$} & \colhead{10$^{14} M_{\odot}$}& \colhead{kpc}} \\
\startdata
Abell~2703&0.114&00:05:23.92&+16:13:09.8&2.72 & 4.64 & 1540\\
Abell~7&0.106&00:11:45.35&+32:24:55.2&4.52 & 6.61 & 1740\\
Abell~21&0.095&00:20:37.27&+28:39:33.6&2.64 & 4.51 & 1540\\
Abell~85&0.055&00:41:50.33&-09:18:11.2&9.41 & 10.27 & 2050\\
Abell~119&0.044&00:56:16.04&-01:15:18.2&3.30 & 5.14 & 1630\\
Abell~133&0.057&01:02:41.68&-21:52:55.8&2.85 & 4.78 & 1580\\
RXCJ0132-08&0.149&01:32:41.07&-08:04:04.5&3.49 &5.61 & 1620 \\
Abell~399&0.072&02:57:53.06&+13:01:51.8&7.06 & 8.15 & 1880\\
RXCJ0352+19&0.109&03:52:58.90&+19:41:00.3&3.89 & 5.96 & 1680\\
Abell~478&0.088&04:13:25.29&+10:27:55.0&16.47 & 19.83 & 2520\\
Abell~553&0.066&06:12:41.06&+48:35:44.3&1.83 & 3.43 & 1410\\
ZwCl0628&0.081&06:31:23.67&+25:01:06.7&3.20 & 5.13 & 1610\\
RXCJ0736+39&0.118&07:36:38.17&+39:24:52.0&4.41& 6.53 & 1720\\
Abell~644&0.070&08:17:25.59&-07:30:45.3&8.33 & 10.02 & 2020\\
Abell~646&0.129&08:22:09.57&+47:05:52.6&4.94 & 6.47 & 1710\\
Abell~655&0.127&08:25:29.02&+47:08:00.1&4.90 & 6.54 & 1720\\
Abell~754&0.054&09:08:32.34&-09:37:47.7&7.00 & 10.11 & 2040\\
Abell~763&0.085&09:12:35.21&+16:00:00.7&2.27 & 4.03 & 1480\\
Abell~780&0.053&09:18:05.67&-12:05:44.0&4.78 & 6.72 & 1780\\
Abell~795&0.136&09:24:05.30&+14:10:21.0&5.70 & 7.89 & 1820\\
Abell~961&0.124&10:16:22.93&+33:38:18.0&3.12 & 5.13 & 1590\\
Abell~990&0.144&10:23:39.86&+49:08:38.0&6.71 & 8.88 & 1890\\
ZwCl1023&0.143&10:25:58.02&+12:41:07.7&4.71& 6.92 & 1740\\
Abell~1033&0.126&10:31:44.31&+35:02:28.7&5.12 & 7.28 & 1780\\
Abell~1068&0.138&10:40:44.46&+39:57:11.4&5.94 & 8.13 & 1840\\
Abell~1132&0.136&10:58:23.71&+56:47:42.1&6.76 & 7.44 & 1790\\
Abell~1285&0.106&11:30:23.79&-14:34:52.8&4.66 & 6.76 & 1750\\
Abell~1348&0.119&11:40:59.30&-12:23:51.9&3.85 & 5.94 & 1670\\
Abell~1361&0.117&11:43:39.57&+46:21:20.2&4.95 & 7.09 & 1770\\
Abell~1413&0.143&11:55:18.01&+23:24:17.4&10.83 & 12.45 & 2120\\
ZwCl1215&0.075&12:17:41.12&+03:39:21.3&5.17 & 7.27 & 1810\\
Abell~1650&0.084&12:58:41.52&-01:45:40.9&5.66 & 9.18 & 1950\\
Abell~1651&0.085&12:59:22.39&-04:11:47.1&6.92 & 10.48 & 2040\\
Abell~1781&0.062&13:44:52.56&+29:46:15.3&3.79 & 5.73 & 1680\\
Abell~1795&0.063&13:48:52.58&+26:35:35.8&10.26 & 12.00 & 2150\\
Abell~1927&0.095&14:31:06.73&+25:38:00.6&2.30 & 4.08 & 1490\\
Abell~1991&0.059&14:54:31.50&+18:38:32.0&1.42 & 2.86 & 1340\\
Abell~2029&0.077&15:10:56.12&+05:44:40.8&17.44 & 16.57 & 2380\\
Abell~2033&0.082&15:11:26.55&+06:20:56.4&2.55& 4.38 & 1530\\
Abell~2050&0.118&15:16:17.94&+00:05:20.8&2.63 & 4.54 & 1530\\
Abell~2055&0.102&15:18:45.75&+06:13:55.9&3.80 & 5.83 & 1670\\
Abell~2064&0.108&15:20:52.23&+48:39:38.8&2.96 & 4.92 & 1570\\
MKW3S&0.045&15:21:51.85&+07:42:31.8&3.45 & 4.09 & 1510 \\
Abell~2065&0.073&15:22:29.16&+27:42:27.0&5.55 & 7.55 & 1840\\
Abell~2069&0.116&15:24:08.44&+29:52:54.6&3.45 & 5.49 & 1630\\
Abell~2142&0.091&15:58:20.08&+27:14:01.1&21.24 & 19.56 & 2510\\
Abell~2319&0.056&19:21:10.20&+43:56:43.8&15.78 & 15.61 & 2350\\
Abell~2409&0.148&22:00:53.51&+20:58:41.8&7.57 & 9.69 & 1950\\
Abell~2420&0.085&22:10:18.60&-12:10:12.3&4.64 & 6.67 & 1760\\
Abell~2426&0.098&22:14:31.60&-10:22:26.6&4.96 & 7.04 & 1780 \\
Abell~2440&0.091&22:23:56.94&-01:34:59.8&3.36 & 5.33 & 1630\\
Abell~2443&0.108&22:26:07.87&+17:21:24.4&3.22 & 5.22 & 1600\\
Abell~2495&0.078&22:50:19.80&+10:54:13.4&2.74 & 4.58 & 1550\\
Abell~2597&0.085&23:25:19.70&-12:07:27.7&6.62 & 8.62 & 1910\\
Abell~2627&0.126&23:36:42.06&+23:55:29.6&3.25 & 5.29 & 1600\\
RXSJ2344-04&0.079&23:44:18.23&-04:22:49.3&3.62 & 5.58 & 1660\\
Abell~2670&0.076&23:54:13.60&-10:25:07.5&2.28 & 4.03 & 1490\\
\enddata

\end{deluxetable}

\clearpage

\begin{deluxetable}{lccccccccccc}
\tiny
\tablecolumns{10}
\tablecaption{MENeaCS cluster SN Ia \label{table:SNItable}}
\tablehead{
\colhead{MENEACS ID} &\colhead{UT Date} & \colhead{Telescope/} & \colhead{Type} & \colhead{Subtype} & \colhead{$z$ (Gal)} & \colhead{$z$ (SNID)} & \colhead{Template} & \colhead{Phase} &\colhead{Exposure Time} \\
\colhead{} &\colhead{}&\colhead{Instrument}&\colhead{}&\colhead{}&\colhead{}&\colhead{}&\colhead{}&\colhead{}&\colhead{(s)}}\\
\startdata
Abell2443\_11\_08\_1 &2009-11-24.26 & MMT/BCS & Ia & ... & 0.111 & 0.1126 (0.0067) &SN95D & 7.5 (3.5) & 900.0 \\
Abell85\_11\_19\_0 & 2009-11-24.15 & MMT/BCS & Ia & norm & ... & 0.0534 (0.0037) & SN89B & 27.0 (14.4) &900.0  \\
Abell2443\_8\_07\_1 & 2009-09-19.13& MMT/BCS & Ia & ... & 0.1135 & 0.1100 (0.0078) & SN03cg& 24.0 (3.0) & 900.0  \\
Abell119\_5\_02\_1 & 2009-07-23.41& MMT/BCS & Ia & norm & 0.0444 & 0.0480 (0.0061) & SN02er& 2.2 (6.6) & 1800.0\\
Abell2319\_6\_04\_2 & 2009-07-23.38& MMT/BCS & Ia & norm & 0.0606 & 0.0631 (0.0053) & SN02er& -0.4 (5.5) & 900.0 \\
Abell1285\_8\_02\_4 & 2009-12-20.51& MMT/Hecto & Ia & norm & ... & 0.1071 (0.0061) & SN99ee & 0.0 (5.2) & 2700.0 \\
Abell1650\_9\_13\_0 & 2009-12-21.45 & MMT/Hecto & Ia & norm & ... & 0.0836 (0.0048) & SN02er & -0.8 (4.7) & 2700.0 \\
Abell1927\_5\_18\_0 & 2009-03-17.26& MMT/Hecto & Ia & norm & ... & 0.0945 (0.0038) & SN03du & 22.5 (16.6) & 3600.0 \\
Abell1991\_5\_23\_0 & 2009-04-27.23 & MMT/Hecto & Ia & norm & ... & 0.0580 (0.0029) & SN89B & 31.0 (13.3) & 3600.0  \\
Abell1991\_8\_13\_1 & 2009-07-16.15 & MMT/Hecto & Ia & norm & ... & 0.0526 (0.0034) & SN94D & 29.7 (11.5) & 2700.0  \\
Abell2409\_13\_14\_0 & 2009-12-20.06 & MMT/Hecto & Ia & norm & ... & 0.1458 (0.0068) & SN02er & 2.2 (6.2) & 3600.0 \\
Abell795\_7\_13\_2 & 2009-03-16.17& MMT/Hecto & Ia & ... & 0.1414 & 0.1388 (0.0079) & SN03cg & 28.4 (9.7) & 3600.0 \\ 
Abell795\_9\_13\_0 &  2009-11-24.40& MMT/BCS & Ia & norm & 0.1363 & 0.1348 (0.0027) & SN03cg & 31.1 (7.4) & 1200.0 \\ 
ZwCl0628\_4\_13\_0 & 2009-01-29.14 & MMT/Hecto & Ia & norm & ... & 0.0822 (0.0052) & SN03cg & -0.3 (5.5) & 3600.0\\
Abell2670\_9\_08\_1 \tablenotemark{$\dagger$} & 2009-12-21.06 & MMT/Hecto & Ia & ... & ... & 0.076 (0.0036) & SN04eo & 57.7 (21.3) & 3600.0 \\
Abell2033\_5\_19\_0 & 2009-05-03.22& MMT/BCS & Ia & norm & ... & 0.0837 (0.0039) & SN03du & 20.2 (23.4) & 1200.0\\
MKW3S\_4\_14\_0  & 2009-05-03.43& MMT/BCS & Ia & norm & ... & 0.0526 (0.0070) & SN94ae & 16.5 (75.8) &1200.0 \\
Abell21\_4\_19\_0 & 2008-11-05.07& MMT/BCS & Ia & norm & 0.0947 & 0.0913 (0.0110) & SN99ee & 7.9 (8.7) & 1200.0 \\
Abell644\_5\_19\_0	& 2008-12-19.32 & MMT/BCS & Ia & norm & ... & 0.0656 (0.0036) & SN89B & 25.4 (17.2) & 2400.0\\
RXCJ0736p39\_2\_09\_0 & 2008-12-29.28 & MMT/BCS & Ia & norm & 0.1225 & 0.1236 (0.0035)& SN94D & 18.9 (9.4) & 1200.0 \\
Abell85\_6\_08\_0 \tablenotemark{$\dagger$} & 2009-07-04.60& Gemini/GMOS & Ia & Ia-91bg & ... & 0.0617 (0.0007) & SN91bg & 34.1 (5.5) & 1800.0  \\
Abell2495\_5\_13\_0 \tablenotemark{$\dagger$}& 2009-06-18.58& Gemini/GMOS & Ia & norm & ... & 0.0796 (0.0032) & SN98aq & 83.3 (15.1) & 3000.0 \\
Abell399\_3\_14\_0 \tablenotemark{$\dagger$}& 2008-11-28.49& Gemini/GMOS & Ia & norm & ... & 0.0613 (0.0025) & SN03du & 14.4 (2.4) & 1200.0 \\
\enddata
\tablenotetext{$\dagger$}{Intracluster SN Ia, as presented in \citet{Sand11}}
\tablenotetext{}{Values in parentheses are the standard deviation as reported by SNID. }
\end{deluxetable}

\clearpage

\begin{deluxetable}{lcccccccc}
\tablefontsize{\tiny}
\tablecolumns{9}
\tablecaption{Cluster Luminosities \& Stellar Masses within $R_{200}$\label{table:clusterlum}}
\tablehead{
\colhead{} &  \multicolumn{2}{c}{g-band Luminosity ($10^{12} L_{\odot}$) } &   \multicolumn{2}{c}{r-band Luminosity ($10^{12} L_{\odot}$) } & \multicolumn{2}{c}{B-band Luminosity ($10^{12} L_{\odot}$) } & \multicolumn{2}{c}{Stellar Mass ($10^{12} M_{\odot}$)} \\
\colhead{Cluster} & \colhead{Total} & \colhead{RS} &   \colhead{Total} & \colhead{RS} &  \colhead{Total} & \colhead{RS} & \colhead{Total} & \colhead{RS}  }\\
\startdata

Abell2703 & $1.67\pm0.51$ & $1.00\pm0.02$ & $2.54\pm0.76$ & $1.53\pm0.04$ &$1.42\pm0.43$ & $0.84\pm0.02$ & $6.19\pm1.71$ & $3.94\pm0.12$ \\
Abell7 & $3.61\pm0.64$ & $1.90\pm0.04$ & $5.08\pm0.94$ & $2.76\pm0.09$ &$3.16\pm0.59$ & $1.65\pm0.04$ & $12.0\pm2.07$ & $6.81\pm0.24$ \\
Abell21 & $1.89\pm0.45$ & $1.20\pm0.03$ & $2.67\pm0.60$ & $1.78\pm0.07$ &$1.66\pm0.40$ & $1.04\pm0.03$ & $6.38\pm1.31$ & $4.44\pm0.18$ \\
Abell85 & $3.16\pm0.55$ & $2.10\pm0.02$ & $4.78\pm0.95$ & $3.09\pm0.10$ &$2.70\pm0.50$ & $1.81\pm0.03$ & $11.2\pm2.05$ & $7.54\pm0.26$ \\
Abell119 & $2.51\pm0.29$ & $2.12\pm0.02$ & $3.87\pm0.59$ & $3.03\pm0.07$ &$2.13\pm0.27$ & $1.85\pm0.02$ & $9.20\pm1.25$ & $7.30\pm0.17$ \\
Abell133 & $1.49\pm0.40$ & $1.24\pm0.01$ & $2.35\pm0.65$ & $1.85\pm0.07$ &$1.25\pm0.34$ & $1.06\pm0.02$ & $5.67\pm1.41$ & $4.57\pm0.18$ \\
RXCJ0132m08 & $1.77\pm0.79$ & $0.74\pm0.04$ & $2.46\pm1.12$ & $1.16\pm0.08$ &$1.56\pm0.66$ & $0.62\pm0.04$ & $5.62\pm2.52$ & $3.00\pm0.21$ \\
Abell399 & $4.90\pm0.48$ & $3.40\pm0.07$ & $7.33\pm0.93$ & $4.97\pm0.11$ &$4.20\pm0.47$ & $2.94\pm0.07$ & $17.4\pm2.02$ & $12.1\pm0.28$ \\
RXCJ0352p19 & $2.04\pm0.62$ & $0.77\pm0.05$ & $2.67\pm0.92$ & $1.13\pm0.11$ &$1.84\pm0.57$ & $0.67\pm0.05$ & $5.75\pm2.01$ & $2.76\pm0.28$ \\
Abell478 & $8.57\pm1.04$ & $2.64\pm0.07$ & $10.5\pm1.65$ & $3.76\pm0.20$ &$7.92\pm1.07$ & $2.31\pm0.08$ & $21.7\pm3.51$ & $8.90\pm0.49$ \\
Abell553 & $1.93\pm0.27$ & $0.83\pm0.01$ & $2.53\pm0.41$ & $1.21\pm0.03$ &$1.74\pm0.26$ & $0.72\pm0.01$ & $5.63\pm0.87$ & $2.97\pm0.08$ \\
ZwCl0628 & $8.20\pm0.49$ & $2.17\pm0.03$ & $9.43\pm0.78$ & $3.08\pm0.09$ &$7.79\pm0.55$ & $1.90\pm0.03$ & $19.1\pm1.70$ & $7.39\pm0.24$ \\
RXCJ0736p39 & $1.66\pm0.80$ & $0.90\pm0.04$ & $2.50\pm1.14$ & $1.38\pm0.07$ &$1.42\pm0.63$ & $0.76\pm0.03$ & $5.99\pm2.53$ & $3.56\pm0.20$ \\
Abell644 & $4.29\pm0.72$ & $2.00\pm0.03$ & $6.12\pm1.17$ & $3.02\pm0.11$ &$3.75\pm0.68$ & $1.70\pm0.04$ & $14.1\pm2.55$ & $7.48\pm0.30$ \\
Abell646 & $2.85\pm0.75$ & $1.35\pm0.04$ & $4.19\pm1.09$ & $2.13\pm0.07$ &$2.46\pm0.65$ & $1.13\pm0.04$ & $10.1\pm2.45$ & $5.54\pm0.20$ \\
Abell655 & $3.95\pm0.72$ & $2.40\pm0.04$ & $5.98\pm1.11$ & $3.72\pm0.07$ &$3.37\pm0.64$ & $2.02\pm0.04$ & $14.7\pm2.49$ & $9.63\pm0.20$ \\
Abell754 & $3.83\pm0.32$ & $3.11\pm0.07$ & $6.29\pm0.98$ & $4.64\pm0.11$ &$3.16\pm0.33$ & $2.67\pm0.07$ & $15.0\pm2.12$ & $11.3\pm0.27$ \\
Abell763 & $0.69\pm0.47$ & $0.45\pm0.05$ & $1.06\pm0.70$ & $0.66\pm0.09$ &$0.58\pm0.35$ & $0.39\pm0.04$ & $2.42\pm1.53$ & $1.61\pm0.24$ \\
Abell780 & $1.52\pm0.48$ & $0.91\pm0.01$ & $2.33\pm0.81$ & $1.35\pm0.08$ &$1.29\pm0.40$ & $0.78\pm0.02$ & $5.24\pm1.74$ & $3.25\pm0.20$ \\
Abell795 & $4.51\pm1.01$ & $2.61\pm0.04$ & $6.99\pm1.43$ & $4.17\pm0.09$ &$3.81\pm0.87$ & $2.18\pm0.04$ & $17.7\pm3.22$ & $10.9\pm0.24$ \\
Abell961 & $2.43\pm0.67$ & $1.75\pm0.04$ & $3.84\pm0.98$ & $2.71\pm0.07$ &$2.04\pm0.56$ & $1.48\pm0.04$ & $9.71\pm2.19$ & $6.99\pm0.20$ \\
Abell990 & $3.04\pm1.05$ & $2.00\pm0.05$ & $4.63\pm1.47$ & $3.12\pm0.10$ &$2.59\pm0.87$ & $1.69\pm0.05$ & $11.4\pm3.30$ & $8.10\pm0.27$ \\
ZwCl1023 & $2.20\pm1.04$ & $0.99\pm0.06$ & $3.25\pm1.58$ & $1.56\pm0.12$ &$1.90\pm0.84$ & $0.84\pm0.06$ & $7.47\pm3.57$ & $4.02\pm0.33$ \\
Abell1033 & $2.88\pm0.83$ & $1.69\pm0.04$ & $4.44\pm1.17$ & $2.63\pm0.08$ &$2.44\pm0.70$ & $1.43\pm0.04$ & $11.1\pm2.60$ & $6.83\pm0.23$ \\
Abell1068 & $3.30\pm1.06$ & $1.03\pm0.05$ & $4.75\pm1.57$ & $1.60\pm0.09$ &$2.87\pm0.92$ & $0.87\pm0.04$ & $11.1\pm3.53$ & $4.14\pm0.26$ \\
Abell1132 & $3.81\pm0.91$ & $2.36\pm0.04$ & $5.78\pm1.30$ & $3.73\pm0.08$ &$3.25\pm0.79$ & $1.98\pm0.04$ & $14.3\pm2.92$ & $9.74\pm0.21$ \\
Abell1285 & $3.62\pm0.71$ & $1.88\pm0.04$ & $5.43\pm1.16$ & $2.87\pm0.07$ &$3.10\pm0.64$ & $1.60\pm0.04$ & $13.1\pm2.59$ & $7.34\pm0.20$ \\
Abell1348 & $3.05\pm0.73$ & $1.20\pm0.05$ & $4.55\pm1.09$ & $1.84\pm0.09$ &$2.62\pm0.64$ & $1.02\pm0.04$ & $11.0\pm2.43$ & $4.73\pm0.24$ \\
Abell1361 & $1.52\pm0.78$ & $0.83\pm0.04$ & $2.19\pm1.16$ & $1.25\pm0.08$ &$1.32\pm0.63$ & $0.71\pm0.04$ & $4.93\pm2.59$ & $3.18\pm0.22$ \\
Abell1413 & $6.18\pm1.41$ & $2.65\pm0.08$ & $8.83\pm2.08$ & $4.09\pm0.15$ &$5.39\pm1.27$ & $2.25\pm0.08$ & $21.0\pm4.67$ & $10.5\pm0.41$ \\
ZwCl1215 & $2.01\pm0.66$ & $1.70\pm0.04$ & $3.31\pm1.05$ & $2.59\pm0.10$ &$1.65\pm0.54$ & $1.44\pm0.04$ & $8.24\pm2.30$ & $6.54\pm0.28$ \\
Abell1650 & $2.59\pm0.75$ & $1.91\pm0.05$ & $4.21\pm1.26$ & $2.85\pm0.12$ &$2.15\pm0.62$ & $1.64\pm0.05$ & $10.1\pm2.76$ & $7.05\pm0.31$ \\
Abell1651 & $3.43\pm0.81$ & $2.24\pm0.04$ & $4.96\pm1.25$ & $3.33\pm0.13$ &$2.98\pm0.72$ & $1.93\pm0.04$ & $11.6\pm2.71$ & $8.23\pm0.34$ \\
Abell1781 & $1.69\pm0.50$ & $0.85\pm0.02$ & $2.50\pm0.79$ & $1.28\pm0.08$ &$1.45\pm0.43$ & $0.72\pm0.02$ & $5.78\pm1.71$ & $3.12\pm0.20$ \\
Abell1795 & $3.52\pm0.71$ & $2.02\pm0.05$ & $4.98\pm1.19$ & $2.91\pm0.13$ &$3.08\pm0.66$ & $1.75\pm0.06$ & $11.0\pm2.55$ & $6.98\pm0.32$ \\
Abell1927 & $2.58\pm0.52$ & $1.33\pm0.04$ & $3.58\pm0.78$ & $2.00\pm0.08$ &$2.28\pm0.48$ & $1.14\pm0.04$ & $8.29\pm1.72$ & $5.05\pm0.23$ \\
Abell1991 & $1.20\pm0.27$ & $0.80\pm0.02$ & $1.83\pm0.47$ & $1.18\pm0.04$ &$1.02\pm0.24$ & $0.68\pm0.02$ & $4.24\pm1.03$ & $2.86\pm0.11$ \\
Abell2029 & $7.12\pm1.20$ & $3.31\pm0.09$ & $10.2\pm1.95$ & $4.96\pm0.19$ &$6.20\pm1.12$ & $2.84\pm0.09$ & $23.9\pm4.29$ & $12.3\pm0.51$ \\
Abell2033 & $3.19\pm0.52$ & $1.88\pm0.04$ & $4.61\pm0.84$ & $2.80\pm0.09$ &$2.77\pm0.48$ & $1.61\pm0.04$ & $10.8\pm1.83$ & $6.97\pm0.24$ \\
Abell2050 & $2.92\pm0.55$ & $1.59\pm0.02$ & $4.28\pm0.78$ & $2.44\pm0.04$ &$2.52\pm0.50$ & $1.35\pm0.02$ & $10.4\pm1.76$ & $6.29\pm0.13$ \\
Abell2055 & $3.72\pm0.60$ & $1.85\pm0.03$ & $5.25\pm0.89$ & $2.78\pm0.07$ &$3.26\pm0.56$ & $1.58\pm0.03$ & $12.4\pm1.94$ & $6.99\pm0.18$ \\
Abell2064 & $1.40\pm0.50$ & $1.27\pm0.04$ & $2.42\pm0.88$ & $1.90\pm0.07$ &$1.13\pm0.39$ & $1.09\pm0.04$ & $5.84\pm1.95$ & $4.72\pm0.20$ \\
MKW3S & $0.86\pm0.25$ & $0.75\pm0.01$ & $1.52\pm0.51$ & $1.13\pm0.05$ &$0.69\pm0.20$ & $0.64\pm0.02$ & $3.58\pm1.11$ & $2.74\pm0.13$ \\
Abell2065 & $3.66\pm0.46$ & $2.40\pm0.03$ & $5.30\pm0.68$ & $3.56\pm0.06$ &$3.18\pm0.43$ & $2.07\pm0.03$ & $12.7\pm1.47$ & $8.87\pm0.17$ \\
Abell2069 & $5.06\pm0.74$ & $3.10\pm0.04$ & $7.33\pm1.09$ & $4.67\pm0.08$ &$4.39\pm0.68$ & $2.65\pm0.04$ & $17.8\pm2.43$ & $11.8\pm0.23$ \\
Abell2142 & $7.06\pm1.20$ & $3.98\pm0.06$ & $10.3\pm1.81$ & $6.07\pm0.15$ &$6.09\pm1.09$ & $3.38\pm0.06$ & $25.2\pm4.00$ & $15.4\pm0.41$ \\
Abell2319 & $15.9\pm0.62$ & $4.69\pm0.02$ & $19.4\pm1.04$ & $6.53\pm0.09$ &$14.8\pm0.68$ & $4.13\pm0.03$ & $41.4\pm2.24$ & $15.5\pm0.22$ \\
Abell2409 & $5.13\pm1.25$ & $2.51\pm0.08$ & $7.50\pm1.75$ & $3.90\pm0.14$ &$4.43\pm1.10$ & $2.12\pm0.07$ & $18.3\pm3.93$ & $10.1\pm0.37$ \\
Abell2420 & $2.62\pm0.62$ & $1.53\pm0.04$ & $3.94\pm0.98$ & $2.31\pm0.11$ &$2.24\pm0.54$ & $1.31\pm0.05$ & $9.37\pm2.15$ & $5.76\pm0.28$ \\
Abell2426 & $4.77\pm0.73$ & $2.47\pm0.04$ & $6.64\pm1.12$ & $3.63\pm0.11$ &$4.20\pm0.69$ & $2.13\pm0.05$ & $15.4\pm2.46$ & $9.04\pm0.30$ \\
Abell2440 & $1.95\pm0.57$ & $1.70\pm0.03$ & $3.20\pm0.93$ & $2.57\pm0.08$ &$1.61\pm0.47$ & $1.45\pm0.03$ & $7.96\pm2.04$ & $6.49\pm0.21$ \\
Abell2443 & $2.81\pm0.56$ & $1.66\pm0.02$ & $4.31\pm0.89$ & $2.53\pm0.06$ &$2.39\pm0.50$ & $1.41\pm0.02$ & $10.6\pm2.00$ & $6.46\pm0.17$ \\
Abell2495 & $1.59\pm0.47$ & $0.90\pm0.03$ & $2.41\pm0.76$ & $1.38\pm0.09$ &$1.36\pm0.40$ & $0.77\pm0.03$ & $5.63\pm1.67$ & $3.42\pm0.23$ \\
Abell2597 & $1.64\pm0.81$ & $1.14\pm0.04$ & $2.72\pm1.31$ & $1.74\pm0.10$ &$1.35\pm0.61$ & $0.97\pm0.04$ & $6.51\pm2.91$ & $4.37\pm0.28$ \\
Abell2627 & $2.87\pm0.58$ & $1.17\pm0.03$ & $3.97\pm0.86$ & $1.79\pm0.06$ &$2.53\pm0.54$ & $0.99\pm0.03$ & $9.25\pm1.92$ & $4.59\pm0.16$ \\
RXSJ2344m04 & $1.52\pm0.60$ & $1.19\pm0.04$ & $2.51\pm0.97$ & $1.82\pm0.08$ &$1.25\pm0.48$ & $1.02\pm0.04$ & $6.05\pm2.12$ & $4.55\pm0.22$ \\
Abell2670 & $3.41\pm0.44$ & $2.38\pm0.02$ & $5.15\pm0.76$ & $3.56\pm0.07$ &$2.91\pm0.41$ & $2.04\pm0.02$ & $12.4\pm1.65$ & $8.85\pm0.18$ \\

\enddata
\end{deluxetable}

\clearpage

\begin{deluxetable}{lccccclc}
\tiny
\tablecolumns{8}
\tablecaption{MENeaCS SN Ia host properties \label{hosttab}}
\tablehead{
\colhead{MENEACS ID} &\colhead{$\alpha$} & \colhead{$\delta$} & \colhead{$r$} & \colhead{$g-r$} & \colhead{$\Delta (g-r)_{rs}$} & \colhead{$R_{clus}$} & \colhead{$R_{clus}/R_{200}$} \\
\colhead{} &\colhead{(J2000.0)}&\colhead{(J2000.0)}&\colhead{(mag)}&\colhead{(mag)} & \colhead{(mag)}&\colhead{(kpc)}&\colhead{}}\\
\startdata
Abell2443\_11\_08\_1 &22:25:59.56 & +17:13:32.8 & $16.70\pm0.02$ & $1.00\pm0.04$ & $-0.02$ & 960 & 0.60 \\
Abell85\_11\_19\_0 & 00:41:12.91 & -09:09:00.3 & $15.88\pm0.02$ & $0.87\pm0.04$ & $-0.03$ & 840 & 0.53 \\
Abell2443\_8\_07\_1 & 22:27:07.87 & +17:14:36.6 & $17.48\pm0.02$ & $0.83\pm0.04$ & $-0.17$ & 1880 & 1.17 \\
Abell119\_5\_02\_1 & 00:56:46.57 & -01:35:47.4 & $15.12\pm0.02$ & $0.93\pm0.03$ & 0.064 & 1140 & 0.70\\
Abell2319\_6\_04\_2 & 19:20:33.97 & +43:33:47.5 & $16.35\pm0.02$ & $0.60\pm0.04$ & $-0.26$ & 1550 & 0.66\\
Abell1285\_8\_02\_4 & 11:31:04.48 & -14:54:59.1 & $17.94\pm0.02$ & $0.88\pm0.03$ & $-0.09$ & 2610 & 1.49 \\
Abell1927\_5\_18\_0 & 14:31:01.13 & +25:46:57.9 & $18.58\pm0.02$ & $0.61\pm0.04$ & $-0.31$ & 960 & 0.64 \\
Abell1991\_5\_23\_0 & 14:54:15.93 & +18:57:51.6 & $15.61\pm0.02$ & $0.86\pm0.04$ & $-0.07$ & 1340 & 0.99 \\
Abell1991\_8\_13\_1 & 14:54:32.91 & +18:39:39.4 & $22.25\pm0.12$ & $0.66\pm0.06$ & $-0.01$  & 80 & 0.06 \\
Abell2409\_13\_14\_0 & 21:59:46.64 & +21:02:20.4 & $16.76\pm0.02$ & $1.10\pm0.04$ & $-0.02$ & 2500 & 1.28\\
Abell795\_7\_13\_2 & 09:23:59.34 & +14:09:50.1 & $17.74\pm0.02$ & $1.07\pm0.03$ & $-0.00$ & 220 & 0.11 \\
Abell795\_9\_13\_0 & 09:23:59.79 & +14:06:47.9 & $19.15\pm0.02$ & $0.97\pm0.03$ & $-0.06$ & 550 & 0.27 \\
ZwCl0628\_4\_13\_0  & 06:31:46.70 & +24:55:13.2 & $18.86\pm0.03$ & $0.43\pm0.05$ & $-0.40$ & 730 & 0.40\\
Abell2670\_9\_08\_1 & 23:53:50.80 & -10:42:40.5 & $19.85\pm0.02$ & $0.56\pm0.04$ & $-0.23$ & 1600 & 1.07 \\
Abell2033\_5\_19\_0 & 15:10:59.45 & +06:33:12.6 & $18.57\pm0.02$ & $0.59\pm0.04$ & $-0.27$ & 1290 & 0.85\\
MKW3S\_4\_14\_0  &15:20:38.05 & +07:39:34.9 & $16.05\pm0.02$ & $0.84\pm0.04$ & $-0.04$ & 980 & 0.60 \\
Abell21\_4\_19\_0 & 00:19:31.13 & +28:51:43.5 & $18.56\pm0.02$ & $0.36\pm0.03$ & $-0.55$ & 2000 & 1.30 \\
Abell644\_5\_19\_0	& 08:16:12.36 & -07:21:30.7 & $21.28\pm0.05$ & $0.66\pm0.08$ & $-0.08$ & 1640 & 0.96 \\
RXCJ0736p39\_2\_09\_0 & 07:35:13.97 & +39:16:53.2 & $18.67\pm0.02$ & $0.48\pm0.03$ & $-0.50$ & 2320 & 1.40 \\
\enddata
\end{deluxetable}

\clearpage


\begin{deluxetable}{lllllll}
\tablecolumns{7}
\tablecaption{MENeaCS Cluster SN\,Ia Rates \label{table:SNrates}}
\tablehead{ 
\colhead{Gal} & \colhead{Rad} & \colhead{$N_{SN}$} & \multicolumn{4}{c}{SN Rate}  \\
\colhead{Type\tablenotemark{a}}   & \colhead{}            & \colhead{}            & \colhead{SNug\tablenotemark{b}} & \colhead{SNur\tablenotemark{c}} & \colhead{SNuB\tablenotemark{d}}     &  \colhead{SNuM\tablenotemark{e}} \\
}
\startdata
\hline
\multicolumn{5}{l}{With the SN\,Ia LF of Li et al. (2010), no subtypes.} \\
\hline
RS &  1 Mpc          & 6 & $\rm 0.153_{-0.055}^{+0.078}$$_{-0.014}^{+0.017}$ & $\rm 0.103_{-0.037}^{+0.051}$$_{-0.009}^{+0.011}$ & $\rm 0.180_{-0.065}^{+0.087}$$_{-0.016}^{+0.020}$ & $\rm 0.041_{-0.015}^{+0.019}$$_{-0.004}^{+0.005}$ \\
All                 &  1 Mpc          & 11 & $\rm 0.168_{-0.048}^{+0.058}$$_{-0.015}^{+0.019}$ & $\rm 0.117_{-0.034}^{+0.039}$$_{-0.011}^{+0.013}$ & $\rm 0.193_{-0.054}^{+0.067}$$_{-0.018}^{+0.021}$ & $\rm 0.049_{-0.014}^{+0.016}$$_{-0.004}^{+0.005}$ \\
RS &  $\rm  R_{200}$  & 9 & $\rm 0.155_{-0.054}^{+0.058}$$_{-0.036}^{+0.017}$ & $\rm 0.103_{-0.036}^{+0.038}$$_{-0.023}^{+0.011}$ & $\rm 0.180_{-0.063}^{+0.066}$$_{-0.041}^{+0.020}$ & $\rm 0.041_{-0.015}^{+0.015}$$_{-0.010}^{+0.005}$ \\
All                 &  $\rm  R_{200}$  & 16 & $\rm 0.144_{-0.035}^{+0.039}$$_{-0.034}^{+0.044}$ & $\rm 0.100_{-0.024}^{+0.027}$$_{-0.021}^{+0.026}$ & $\rm 0.164_{-0.041}^{+0.048}$$_{-0.039}^{+0.056}$ & $\rm 0.042_{-0.010}^{+0.012}$$_{-0.008}^{+0.010}$ \\
\hline
\multicolumn{5}{l}{With the SN\,Ia LF of Li et al. (2010), including subtypes.} \\
\hline
RS\tablenotemark{e} &  1 Mpc          & 6 & $\rm 0.194_{-0.086}^{+0.143}$$_{-0.018}^{+0.022}$ & $\rm 0.127_{-0.056}^{+0.103}$$_{-0.012}^{+0.014}$ & $\rm 0.215_{-0.092}^{+0.166}$$_{-0.020}^{+0.024}$ & $\rm 0.052_{-0.024}^{+0.036}$$_{-0.005}^{+0.006}$ \\
All                 &  1 Mpc          & 11 & $\rm 0.173_{-0.053}^{+0.094}$$_{-0.016}^{+0.019}$ & $\rm 0.124_{-0.040}^{+0.083}$$_{-0.011}^{+0.014}$ & $\rm 0.204_{-0.063}^{+0.131}$$_{-0.019}^{+0.023}$ & $\rm 0.050_{-0.015}^{+0.024}$$_{-0.005}^{+0.006}$ \\
RS\tablenotemark{e} &  $\rm  R_{200}$  & 9 & $\rm 0.189_{-0.074}^{+0.127}$$_{-0.044}^{+0.021}$ & $\rm 0.125_{-0.048}^{+0.094}$$_{-0.028}^{+0.014}$ & $\rm 0.212_{-0.079}^{+0.151}$$_{-0.050}^{+0.024}$ & $\rm 0.049_{-0.018}^{+0.041}$$_{-0.010}^{+0.005}$ \\
All                 &  $\rm  R_{200}$  & 16 & $\rm 0.149_{-0.041}^{+0.068}$$_{-0.034}^{+0.046}$ & $\rm 0.106_{-0.029}^{+0.061}$$_{-0.022}^{+0.030}$ & $\rm 0.177_{-0.051}^{+0.098}$$_{-0.046}^{+0.056}$ & $\rm 0.043_{-0.011}^{+0.020}$$_{-0.008}^{+0.012}$ \\
\hline
\multicolumn{5}{l}{With the SN\,Ia LF of Cappellaro et al. (1999).} \\
\hline
RS\tablenotemark{e} &  1 Mpc          & 6 & $\rm 0.158_{-0.058}^{+0.080}$$_{-0.014}^{+0.018}$ & $\rm 0.104_{-0.037}^{+0.053}$$_{-0.009}^{+0.012}$ & $\rm 0.183_{-0.066}^{+0.094}$$_{-0.017}^{+0.020}$ & $\rm 0.042_{-0.015}^{+0.021}$$_{-0.004}^{+0.005}$ \\
All                 &  1 Mpc          & 11 & $\rm 0.171_{-0.049}^{+0.066}$$_{-0.016}^{+0.019}$ & $\rm 0.122_{-0.036}^{+0.041}$$_{-0.011}^{+0.014}$ & $\rm 0.199_{-0.059}^{+0.070}$$_{-0.018}^{+0.022}$ & $\rm 0.050_{-0.015}^{+0.018}$$_{-0.005}^{+0.006}$ \\
RS\tablenotemark{e} &  $\rm  R_{200}$  & 9 & $\rm 0.160_{-0.055}^{+0.056}$$_{-0.037}^{+0.018}$ & $\rm 0.107_{-0.037}^{+0.040}$$_{-0.023}^{+0.012}$ & $\rm 0.189_{-0.065}^{+0.067}$$_{-0.044}^{+0.021}$ & $\rm 0.043_{-0.016}^{+0.015}$$_{-0.010}^{+0.005}$ \\
All                 &  $\rm  R_{200}$  & 16 & $\rm 0.148_{-0.035}^{+0.044}$$_{-0.034}^{+0.049}$ & $\rm 0.104_{-0.026}^{+0.028}$$_{-0.023}^{+0.027}$ & $\rm 0.170_{-0.041}^{+0.049}$$_{-0.041}^{+0.059}$ & $\rm 0.044_{-0.011}^{+0.012}$$_{-0.009}^{+0.010}$ \\
\hline
\enddata
\tablenotetext{}{{\bf Note} -- The first and second uncertainty intervals are the statistical and systematic errors, respectively. }
\tablenotetext{a}{We present our SN Ia rates in both our red sequence (RS) galaxies and over all cluster galaxies (All).}
\tablenotetext{b}{SNug $\equiv$ SNe(100 yr $10^{10}$ $L_{g,\odot}$)$^{-1}$}
\tablenotetext{c}{SNur $\equiv$ SNe(100 yr $10^{10}$ $L_{r,\odot}$)$^{-1}$}
\tablenotetext{d}{SNuB $\equiv$ SNe(100 yr $10^{10}$ $L_{B,\odot}$)$^{-1}$}
\tablenotetext{e}{SNuM $\equiv$ SNe(100 yr $10^{10}$ $M_{\odot}$)$^{-1}$}
\end{deluxetable}

\clearpage

\begin{deluxetable}{lll}
\tablecolumns{3}
\tablecaption{Summary of Cluster SN Rate Measurements  \label{table:SNsumm}}
\tablehead{
\colhead{Reference} & \colhead{Redshift}& \colhead{SN Rate\tablenotemark{$\dagger$}} \\
\colhead{} & \colhead{(Range)}  &\colhead{SNuM} \\
}
\startdata
Overall Cluster Rate\\
\hline
This work \tablenotemark{a}& 0.10 (0.05 -- 0.15) & $0.049^{+0.021}_{-0.018}$\\
Mannucci et al. 2008\tablenotemark{b} & 0.020 (0.005 -- 0.04)  & $0.057^{+0.021}_{-0.016}$\\
 Dilday et al. 2010 & 0.084 (0.03 -- 0.17)  & $0.060^{+0.029}_{-0.021}$\\
Sharon et al. 2007 & 0.15 (0.06 -- 0.19) &  $0.098^{+0.068}_{-0.048}$\\
 Dilday et al. 2010 & 0.225 (0.100 -- 0.300)  & $0.088^{+0.027}_{-0.020}$\\
 Graham et al. 2008 \tablenotemark{c}& 0.46 (0.20 -- 0.60) & $0.177^{+0.212}_{-0.124}$\\
 Sharon et al. 2010\tablenotemark{d} & 0.60 (0.50 -- 0.89)  & $0.151^{+0.138}_{-0.116}$ \\
 Barbary et al. 2010 & 1.12 (0.90 -- 1.45) &  $0.36^{+0.23}_{-0.19}$ \\
 
\hline
\hline
Cluster early-type galaxies or red sequence galaxies\\
\hline
This work\tablenotemark{a} & 0.10 (0.05 -- 0.15) & $0.041^{+0.024}_{-0.019}$ \\
Mannucci et al. 2008 \tablenotemark{b}& 0.020 (0.005 -- 0.04)  & $0.066^{+0.027}_{-0.020}$\\
 Dilday et al. 2010\tablenotemark{e} & 0.084 (0.03 -- 0.17)  & $0.051^{+0.031}_{-0.021}$\\
 Dilday et al. 2010\tablenotemark{e} & 0.225 (0.100 -- 0.300)  & $0.080^{+0.028}_{-0.020}$\\
Graham et al. 2008 \tablenotemark{c}&  0.46 (0.20 -- 0.60) & $0.085^{+0.205}_{-0.051}$\\
Barbary et al. 2010 \tablenotemark{f}& 1.12 (0.90 -- 1.45) & $0.37^{+0.22}_{-0.18}$\\

 \enddata
\tablenotetext{$\dagger$}{Quoted SN rate errors are the sum of statistical and systematic components, as reported}
\tablenotetext{a}{Our values are those within 1 Mpc, and without considering SN Ia subtypes, as reported in Table~\ref{table:SNrates}.}
\tablenotetext{b}{We use the cluster SN rate values within $R<0.5$ Mpc presented by \citet{Mannucci08}}
\tablenotetext{c}{We present the SN Ia rate numbers of \citet{Graham08} at a clustercentric distance of 1.5 Mpc.  We also rescale their values by a factor of 1.77 for a `diet' Salpeter IMF, as determined by \citet{Maoz10clus}}
\tablenotetext{d}{We present the re-scaled SN rate values from \citet{Sharon10}, as calculated by \citet{Barbary10}.}
\tablenotetext{e}{To get the Dilday et al. (2010) early-type rates, we used their 'early type' SNur rates, and divided by 3, their chosen $M/L_{r}$.}
\tablenotetext{f}{The SN Ia rate in 'red sequence' galaxies from \citet{Barbary10}.}
\end{deluxetable}

\clearpage

\clearpage

\clearpage

\clearpage

\clearpage

\clearpage

\clearpage

\clearpage

\clearpage

\clearpage

\clearpage

\clearpage

\clearpage

\clearpage

\end{document}